\documentclass[11pt]{article}
\usepackage{amsmath,amsfonts,amssymb,graphicx,epsfig,pdflscape,multirow,theorem,gensymb,nccmath}
\usepackage[normalem]{ulem}
\usepackage[matrix,arrow,curve]{xy} 
\numberwithin{equation}{section}
\graphicspath{{./eps/}}
\allowdisplaybreaks

\usepackage[nosort]{cite}
\usepackage[usenames]{color}
\usepackage{pstricks}
\usepackage[backref=false]{hyperref}
\hypersetup{
colorlinks=true,
citecolor=red,
linkcolor=darkblue
}
\definecolor{darkblue}{rgb}{0,0,.8}
\definecolor{red}{rgb}{1,0,0}
\newcommand{\nc}{\newcommand}
\def\tmat#1{\mbox{\tiny{\mbox{$\mypmatrix{#1}$}}}}
\def\smat#1{\mbox{\scriptsize{\mbox{$\mypmatrix{#1}$}}}}
\def\mypmatrix#1{\begin{pmatrix}#1\end{pmatrix}}

\nc{\fh}{\hat{f}}
\nc{\muh}{\hat{\mu}}
\nc{\nuh}{\hat{\nu}}
\long\def\ignore#1{}
\def\calN{{\cal N}}
\def\calV{{\cal V}}
\def\dcol{\facegridyellow{1}{2}}
\nc{\bib}{\bibitem}
\nc{\al}{\alpha}
\nc{\mch}{\mathrm{ch}}
\nc{\g}{\gamma}
\nc{\G}{\Gamma}
\nc{\D}{\Delta}
\nc{\eps}{\epsilon}
\nc{\la}{\lambda}
\nc{\La}{\Lambda}
\nc{\var}{\varphi}
\nc{\pa}{\partial}
\nc{\nn}{\nonumber \\ }
\nc{\hf}{\frac{1}{2}}
\nc{\dz}{\frac{dz}{2\pi i}}
\nc{\bin}[2]{\left(\!\!\!\begin{array}{c} {#1}\\ {#2} \end{array}\!\!\!\right)}
\nc{\be}{\begin{equation}}
\nc{\ee}{\end{equation}}
\nc{\bea}{\begin{eqnarray}}
\nc{\eea}{\end{eqnarray}}
\nc{\bra}[1]{\langle {#1}|}
\nc{\ket}[1]{|{#1}\rangle}
\def\Re{\mathop{\rm Re}\nolimits}

\nc{\chit}{\raisebox{0.25ex}{$\chi$}}
\nc{\ch}{{\rm ch}}
\nc{\Bb}{\mbox{\boldmath $B$}}
\nc{\Fb}{\mbox{\boldmath $F$}}
\nc{\Hb}{\mbox{\boldmath $H$}}
\nc{\Ib}{\mbox{\boldmath $I$}}
\nc{\Jb}{\mbox{\boldmath $J$}}
\nc{\Rb}{\mbox{\boldmath $R$}}
\nc{\Tb}{\mbox{\boldmath $T$}}

\nc{\Hc}{{\cal H}}
\nc{\Rc}{{\cal R}}
\nc{\Lc}{{\cal L}}
\nc{\Vc}{{\cal V}}
\nc{\rbar}{\bar{r}}
\nc{\qbar}{\bar{q}}
\nc{\kbar}{\bar{k}}
\def\floor#1{\lfloor #1\rfloor}
\nc{\sbin}[2]{\Big\{\!\begin{array}{c} {#1}\\ {#2} \end{array}\!\Big\}}
\nc{\sbinlr}[2]{\Big\langle\!\!\begin{array}{c} {#1}\\ {#2} \end{array}\!\!\Big\rangle}
\def\vvdots{\mathinner{\mkern1mu\raise1pt\vbox{\kern7pt\hbox{.}}\mkern2mu
  \raise4pt\hbox{.}\mkern2mu\raise7pt\hbox{.}\mkern1mu}}
\nc{\gauss}[2]{\left[\!\!\begin{array}{c} {#1}\\ {#2} \end{array}\!\!\right]_{\!q}}
\nc{\gaussbar}[2]{\left[\!\!\begin{array}{c} {#1}\\ {#2} \end{array}\!\!\right]_{\!\qbar}}
\nc{\perm}[2]{\left[\!\!\begin{array}{c} {#1}\\ \\ {#2} \end{array}\!\!\right]}
\nc{\bino}[2]{\left(\!\!\begin{array}{c} {#1}\\ {#2} \end{array}\!\!\right)}
\def\half {\mbox{$\textstyle \frac{1}{2}$}}
\def\vec#1{\mbox {\boldmath $#1$}}
\def\svec#1{\mbox {\scriptsize\boldmath $#1$}}
\definecolor{lightblue}{rgb}{.7,.7,1}
\definecolor{purple}{rgb}{1,0,1}
\definecolor{lightlightblue}{rgb}{.92,.92,1}
\definecolor{lightyellow}{rgb}{.99,.99,.85}

\def\facegrid#1#2{
\psset{linewidth=.25pt}
\psframe[fillstyle=solid,fillcolor=lightlightblue,linewidth=0pt]#1#2
\psset{linewidth=.25pt}
\multirput(0,0)(1,0){6}{\psline[linewidth=.5pt](1,1)(1,2)}}

\def\facegridpink#1#2{
\psframe[fillstyle=solid,fillcolor=lightpurple,linewidth=0pt]#1#2
\multirput(0,1)(1,0){6}{\psline[linewidth=.5pt](1,1)(1,2)}}

\definecolor{sky}{rgb}{.4,.8,1}
\definecolor{orange}{rgb}{1,.4,0}
\definecolor{green}{rgb}{0,1,0}
\definecolor{brightblue}{rgb}{0,1,1}
\definecolor{mediumblue}{rgb}{.34,.446,1}
\definecolor{greyblue}{rgb}{.345,.345,.7424}
\definecolor{bglightblue}{rgb}{.625,.845,.917}
\definecolor{graphblue}{rgb}{.719,.918,.996}
\definecolor{lightlightblue}{rgb}{.89,.89,1}
\definecolor{lightpurple}{rgb}{1,.88,1}
\definecolor{medpurple}{rgb}{1,.65,1}
\definecolor{WeekColor}{rgb}{1,.7,.9}
\definecolor{lightblue}{rgb}{.55,.55,1}
\definecolor{midblue}{rgb}{.7,.7,1}
\definecolor{lightlightblue}{rgb}{.89,.89,1}
\definecolor{lightestblue}{rgb}{.96,.96,1}

\def\sm#1{\mbox{\small $#1$}}
\def\disp{\displaystyle}

\definecolor{apricot}{rgb}{.99,.87,.77}
\def\var{
\pspolygon[linewidth=.25pt,fillstyle=solid,fillcolor=lightlightblue](0,0)(1,0)(1,1)(0,1)(0,0)
\psline[linewidth=1.pt,arrowsize=9pt,linecolor=blue]{>->}(-.15,.5)(1.15,.5)
\psline[linewidth=1.pt,arrowsize=9pt,linecolor=blue]{>->}(.5,-.15)(.5,1.15)}

\def\vcs{
\pspolygon[linewidth=.25pt,fillstyle=solid,fillcolor=lightpurple](0,0)(1,0)(1,1)(0,1)(0,0)
\psline[linewidth=1.pt,arrowsize=9pt,linecolor=blue]{<-<}(-.15,.5)(1.15,.5)
\psline[linewidth=1.pt,arrowsize=9pt,linecolor=blue]{>->}(.5,-.15)(.5,1.15)}

\def\va{
\pspolygon[linewidth=.25pt,fillstyle=solid,fillcolor=lightlightblue](0,0)(1,0)(1,1)(0,1)(0,0)
\psline[linewidth=1.pt,arrowsize=9pt]{>->}(-.15,.5)(1.15,.5)
\psline[linewidth=1.pt,arrowsize=9pt]{>->}(.5,-.15)(.5,1.15)}
\def\vb{
\pspolygon[linewidth=.25pt,fillstyle=solid,fillcolor=lightlightblue](0,0)(1,0)(1,1)(0,1)(0,0)
\psline[linewidth=1.pt,arrowsize=9pt]{<-<}(-.15,.5)(1.15,.5)
\psline[linewidth=1.pt,arrowsize=9pt]{<-<}(.5,-.15)(.5,1.15)}
\def\vc{
\pspolygon[linewidth=.25pt,fillstyle=solid,fillcolor=lightlightblue](0,0)(1,0)(1,1)(0,1)(0,0)
\psline[linewidth=1.pt,arrowsize=9pt]{<-<}(-.15,.5)(1.15,.5)
\psline[linewidth=1.pt,arrowsize=9pt]{>->}(.5,-.15)(.5,1.15)}
\def\vd{
\pspolygon[linewidth=.25pt,fillstyle=solid,fillcolor=lightlightblue](0,0)(1,0)(1,1)(0,1)(0,0)
\psline[linewidth=1.pt,arrowsize=9pt]{>->}(-.15,.5)(1.15,.5)
\psline[linewidth=1.pt,arrowsize=9pt]{<-<}(.5,-.15)(.5,1.15)}
\def\ve{
\pspolygon[linewidth=.25pt,fillstyle=solid,fillcolor=lightlightblue](0,0)(1,0)(1,1)(0,1)(0,0)
\psline[linewidth=1.pt,arrowsize=9pt]{<->}(-.15,.5)(1.15,.5)
\psline[linewidth=1.pt,arrowsize=9pt]{>-<}(.5,-.15)(.5,1.15)}
\def\vf{
\pspolygon[linewidth=.25pt,fillstyle=solid,fillcolor=lightlightblue](0,0)(1,0)(1,1)(0,1)(0,0)
\psline[linewidth=1.pt,arrowsize=9pt]{>-<}(-.15,.5)(1.15,.5)
\psline[linewidth=1.pt,arrowsize=9pt]{<->}(.5,-.15)(.5,1.15)}
\def\vva{
\pspolygon[linewidth=.25pt,fillstyle=solid,fillcolor=lightpurple](0,0)(1,0)(1,1)(0,1)(0,0)
\psline[linewidth=1.pt,arrowsize=9pt]{>->}(-.15,.5)(1.15,.5)
\psline[linewidth=1.pt,arrowsize=9pt]{>->}(.5,-.15)(.5,1.15)}
\def\vvb{
\pspolygon[linewidth=.25pt,fillstyle=solid,fillcolor=lightpurple](0,0)(1,0)(1,1)(0,1)(0,0)
\psline[linewidth=1.pt,arrowsize=9pt]{<-<}(-.15,.5)(1.15,.5)
\psline[linewidth=1.pt,arrowsize=9pt]{<-<}(.5,-.15)(.5,1.15)}
\def\vvc{
\pspolygon[linewidth=.25pt,fillstyle=solid,fillcolor=lightpurple](0,0)(1,0)(1,1)(0,1)(0,0)
\psline[linewidth=1.pt,arrowsize=9pt]{<-<}(-.15,.5)(1.15,.5)
\psline[linewidth=1.pt,arrowsize=9pt]{>->}(.5,-.15)(.5,1.15)}
\def\vvd{
\pspolygon[linewidth=.25pt,fillstyle=solid,fillcolor=lightpurple](0,0)(1,0)(1,1)(0,1)(0,0)
\psline[linewidth=1.pt,arrowsize=9pt]{>->}(-.15,.5)(1.15,.5)
\psline[linewidth=1.pt,arrowsize=9pt]{<-<}(.5,-.15)(.5,1.15)}
\def\vve{
\pspolygon[linewidth=.25pt,fillstyle=solid,fillcolor=lightpurple](0,0)(1,0)(1,1)(0,1)(0,0)
\psline[linewidth=1.pt,arrowsize=9pt]{<->}(-.15,.5)(1.15,.5)
\psline[linewidth=1.pt,arrowsize=9pt]{>-<}(.5,-.15)(.5,1.15)}
\def\vvf{
\pspolygon[linewidth=.25pt,fillstyle=solid,fillcolor=lightpurple](0,0)(1,0)(1,1)(0,1)(0,0)
\psline[linewidth=1.pt,arrowsize=9pt]{>-<}(-.15,.5)(1.15,.5)
\psline[linewidth=1.pt,arrowsize=9pt]{<->}(.5,-.15)(.5,1.15)}
\def\pa{
\pspolygon[linewidth=.25pt,fillstyle=solid,fillcolor=lightlightblue](0,0)(1,0)(1,1)(0,1)(0,0)
\psarc[linewidth=.5pt,linecolor=red](0,0){.1}{0}{90}}
\def\pb{
\pspolygon[linewidth=.25pt,fillstyle=solid,fillcolor=lightlightblue](0,0)(1,0)(1,1)(0,1)(0,0)
\psline[linewidth=2pt](0,.5)(.4,.5)
\psline[linewidth=2pt](.5,0)(.5,.4)
\psline[linewidth=2pt](.6,.5)(1,.5)
\psline[linewidth=2pt](.5,.6)(.5,1)
\psarc[linewidth=2pt](.6,.4){.1}{90}{180}
\psarc[linewidth=2pt](.4,.6){.1}{270}{0}
\psarc[linewidth=.5pt,linecolor=red](0,0){.1}{0}{90}}
\def\pc{
\pspolygon[linewidth=.25pt,fillstyle=solid,fillcolor=lightlightblue](0,0)(1,0)(1,1)(0,1)(0,0)
\psline[linewidth=2pt](0,.5)(1,.5)
\psarc[linewidth=.5pt,linecolor=red](0,0){.1}{0}{90}}
\def\pd{
\pspolygon[linewidth=.25pt,fillstyle=solid,fillcolor=lightlightblue](0,0)(1,0)(1,1)(0,1)(0,0)
\psline[linewidth=2pt](.5,0)(.5,1)
\psarc[linewidth=.5pt,linecolor=red](0,0){.1}{0}{90}}
\def\pe{
\pspolygon[linewidth=.25pt,fillstyle=solid,fillcolor=lightlightblue](0,0)(1,0)(1,1)(0,1)(0,0)
\psline[linewidth=2pt](0,.5,)(.4,.5)
\psline[linewidth=2pt](.5,.6)(.5,1)
\psarc[linewidth=2pt](.4,.6){.1}{270}{0}
\psarc[linewidth=.5pt,linecolor=red](0,0){.1}{0}{90}}
\def\pf{
\pspolygon[linewidth=.25pt,fillstyle=solid,fillcolor=lightlightblue](0,0)(1,0)(1,1)(0,1)(0,0)
\psline[linewidth=2pt](.6,.5)(1,.5)
\psline[linewidth=2pt](.5,0)(.5,.4)
\psarc[linewidth=2pt](.6,.4){.1}{90}{180}
\psarc[linewidth=.5pt,linecolor=red](0,0){.1}{0}{90}}

\def\qa{
\pspolygon[linewidth=.25pt,fillstyle=solid,fillcolor=lightpurple](0,0)(1,0)(1,1)(0,1)(0,0)
\psarc[linewidth=.5pt,linecolor=red](1,0){.1}{90}{180}}
\def\qb{
\pspolygon[linewidth=.25pt,fillstyle=solid,fillcolor=lightpurple](0,0)(1,0)(1,1)(0,1)(0,0)
\psline[linewidth=2pt](0,.5)(.4,.5)
\psline[linewidth=2pt](.5,0)(.5,.4)
\psline[linewidth=2pt](.6,.5)(1,.5)
\psline[linewidth=2pt](.5,.6)(.5,1)
\psarc[linewidth=2pt](.4,.4){.1}{0}{90}
\psarc[linewidth=2pt](.6,.6){.1}{180}{270}
\psarc[linewidth=.5pt,linecolor=red](1,0){.1}{90}{180}}
\def\qc{
\pspolygon[linewidth=.25pt,fillstyle=solid,fillcolor=lightpurple](0,0)(1,0)(1,1)(0,1)(0,0)
\psline[linewidth=2pt](0,.5)(1,.5)
\psarc[linewidth=.5pt,linecolor=red](1,0){.1}{90}{180}}
\def\qd{
\pspolygon[linewidth=.25pt,fillstyle=solid,fillcolor=lightpurple](0,0)(1,0)(1,1)(0,1)(0,0)
\psline[linewidth=2pt](.5,0)(.5,1)
\psarc[linewidth=.5pt,linecolor=red](1,0){.1}{90}{180}}

\def\qg{
\pspolygon[linewidth=.25pt,fillstyle=solid,fillcolor=lightpurple](0,0)(1,0)(1,1)(0,1)(0,0)
\psline[linewidth=2pt](.6,.5,)(1,.5)
\psline[linewidth=2pt](.5,.6)(.5,1)
\psarc[linewidth=2pt](.6,.6){.1}{180}{270}
\psarc[linewidth=.5pt,linecolor=red](1,0){.1}{90}{180}}
\def\qh{
\pspolygon[linewidth=.25pt,fillstyle=solid,fillcolor=lightpurple](0,0)(1,0)(1,1)(0,1)(0,0)
\psline[linewidth=2pt](.5,0)(.5,.4)
\psline[linewidth=2pt](0,.5)(.4,.5)
\psarc[linewidth=2pt](.4,.4){.1}{0}{90}
\psarc[linewidth=.5pt,linecolor=red](1,0){.1}{90}{180}}

\def\daDimer{
\pspolygon[linewidth=1.5pt,fillstyle=solid,fillcolor=lightyellow](0,0)(.5,-.5)(1.5,.5)(1,1)(0,0)
\pspolygon[linewidth=.25pt,fillstyle=solid,fillcolor=lightlightblue](0,0)(1,0)(1,1)(0,1)(0,0)
\psline[linewidth=1.5pt](0,0)(1,1)
\psline[linewidth=1.5pt](0,1)(.5,.5)}
\def\dbDimer{
\pspolygon[linewidth=1.5pt,fillstyle=solid,fillcolor=lightyellow](0,0)(-.5,.5)(.5,1.5)(1,1)(0,0)
\pspolygon[linewidth=.25pt,fillstyle=solid,fillcolor=lightlightblue](0,0)(1,0)(1,1)(0,1)(0,0)
\psline[linewidth=1.5pt](0,0)(1,1)
\psline[linewidth=1.5pt](.5,.5)(1,0)}
\def\dcDimer{
\pspolygon[linewidth=1.5pt,fillstyle=solid,fillcolor=lightyellow](0,1)(-.5,.5)(.5,-.5)(1,0)(0,0)
\pspolygon[linewidth=.25pt,fillstyle=solid,fillcolor=lightlightblue](0,0)(1,0)(1,1)(0,1)(0,0)
\psline[linewidth=1.5pt](0,1)(1,0)
\psline[linewidth=1.5pt](1,1)(.5,.5)}
\def\ddDimer{
\pspolygon[linewidth=1.5pt,fillstyle=solid,fillcolor=lightyellow](0,1)(.5,1.5)(1.5,.5)(1,0)(0,0)
\pspolygon[linewidth=.25pt,fillstyle=solid,fillcolor=lightlightblue](0,0)(1,0)(1,1)(0,1)(0,0)
\psline[linewidth=1.5pt](0,1)(1,0)
\psline[linewidth=1.5pt](.5,.5)(0,0)}
\def\deDimer{
\pspolygon[linewidth=1.5pt,fillstyle=solid,fillcolor=lightyellow](0,0)(.5,-.5)(1.5,.5)(1,1)(0,0)
\pspolygon[linewidth=1.5pt,fillstyle=solid,fillcolor=lightyellow](0,0)(-.5,.5)(.5,1.5)(1,1)(0,0)
\pspolygon[linewidth=.25pt,fillstyle=solid,fillcolor=apricot](0,0)(1,0)(1,1)(0,1)(0,0)
\psline[linewidth=1.5pt](0,0)(1,1)}
\def\deeDimer{
\pspolygon[linewidth=1.5pt,fillstyle=solid,fillcolor=lightyellow](0,1)(-.5,.5)(.5,-.5)(1,0)(0,0)
\pspolygon[linewidth=1.5pt,fillstyle=solid,fillcolor=lightyellow](0,1)(.5,1.5)(1.5,.5)(1,0)(0,0)
\pspolygon[linewidth=.25pt,fillstyle=solid,fillcolor=apricot](0,0)(1,0)(1,1)(0,1)(0,0)
\psline[linewidth=1.5pt](1,0)(0,1)}

\def\da{
\pspolygon[linewidth=.25pt,fillstyle=solid,fillcolor=lightlightblue](0,0)(1,0)(1,1)(0,1)(0,0)
\psline[linewidth=1.5pt](0,0)(1,1)
\psline[linewidth=1.5pt](0,1)(.5,.5)}
\def\db{
\pspolygon[linewidth=.25pt,fillstyle=solid,fillcolor=lightlightblue](0,0)(1,0)(1,1)(0,1)(0,0)
\psline[linewidth=1.5pt](0,0)(1,1)
\psline[linewidth=1.5pt](.5,.5)(1,0)}
\def\dc{
\pspolygon[linewidth=.25pt,fillstyle=solid,fillcolor=lightlightblue](0,0)(1,0)(1,1)(0,1)(0,0)
\psline[linewidth=1.5pt](0,1)(1,0)
\psline[linewidth=1.5pt](1,1)(.5,.5)}
\def\dd{
\pspolygon[linewidth=.25pt,fillstyle=solid,fillcolor=lightlightblue](0,0)(1,0)(1,1)(0,1)(0,0)
\psline[linewidth=1.5pt](0,1)(1,0)
\psline[linewidth=1.5pt](.5,.5)(0,0)}
\def\de{
\pspolygon[linewidth=.25pt,fillstyle=solid,fillcolor=apricot](0,0)(1,0)(1,1)(0,1)(0,0)
\psline[linewidth=1.5pt](0,0)(1,1)}
\def\dee{
\pspolygon[linewidth=.25pt,fillstyle=solid,fillcolor=apricot](0,0)(1,0)(1,1)(0,1)(0,0)
\psline[linewidth=1.5pt](1,0)(0,1)}
\def\df{
\pspolygon[linewidth=.25pt,fillstyle=solid,fillcolor=lightlightblue](0,0)(1,0)(1,1)(0,1)(0,0)
\psline[linewidth=1.5pt](0,0)(1,1)
\psline[linewidth=1.5pt](0,1)(1,0)}

\def\oface#1{
\pspolygon[linewidth=.25pt,fillstyle=solid,fillcolor=lightlightblue](0,0)(1,0)(1,1)(0,1)(0,0)
\rput(.5,.5){\small $#1$}
\psarc[linewidth=.5pt,linecolor=red](0,0){.125}{0}{90}}
\def\eface#1{
\pspolygon[linewidth=.25pt,fillstyle=solid,fillcolor=lightpurple](0,0)(1,0)(1,1)(0,1)(0,0)
\rput(.5,.5){\small $#1$}
\psarc[linewidth=.5pt,linecolor=red](1,0){.125}{90}{180}}

\nc{\spos}[2]{\makebox(0,0)[#1]{$\sm{#2}$}}
\nc{\botrightrefnodots}[4]{
\begin{pspicture}(1.5,2)
\rput(-.5,0){\psline[linewidth=.5pt,fillstyle=solid,fillcolor=lightlightblue](2,0)(2,2)(1,1)}
\rput(0,0){\psline[linewidth=.5pt,fillstyle=solid,fillcolor=lightlightblue](0,.5)(.5,0)(1,.5)(.5,1)}
\put(1.5,0){\line(0,1){2}}
\put(1.5,1){\line(-1,1){0.5}}
\put(1.5,1){\line(-1,-1){1}}\put(1.5,0){\line(-1,1){1}}
\put(1.5,2){\line(-1,-1){1.5}}\put(0,0.5){\line(1,-1){0.5}}
\put(1.4,0.5){\spos{r}{#1}}\put(1.4,1.5){\spos{r}{#2}}
\put(0.5,0.5){\spos{}{#3}}
\put(1,1){\spos{}{#4}}
\psarc[linecolor=red](0,.5){.1}{-45}{45}
\psarc[linecolor=red](1,.5){.1}{45}{135}
\end{pspicture}}
\nc{\toprightrefnodots}[4]{
\begin{pspicture}(1.5,2)
\rput(-.5,0){\psline[linewidth=.5pt,fillstyle=solid,fillcolor=lightlightblue](2,0)(2,2)(1,1)}
\rput(0,1){\psline[linewidth=.5pt,fillstyle=solid,fillcolor=lightlightblue](0,.5)(.5,0)(1,.5)(.5,1)}
\put(1.5,0){\line(0,1){2}}
\put(1.5,1){\line(-1,1){1}}\put(1.5,1){\line(-1,-1){0.5}}
\put(1.5,0){\line(-1,1){1.5}}\put(1.5,2){\line(-1,-1){1}}
\put(0,1.5){\line(1,1){0.5}}
\put(1.4,0.5){\spos{r}{#1}}\put(1.4,1.5){\spos{r}{#2}}
\put(1,1){\spos{}{#3}}\put(0.5,1.5){\spos{}{#4}}
\psarc[linecolor=red](0,1.5){.1}{-45}{45}
\psarc[linecolor=red](1,.5){.1}{45}{135}
\end{pspicture}}

\def\righttri#1#2#3{
\begin{pspicture}(0,0)
\pspolygon[linewidth=.25pt,fillstyle=solid,fillcolor=lightlightblue](0,1)(1,0)(1,2)
\rput[br](.5,1.5){\small $#1$}
\rput[tr](.5,.5){\small $#2$}
\rput(.6,1){\small $#3$}
\end{pspicture}}

\def\diamodd#1{
\begin{pspicture}(0,0)(2,2)
\pspolygon[linewidth=.25pt,fillstyle=solid,fillcolor=lightlightblue](1,0)(2,1)(1,2)(0,1)
\rput(1,1){\small $#1$}
\end{pspicture}}
\def\diamoddleft#1{
\begin{pspicture}(0,0)(2,2)
\pspolygon[linewidth=.25pt,fillstyle=solid,fillcolor=lightlightblue](1,0)(2,1)(1,2)(0,1)
\rput(1,1){\small $#1$}
\psarc[linewidth=.5pt,linecolor=red](0,1){.15}{-45}{45}
\end{pspicture}}
\def\diamoddy#1{
\begin{pspicture}(0,0)(2,2)
\pspolygon[linewidth=.25pt,fillstyle=solid,fillcolor=yellow!40!white](1,0)(2,1)(1,2)(0,1)
\rput(1,1){\small $#1$}
\end{pspicture}}

\def\rbfacepink#1{
\begin{pspicture}(0,0)(2,2)
\pspolygon[linewidth=.25pt,fillstyle=solid,fillcolor=lightpurple](0,0)(1.4,0)(1.4,1.4)(0,1.4)(0,0)
\rput(.7,.7){\small $#1$}
\psarc[linewidth=1pt,linecolor=red](1.4,0){.15}{90}{180}
\end{pspicture}}

\def\lbface#1{
\begin{pspicture}(0,0)(2,2)
\pspolygon[linewidth=.25pt,fillstyle=solid,fillcolor=lightlightblue](0,0)(1.4,0)(1.4,1.4)(0,1.4)(0,0)
\rput(.7,.7){\small $#1$}
\psarc[linewidth=1pt,linecolor=red](0,0){.15}{0}{90}
\end{pspicture}}

\def\righttri#1#2#3{
\begin{pspicture}(0,0)
\pspolygon[linewidth=.25pt,fillstyle=solid,fillcolor=lightlightblue](0,1)(1,0)(1,2)
\rput[br](.5,1.5){\small $#1$}
\rput[tr](.5,.5){\small $#2$}
\rput(.65,1){\small $#3$}
\end{pspicture}}
\def\lefttri#1#2#3{
\begin{pspicture}(0,0)
\pspolygon[linewidth=.25pt,fillstyle=solid,fillcolor=lightlightblue](0,0)(0,2)(1,1)
\rput[br](.5,1.5){\small $#1$}
\rput[tr](.5,.5){\small $#2$}
\rput(.45,1){\small $#3$}
\end{pspicture}}
\def\diam#1#2#3#4#5{
\begin{pspicture}(0,0)(2,2)
\pspolygon[linewidth=.25pt,fillstyle=solid,fillcolor=lightlightblue](1,0)(2,1)(1,2)(0,1)
\rput[tr](.5,.5){\small $#1$}
\rput[tl](1.5,.5){\small $#2$}
\rput[bl](1.5,1.5){\small $#3$}
\rput[br](.5,1.5){\small $#4$}
\rput(1,1){\small $#5$}
\psarc[linewidth=.5pt,linecolor=red](1,0){.15}{45}{135}
\end{pspicture}}
\def\diampink#1#2#3#4#5{
\begin{pspicture}(0,0)(2,2)
\pspolygon[linewidth=.25pt,fillstyle=solid,fillcolor=lightpurple](1,0)(2,1)(1,2)(0,1)
\rput[tr](.5,.5){\small $#1$}
\rput[tl](1.5,.5){\small $#2$}
\rput[bl](1.5,1.5){\small $#3$}
\rput[br](.5,1.5){\small $#4$}
\rput(1,1){\small $#5$}
\psarc[linewidth=1pt,linecolor=red](1,0){.15}{45}{135}
\end{pspicture}}

\def\ldiam#1#2#3#4#5{
\begin{pspicture}(0,0)(2,2)
\pspolygon[linewidth=.25pt,fillstyle=solid,fillcolor=lightlightblue](1,0)(2,1)(1,2)(0,1)
\rput[tr](.5,.5){\small $#1$}
\rput[tl](1.5,.5){\small $#2$}
\rput[bl](1.5,1.5){\small $#3$}
\rput[br](.5,1.5){\small $#4$}
\rput(1,1){\small $#5$}
\psarc[linewidth=.5pt,linecolor=red](0,1){.15}{-45}{45}
\end{pspicture}}
\def\rdiam#1#2#3#4#5{
\begin{pspicture}(0,0)(2,2)
\pspolygon[linewidth=.25pt,fillstyle=solid,fillcolor=lightlightblue](1,0)(2,1)(1,2)(0,1)
\rput[tr](.5,.5){\small $#1$}
\rput[tl](1.5,.5){\small $#2$}
\rput[bl](1.5,1.5){\small $#3$}
\rput[br](.5,1.5){\small $#4$}
\rput(1,1){\small $#5$}
\psarc[linewidth=.5pt,linecolor=red](2,1){.15}{135}{225}
\end{pspicture}}
\def\rdiampink#1#2#3#4#5{
\begin{pspicture}(0,0)(2,2)
\pspolygon[linewidth=.25pt,fillstyle=solid,fillcolor=lightpurple](1,0)(2,1)(1,2)(0,1)
\rput[tr](.5,.5){\small $#1$}
\rput[tl](1.5,.5){\small $#2$}
\rput[bl](1.5,1.5){\small $#3$}
\rput[br](.5,1.5){\small $#4$}
\rput(1,1){\small $#5$}
\psarc[linewidth=1pt,linecolor=red](2,1){.15}{135}{225}
\end{pspicture}}
\def\sdiam#1#2#3#4#5{
\begin{pspicture}(0,0)(2,2)
\pspolygon[linewidth=.25pt,fillstyle=solid,fillcolor=yellow!40!white](1,0)(2,1)(1,2)(0,1)
\rput[tr](.5,.5){\small $#1$}
\rput[tl](1.5,.5){\small $#2$}
\rput[bl](1.5,1.5){\small $#3$}
\rput[br](.5,1.5){\small $#4$}
\rput(1,1){\small $#5$}
\psarc[linewidth=.5pt,linecolor=red](1,0){.15}{45}{135}
\end{pspicture}}

\def\WGblue#1#2#3#4#5#6#7{W\Big(\mbox{\scriptsize $
\begingroup 
\setlength\arraycolsep{#6pt}
\begin{matrix}
&#4&\\[-2pt] #1&&#3\\[-2pt] &#2&\end{matrix}\endgroup$}\,\Big|#5,#7\!\Big)}

\definecolor{tile}{rgb}{1,.9,.7}

\def\tilea{\;
\begin{pspicture}[shift=-.65](0,0)(1.5,1.5)
\pspolygon[linewidth=.5pt,fillstyle=solid,fillcolor=tile](.75,0)(1.5,.75)(.75,1.5)(0,.75)(.75,0)
\end{pspicture}\;}

\def\tileb{\;
\begin{pspicture}[shift=-.65](0,0)(1.5,1.5)
\pspolygon[linewidth=.5pt,fillstyle=solid,fillcolor=tile](.75,0)(1.5,.75)(.75,1.5)(0,.75)(.75,0)
\psarc[linewidth=1.25pt](0,.75){.575}{-45}{45}
\psarc[linewidth=1.25pt](1.5,.75){.575}{135}{225}
\end{pspicture}\;}

\def\tilec{\;
\begin{pspicture}[shift=-.65](0,0)(1.5,1.5)
\pspolygon[linewidth=.5pt,fillstyle=solid,fillcolor=tile](.75,0)(1.5,.75)(.75,1.5)(0,.75)(.75,0)
\psline[linewidth=1.25pt](.375,.375)(1.125,1.125)
\end{pspicture}\;}

\def\tiled{\;
\begin{pspicture}[shift=-.65](0,0)(1.5,1.5)
\pspolygon[linewidth=.5pt,fillstyle=solid,fillcolor=tile](.75,0)(1.5,.75)(.75,1.5)(0,.75)(.75,0)
\psline[linewidth=1.25pt](1.125,.375)(.375,1.125)
\end{pspicture}\;}

\def\tilecd{\;
\begin{pspicture}[shift=-.65](0,0)(1.5,1.5)
\pspolygon[linewidth=.5pt,fillstyle=solid,fillcolor=tile](.75,0)(1.5,.75)(.75,1.5)(0,.75)(.75,0)
\psline[linewidth=1.25pt](.375,.375)(1.125,1.125)
\psline[linewidth=1.25pt](1.125,.375)(.375,1.125)
\end{pspicture}\;}

\def\tilee{\;
\begin{pspicture}[shift=-.65](0,0)(1.5,1.5)
\pspolygon[linewidth=.5pt,fillstyle=solid,fillcolor=tile](.75,0)(1.5,.75)(.75,1.5)(0,.75)(.75,0)
\psarc[linewidth=1.25pt](0,.75){.575}{-45}{45}
\end{pspicture}\;}

\def\tilef{\;
\begin{pspicture}[shift=-.65](0,0)(1.5,1.5)
\pspolygon[linewidth=.5pt,fillstyle=solid,fillcolor=tile](.75,0)(1.5,.75)(.75,1.5)(0,.75)(.75,0)
\psarc[linewidth=1.25pt](1.5,.75){.575}{135}{225}
\end{pspicture}\;}

\def\tileg{\;
\begin{pspicture}[shift=-.65](0,0)(1.5,1.5)
\pspolygon[linewidth=.5pt,fillstyle=solid,fillcolor=tile](.75,0)(1.5,.75)(.75,1.5)(0,.75)(.75,0)
\psarc[linewidth=1.25pt](.75,0){.575}{45}{135}
\end{pspicture}\;}

\def\tileh{\;
\begin{pspicture}[shift=-.65](0,0)(1.5,1.5)
\pspolygon[linewidth=.5pt,fillstyle=solid,fillcolor=tile](.75,0)(1.5,.75)(.75,1.5)(0,.75)(.75,0)
\psarc[linewidth=1.25pt](.75,1.5){.575}{225}{315}
\end{pspicture}\;}

\def\tilek{\;
\begin{pspicture}[shift=-.65](0,0)(1.5,1.5)
\pspolygon[linewidth=.5pt,fillstyle=solid,fillcolor=tile](.75,0)(1.5,.75)(.75,1.5)(0,.75)(.75,0)
\psarc[linewidth=1.25pt](.75,0){.575}{45}{135}
\psarc[linewidth=1.25pt](.75,1.5){.575}{225}{315}
\end{pspicture}\;}

\begin{document}

\topmargin -5mm
\oddsidemargin 5mm

\begin{titlepage}
\setcounter{page}{0}

\vspace{8mm}
\begin{center}
{\huge {\bf Yang-Baxter Integrable Dimers on a Strip}}

\vspace{10mm}
{\Large  Paul A. Pearce${}^\dagger{}^\ddagger$, J\o rgen Rasmussen${}^\ddagger$, Alessandra Vittorini-Orgeas${}^\dagger$}\\[.3cm]
{\em ${}^\dagger$School of Mathematics and Statistics, University of Melbourne}\\
{\em Parkville, Victoria 3010, Australia}\\[.4cm]
{\em ${}^\ddagger$School of Mathematics and Physics, University of Queensland}\\
{\em St Lucia, Brisbane, Queensland 4072, Australia}\\[.4cm]
papearce@unimelb.edu.au, j.rasmussen@uq.edu.au, alessandra.vittorini@unimelb.edu.au

\end{center}

\vspace{10mm}
\centerline{{\bf{Abstract}}}
\vskip.4cm
\noindent
The dimer model on a strip is considered as a Yang-Baxter \mbox{integrable} six vertex model at the free-fermion point with crossing parameter $\lambda=\tfrac{\pi}{2}$ and quantum group invariant boundary conditions. A one-to-many mapping of vertex onto dimer configurations allows for the solution of the free-fermion model to be applied to the anisotropic dimer model on a square lattice where the dimers are rotated by $45\degree$ compared to their usual orientation. 
In a suitable gauge, the dimer model is described by the Temperley-Lieb algebra with loop fugacity $\beta=2\cos\lambda=0$. 
It follows that the model is exactly solvable in geometries of arbitrary finite size. 
We establish and solve transfer matrix inversion identities
on the strip with arbitrary finite width $N$. In the continuum scaling limit, in sectors with magnetization $S_z$, we obtain the conformal weights $\Delta_{s}=\big((2-s)^2-1\big)/8$ where $s=|S_z|+1=1,2,3,\ldots$. We further show that the corresponding finitized characters $\chit_s^{(N)}(q)$ decompose into sums of $q$-Narayana numbers or, equivalently, skew $q$-binomials. 
In the particle representation, the local face tile operators give a representation of the fermion algebra and the fermion particle trajectories play the role of nonlocal degrees of freedom. 
We argue that, in the continuum scaling limit, there exist nontrivial Jordan blocks of rank 2 in the Virasoro dilatation operator $L_0$.
This confirms that, with quantum group invariant boundary conditions, the dimer model gives rise to a {\em logarithmic} conformal field theory with central charge $c=-2$, minimal conformal weight $\Delta_{\text{min}}=-\frac{1}{8}$ and effective central charge $c_{\text{eff}}=1$.
Our analysis of the structure of the ensuing rank 2 modules indicates that the familiar staggered $c=-2$ modules appear as submodules.

\end{titlepage}
\newpage
\renewcommand{\thefootnote}{\arabic{footnote}}
\setcounter{footnote}{0}

\tableofcontents

\newpage
\section{Introduction}

The dimer model~\cite{Roberts,FowlerRush} was solved exactly~\cite{Kasteleyn1961,TempFisher,Fisher1961,LiebTransfer} in the early sixties. 
After more than 50 years, the model continues to be the subject of extensive 
study~\cite{IzOganHu2003,IzPRHu2005,IzPR2007,RasRuelle2012,Nigro2012,Allegra2015,MDRR2015,MDRR2016}. 
The current interest is twofold: (i)~to understand the finite-size effects of boundary conditions and steric 
effects~\cite{Aztec,DWBC,KorepinZJ} under the influence of infinitely repulsive hard-core local interactions and (ii)~to understand the conformal description of the model in the continuum scaling limit.
Traditionally, it is asserted that the dimer model is described~\cite{Kenyon} by a $c=1$ Gaussian free field. But without full access to the various sectors and boundary conditions on the strip, it is difficult to distinguish between a $c=1$ theory and a $c_\text{eff}=1$ theory and a number of authors~\cite{IzPRHu2005,IzPR2007} have suggested that 
the model is described by a {\em logarithmic} Conformal Field Theory (CFT) with $c=-2$. 

Recently, the dimer model was shown~\cite{PVO2017} to be Yang-Baxter integrable~\cite{BaxBook} by mapping~\cite{Baxter1972,KorepinZJ,FerrariSpohn} it onto the free-fermion six vertex model~\cite{Pauling,Lieb,Sutherland,LiebWu,FanWu}. Notably, this maps six vertex configurations onto dimer configurations where the dimers are rotated by $45\degree$, as shown in Figures~\ref{vpd}, \ref{TheDimers} and \ref{vacBdy}, compared to their usual orientation parallel to the bonds of the square lattice. This technique combined with inversion identities~\cite{Felderhof,BaxBook,OPW1996} enables 
the model
to be solved exactly for finite lattices with various boundary conditions and topologies. The conformal properties can therefore be readily extracted from the finite-size scaling behaviour. On this basis, it was argued in \cite{PVO2017} that
the dimer model
is best described as a logarithmic CFT with effective central charge $c_\text{eff}=1$ but central charge $c=-2$, 
in agreement with the findings of \cite{IzPRHu2005,IzPR2007,RasRuelle2012,MDRR2016}. The primary characterization of logarithmic CFTs is the appearance of nontrivial Jordan blocks in the Virasoro dilatation operator $L_0$. 
Indeed, for simple dimer boundary conditions on the strip, corresponding to the $U_q(sl(2))$-invariant XX Hamiltonian $\cal H$ of the free-fermion six vertex model, the preliminary results of \cite{PVO2017} indicate that $\cal H$ admits nontrivial Jordan blocks for finite systems. Since the appearance of these blocks is stable, as the system size increases, these blocks are expected to persist for large sizes and appear in the Virasoro operator $L_0$.

In this paper, we solve exactly the anisotropic square lattice dimer model with the $45\degree$ rotated orientation on the strip in sectors labelled by the magnetization $S_z$ of the related free-fermion six vertex model.
This is achieved, using Yang-Baxter integrability, by mapping the model with given boundary conditions onto a free-fermion six vertex model and solving the associated inversion identities~\cite{Felderhof,BaxBook,OPW1996} satisfied by the double row transfer matrices. 
The solution of the inversion identities allows to obtain the exact finite spectra in the various sectors. Finite-size scaling then yields the central charge and the conformal weights. 
In addition, combinatorial analysis of the patterns of zeros, in the complex spectral parameter plane, of the double row transfer matrix eigenvalues allows us to obtain finitized characters. 
We confirm the central charge $c=-2$ and the conformal weights 
$\Delta_{s}=\big((2-s)^2-1\big)/8$ with $s=1,2,3,\ldots$. Remarkably, although the characters are different, the conformal weights coincide with those in the first column of the infinitely extended Kac table of critical dense polymers~\cite{SaleurSuper,PR2007,PR2007b,PRV,PRVKac,MDPR2013}, as shown in Figure~\ref{Kac}.

\begin{figure}[t]
{\vspace{0in}\psset{unit=1cm}
{
\small
\begin{center}
\qquad
\begin{pspicture}(0,0)(7,11)
\psframe[linewidth=0pt,fillstyle=solid,fillcolor=lightlightblue](0,0)(7,11)
\multiput(0,0)(0,2){5}{\psframe[linewidth=0pt,fillstyle=solid,fillcolor=midblue](0,1)(7,2)}
\psframe[linewidth=0pt,fillstyle=solid,fillcolor=lightpurple](0,0)(1,11)
\multiput(0,0)(0,2){5}{\psframe[linewidth=0pt,fillstyle=solid,fillcolor=medpurple](0,1)(1,2)}
\psgrid[gridlabels=0pt,subgriddiv=1]
\rput(.5,10.65){$\vdots$}\rput(1.5,10.65){$\vdots$}\rput(2.5,10.65){$\vdots$}
\rput(3.5,10.65){$\vdots$}\rput(4.5,10.65){$\vdots$}\rput(5.5,10.65){$\vdots$}
\rput(6.5,10.5){$\vvdots$}\rput(.5,9.5){$\frac{63}8$}\rput(1.5,9.5){$\frac{35}8$}
\rput(2.5,9.5){$\frac{15}8$}\rput(3.5,9.5){$\frac{3}8$}\rput(4.5,9.5){$-\frac 18$}
\rput(5.5,9.5){$\frac{3}8$}\rput(6.5,9.5){$\cdots$}
\rput(.5,8.5){$6$}\rput(1.5,8.5){$3$}\rput(2.5,8.5){$1$}\rput(3.5,8.5){$0$}
\rput(4.5,8.5){$0$}\rput(5.5,8.5){$1$}\rput(6.5,8.5){$\cdots$}
\rput(.5,7.5){$\frac{35}8$}\rput(1.5,7.5){$\frac {15}8$}\rput(2.5,7.5){$\frac 38$}
\rput(3.5,7.5){$-\frac{1}8$}\rput(4.5,7.5){$\frac 38$}\rput(5.5,7.5){$\frac{15}8$}
\rput(6.5,7.5){$\cdots$}\rput(.5,6.5){$3$}\rput(1.5,6.5){$1$}\rput(2.5,6.5){$0$}\rput(3.5,6.5){$0$}
\rput(4.5,6.5){$1$}\rput(5.5,6.5){$3$}\rput(6.5,6.5){$\cdots$}
\rput(.5,5.5){$\frac{15}8$}\rput(1.5,5.5){$\frac {3}{8}$}\rput(2.5,5.5){$-\frac 18$}
\rput(3.5,5.5){$\frac{3}{8}$}\rput(4.5,5.5){$\frac {15}8$}\rput(5.5,5.5){$\frac{35}{8}$}
\rput(6.5,5.5){$\cdots$}\rput(.5,4.5){$1$}\rput(1.5,4.5){$0$}\rput(2.5,4.5){$0$}
\rput(3.5,4.5){$1$}\rput(4.5,4.5){$3$}\rput(5.5,4.5){$6$}\rput(6.5,4.5){$\cdots$}
\rput(.5,3.5){$\frac 38$}\rput(1.5,3.5){$-\frac 18$}\rput(2.5,3.5){$\frac 38$}
\rput(3.5,3.5){$\frac{15}8$}\rput(4.5,3.5){$\frac{35}8$}\rput(5.5,3.5){$\frac{63}8$}
\rput(6.5,3.5){$\cdots$}\rput(.5,2.5){$0$}\rput(1.5,2.5){$0$}\rput(2.5,2.5){$1$}\rput(3.5,2.5){$3$}
\rput(4.5,2.5){$6$}\rput(5.5,2.5){$10$}\rput(6.5,2.5){$\cdots$}
\rput(.5,1.5){$-\frac 18$}\rput(1.5,1.5){$\frac 38$}\rput(2.5,1.5){$\frac{15}8$}
\rput(3.5,1.5){$\frac{35}8$}\rput(4.5,1.5){$\frac{63}8$}\rput(5.5,1.5){$\frac{99}8$}
\rput(6.5,1.5){$\cdots$}\rput(.5,.5){$0$}\rput(1.5,.5){$1$}\rput(2.5,.5){$3$}\rput(3.5,.5){$6$}
\rput(4.5,.5){$10$}\rput(5.5,.5){$15$}\rput(6.5,.5){$\cdots$}
{\color{blue}
\rput(.5,-.5){$1$}
\rput(1.5,-.5){$2$}
\rput(2.5,-.5){$3$}
\rput(3.5,-.5){$4$}
\rput(4.5,-.5){$5$}
\rput(5.5,-.5){$6$}
\rput(6.5,-.5){$r$}
\rput(-.5,.5){$1$}
\rput(-.5,1.5){$2$}
\rput(-.5,2.5){$3$}
\rput(-.5,3.5){$4$}
\rput(-.5,4.5){$5$}
\rput(-.5,5.5){$6$}
\rput(-.5,6.5){$7$}
\rput(-.5,7.5){$8$}
\rput(-.5,8.5){$9$}
\rput(-.5,9.5){$10$}
\rput(-.5,10.5){$s$}}
\end{pspicture}
\end{center}}}
\caption{\label{Kac}Kac table of conformal weights $\Delta_{r,s}$ of critical dense polymers taken from \cite{PRVKac}. 
The conformal weights of dimers coincide with the conformal weights in the first ($r=1$) column of this Kac table. Both theories are described by CFTs with $c=-2$, although their conformal characters are different.}
\end{figure}

The layout of the paper is as follows. In Section~\ref{SecLattice}, we recall the rotated dimer model on the square lattice and review its relation to the free-fermion six vertex model. We also describe the underlying free-fermion and Temperley-Lieb algebras. In Section~\ref{secSixVertex}, we present the local Yang-Baxter relations of the six vertex model using the particle representation of the planar algebra and establish the commutation of the double row transfer matrices. In Section~\ref{SecStrip}, we specialise to the dimer model and solve the associated inversion identities on the strip for the finite size spectra. This involves the combinatorial analysis of the patterns of zeros of the eigenvalues and the empirical determination of selection rules to fix the eigenvalue degeneracies which are not fixed by the functional equations alone. Jordan decompositions of the isotropic double row transfer matrices and their quantum Hamiltonians, for some small system sizes, are presented in Section~\ref{secJordan} to reveal the existence of nontrivial Jordan blocks of rank 2. In the continuum scaling limit, the Hamiltonian gives rise to
the Virasoro dilatation operator $L_0$. Since the indications are that these Jordan blocks persist in the continuum scaling limit, we are led to conclude that the CFT describing the dimer model is logarithmic. 
We also analyse the structure of the ensuing 
rank $2$ modules and determine the finitized characters of their irreducible sub-quotients. We finish with some concluding remarks in Section~\ref{secConclusion}, comparing dimers with critical dense polymers. Details of the proof of the inversion identities and the properties of the skew $q$-binomials appearing in the selection rules are relegated to Appendices.

\def\mydiam#1#2#3#4#5{
\begin{pspicture}[shift=-.65](0,0)(1.5,1.5)
\pspolygon[linewidth=.5pt,fillstyle=solid,fillcolor=lightlightblue](.75,0)(1.5,.75)(.75,1.5)(0,.75)(.75,0)
\rput[tr](.375,.375){\small $#1$}
\rput[tl](1.125,.375){\small $#2$}
\rput[bl](1.125,1.1255){\small $#3$}
\rput[br](.375,1.125){\small $#4$}
\rput(.75,.75){\small $#5$}
\psarc[linewidth=.5pt,linecolor=red](.75,0){.15}{45}{135}
\end{pspicture}}
\def\mydiampink#1#2#3#4#5{
\begin{pspicture}[shift=-.65](0,0)(1.5,1.5)
\pspolygon[linewidth=.5pt,fillstyle=solid,fillcolor=lightpurple](.75,0)(1.5,.75)(.75,1.5)(0,.75)(.75,0)
\rput[tr](.375,.375){\small $#1$}
\rput[tl](1.125,.375){\small $#2$}
\rput[bl](1.125,1.1255){\small $#3$}
\rput[br](.375,1.125){\small $#4$}
\rput(.75,.75){\small $#5$}
\psarc[linewidth=.5pt,linecolor=red](1.5,.75){.15}{135}{215}
\end{pspicture}}
\def\mydiamtop#1#2#3#4#5{
\begin{pspicture}[shift=-.65](0,0)(1.5,1.5)
\pspolygon[linewidth=.5pt,fillstyle=solid,fillcolor=lightlightblue](.75,0)(1.5,.75)(.75,1.5)(0,.75)(.75,0)
\rput[tr](.375,.375){\small $#1$}
\rput[tl](1.125,.375){\small $#2$}
\rput[bl](1.125,1.1255){\small $#3$}
\rput[br](.375,1.125){\small $#4$}
\rput(.75,.75){\small $#5$}
\psarc[linewidth=.5pt,linecolor=red](.75,1.5){.15}{225}{315}
\end{pspicture}}
\def\mydiampinkleft#1#2#3#4#5{
\begin{pspicture}[shift=-.65](0,0)(1.5,1.5)
\pspolygon[linewidth=.5pt,fillstyle=solid,fillcolor=lightpurple](.75,0)(1.5,.75)(.75,1.5)(0,.75)(.75,0)
\rput[tr](.375,.375){\small $#1$}
\rput[tl](1.125,.375){\small $#2$}
\rput[bl](1.125,1.1255){\small $#3$}
\rput[br](.375,1.125){\small $#4$}
\rput(.75,.75){\small $#5$}
\psarc[linewidth=.5pt,linecolor=red](0,.75){.15}{-45}{45}
\end{pspicture}}

\section{Dimers as a Free-Fermion Six Vertex Model}
\label{SecLattice}

\psset{unit=1.5cm}
\begin{figure}[p]
\begin{center}
\begin{pspicture}(0,0)(8.5,7.2)
\rput(0,6){\va}
\rput(1.5,6){\vb}
\rput(3,6){\vc}
\rput(4.5,6){\vd}
\rput(6,6){\ve}
\rput(7.5,6){\vf}
\rput(0,4.5){\pa}
\rput(1.5,4.5){\pb}
\rput(3,4.5){\pc}
\rput(4.5,4.5){\pd}
\rput(6,4.5){\pe}
\rput(7.5,4.5){\pf}
\rput(0,3){\qc}
\rput(1.5,3){\qd}
\rput(3,3){\qa}
\rput(4.5,3){\qb}
\rput(6,3){\qg}
\rput(7.5,3){\qh}
\rput(0,1){\da}
\rput(1.5,1){\db}
\rput(3,1){\dc}
\rput(4.5,1){\dd}
\rput(6.5,1.5){\mbox{or}}
\rput(6,1.7){\de}
\rput(6,.3){\dee}
\rput(7.5,1){\df}
\rput(1.25,-.1){$\underbrace{\hspace{3.75cm}}_{\mbox{\large $a(u)$}}$}
\rput(4.25,-.1){$\underbrace{\hspace{3.75cm}}_{\mbox{\large $b(u)$}}$}
\rput(6.5,-.1){$\underbrace{\hspace{1.5cm}}_{\mbox{\large $c_1(u)$}}$}
\rput(8,-.1){$\underbrace{\hspace{1.5cm}}_{\mbox{\large $c_2(u)$}}$}
\end{pspicture}
\end{center}
\caption{\label{vpd}Equivalent face tiles of the six vertex model in the vertex, particle (even and odd rows) and dimer representations. On the strip, the odd and even rows alternate. For periodic boundary conditions, all rows are odd. 
The heavy particle lines are drawn whenever the arrows disagree with the reference state, as shown in Figure~\ref{RefStates}. 
The particles move up and to the right on odd rows and up and to the left on even rows. 
}
\end{figure}
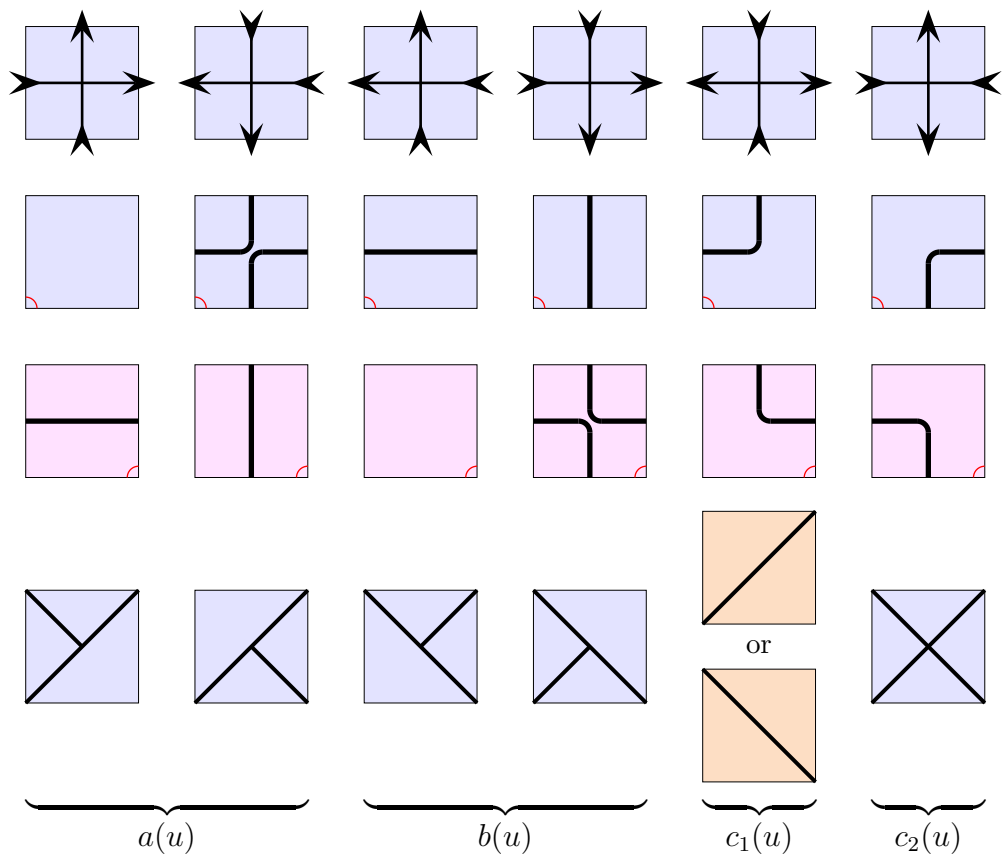

\psset{unit=1.2cm}
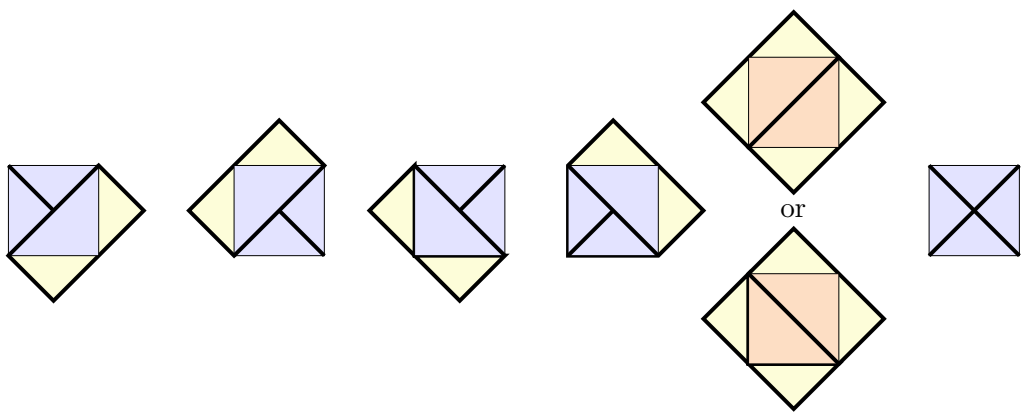
\begin{figure}
\begin{center}
\begin{pspicture}(0,-.4)(10.5,3.6)
\rput(-.2,1){\daDimer}
\rput(2.3,1){\dbDimer}
\rput(4.3,1){\dcDimer}
\rput(6,1){\ddDimer}
\rput(8.5,1.5){\mbox{or}}
\rput(8,2.2){\deDimer}
\rput(8,-.2){\deeDimer}
\rput(10,1){\df}
\end{pspicture}\qquad\mbox{}
\end{center}
\caption{\label{TheDimers} Face configurations showing (in light yellow) the one or two dimers associated with each face. No dimers are associated with the last face.}
\end{figure}

\definecolor{apricot}{rgb}{1,0.9,0.7}
\definecolor{darkapricot}{rgb}{1.2,0.95,0.9}
%
\def\da{
\pspolygon[linewidth=.25pt,fillstyle=solid,fillcolor=lightlightblue](0,0)(1,0)(1,1)(0,1)(0,0)
\psline[linewidth=1.5pt](0,0)(1,1)
\psline[linewidth=1.5pt](0,1)(.5,.5)}
\def\db{
\pspolygon[linewidth=.25pt,fillstyle=solid,fillcolor=lightlightblue](0,0)(1,0)(1,1)(0,1)(0,0)
\psline[linewidth=1.5pt](0,0)(1,1)
\psline[linewidth=1.5pt](.5,.5)(1,0)}
\def\dc{
\pspolygon[linewidth=.25pt,fillstyle=solid,fillcolor=lightlightblue](0,0)(1,0)(1,1)(0,1)(0,0)
\psline[linewidth=1.5pt](0,1)(1,0)
\psline[linewidth=1.5pt](1,1)(.5,.5)}
\def\dd{
\pspolygon[linewidth=.25pt,fillstyle=solid,fillcolor=lightlightblue](0,0)(1,0)(1,1)(0,1)(0,0)
\psline[linewidth=1.5pt](0,1)(1,0)
\psline[linewidth=1.5pt](.5,.5)(0,0)}
\def\de{
\pspolygon[linewidth=.25pt,fillstyle=solid,fillcolor=apricot](0,0)(1,0)(1,1)(0,1)(0,0)
\psline[linewidth=1.5pt](0,0)(1,1)}
\def\dee{
\pspolygon[linewidth=.25pt,fillstyle=solid,fillcolor=apricot](0,0)(1,0)(1,1)(0,1)(0,0)
\psline[linewidth=1.5pt](1,0)(0,1)}
\def\df{
\pspolygon[linewidth=.25pt,fillstyle=solid,fillcolor=lightlightblue](0,0)(1,0)(1,1)(0,1)(0,0)
\psline[linewidth=1.5pt](0,0)(1,1)
\psline[linewidth=1.5pt](0,1)(1,0)}

\def\da{
\pspolygon[linewidth=.25pt,fillstyle=solid,fillcolor=lightlightblue](0,0)(1,0)(1,1)(0,1)(0,0)
\psline[linewidth=1.5pt](0,0)(1,1)
\psline[linewidth=1.5pt](0,1)(.5,.5)}
\def\db{
\pspolygon[linewidth=.25pt,fillstyle=solid,fillcolor=lightlightblue](0,0)(1,0)(1,1)(0,1)(0,0)
\psline[linewidth=1.5pt](0,0)(1,1)
\psline[linewidth=1.5pt](.5,.5)(1,0)}
\def\dc{
\pspolygon[linewidth=.25pt,fillstyle=solid,fillcolor=lightlightblue](0,0)(1,0)(1,1)(0,1)(0,0)
\psline[linewidth=1.5pt](0,1)(1,0)
\psline[linewidth=1.5pt](1,1)(.5,.5)}
\def\dd{
\pspolygon[linewidth=.25pt,fillstyle=solid,fillcolor=lightlightblue](0,0)(1,0)(1,1)(0,1)(0,0)
\psline[linewidth=1.5pt](0,1)(1,0)
\psline[linewidth=1.5pt](.5,.5)(0,0)}
\def\de{
\pspolygon[linewidth=.25pt,fillstyle=solid,fillcolor=apricot](0,0)(1,0)(1,1)(0,1)(0,0)
\psline[linewidth=1.5pt](0,0)(1,1)}
\def\dee{
\pspolygon[linewidth=.25pt,fillstyle=solid,fillcolor=apricot](0,0)(1,0)(1,1)(0,1)(0,0)
\psline[linewidth=1.5pt](1,0)(0,1)}
\def\df{
\pspolygon[linewidth=.25pt,fillstyle=solid,fillcolor=lightlightblue](0,0)(1,0)(1,1)(0,1)(0,0)
\psline[linewidth=1.5pt](0,0)(1,1)
\psline[linewidth=1.5pt](0,1)(1,0)}
\def\fracdb{
\pspolygon[linewidth=.25pt,fillstyle=solid,fillcolor=lightlightblue](0,0)(0.5,-0.5)(1,0)(0,0)}
\def\fracdl{
\pspolygon[linewidth=.25pt,fillstyle=solid,fillcolor=lightlightblue](0,0)(-0.5,0.5)(0,1)(0,0)}
\def\fracdr{
\pspolygon[linewidth=.25pt,fillstyle=solid,fillcolor=lightlightblue](0,0)(0.5,0.5)(0,1)(0,0)}
\def\fracdt{
\pspolygon[linewidth=.25pt,fillstyle=solid,fillcolor=lightlightblue](0,0)(0.5,0.5)(1,0)(0,0)}
\def\fracdby{
\pspolygon[linewidth=.25pt,fillstyle=solid,fillcolor=apricot](0,0)(0.5,-0.5)(1,0)(0,0)}
\def\fracdly{
\pspolygon[linewidth=.25pt,fillstyle=solid,fillcolor=apricot](0,0)(-0.5,0.5)(0,1)(0,0)}
\def\fracdry{
\pspolygon[linewidth=.25pt,fillstyle=solid,fillcolor=apricot](0,0)(0.5,0.5)(0,1)(0,0)}
\def\fracdty{
\pspolygon[linewidth=.25pt,fillstyle=solid,fillcolor=apricot](0,0)(0.5,0.5)(1,0)(0,0)}
\def\drightup{
\pspolygon[linewidth=2.0pt,fillstyle=solid,fillcolor=lightyellow](0,0)(0.5,-0.5)(1.5,0.5)(1,1)(0,0)}
\def\dleftup{
\pspolygon[linewidth=2.0pt,fillstyle=solid,fillcolor=lightyellow](1,0)(0.5,-0.5)(-0.5,0.5)(0,1)(1,0)}
\def\drightupy{
\pspolygon[linewidth=2.0pt,fillstyle=solid,fillcolor=apricot](0,0)(0.5,-0.5)(1.5,0.5)(1,1)(0,0)}
\def\dleftupy{
\pspolygon[linewidth=2.0pt,fillstyle=solid,fillcolor=apricot](1,0)(0.5,-0.5)(-0.5,0.5)(0,1)(1,0)}
\def\drightupyy{
\pspolygon[linewidth=2.0pt,fillstyle=solid,fillcolor=darkapricot](0,0)(0.5,-0.5)(1.5,0.5)(1,1)(0,0)}
\def\dleftupyy{
\pspolygon[linewidth=2.0pt,fillstyle=solid,fillcolor=darkapricot](1,0)(0.5,-0.5)(-0.5,0.5)(0,1)(1,0)}

\subsection{Face tiles and equivalence of vertex, particle and dimer representations}

A mapping between the free-fermion six vertex model and dimer configurations was given in \cite{PVO2017}. 
The allowed six vertex (arrow conserving) face configurations and the equivalent tiles in the particle (even and odd rows) and dimer~\cite{KorepinZJ} representations are shown in Figure~\ref{vpd}. The vertex (arrow) degrees of freedom $\sigma_j=\pm 1$ and the particle occupation numbers $a_j=\half(1-\sigma_j)=0,1$ live on the medial lattice.  
The Boltzmann weights of the six vertex tiles are
\bea
a(u)=\rho\,\frac{\sin(\lambda-u)}{\sin\lambda},\quad b(u)=\rho\,\frac{\sin u}{\sin\lambda},\quad c_1(u)=\rho g,\quad c_2(u)=\frac{\rho}{g},\qquad \lambda\in (0,\pi),\quad \rho\in{\Bbb R}\label{6Vwts}
\eea
The spectral parameter $u$ plays the role of spatial anisotropy with $u=\frac{\lambda}{2}$ being the isotropic point. 
Geometrically~\cite{KimP}, varying $u$ effectively distorts a square tile into a rhombus with an opening anisotropy angle $\vartheta=\frac{\pi u}{\lambda}$. 
The arbitrary parameter $\rho$ is an overall normalization. Assuming boundary conditions such that there are an equal number of sources and sinks of horizontal arrows (vertices $c_1$ and $c_2$) along any row, the transfer matrix entries are all independent of the gauge factor~$g$ which may depend on $u$. 

At the free-fermion point ($\lambda=\frac{\pi}{2}$), the six vertex face weights reduce to
\bea
a(u)=\rho\cos u,\quad b(u)=\rho\sin u,\quad c_1(u)=\rho g,\quad c_2(u)=\frac{\rho}{g},\qquad  \rho\in{\Bbb R}\label{FreeFermion}
\eea
These weights satisfy the free-fermion condition 
\bea
a(u)^2+b(u)^2=c_1(u)c_2(u)\label{FreeFermionCond}
\eea
As shown in Section~\ref{SectTL}, with the special choice of gauge $g=z:=e^{iu}$, the tiles 
give a representation of the free-fermion algebra with generators $f_j$, $f_j^\dagger$ and, consequently, also a representation of the Temperley-Lieb algebra~\cite{TempLieb} with generators $e_j$ and loop fugacity $\beta=2\cos\lambda=0$. Explicitly, the face operators are
\bea
X_j(u)=\rho(\cos u\,I + \sin u\,e_j)\label{faceOps}
\eea
This Temperley-Lieb model is directly equivalent to an anisotropic dimer model as shown in Figures~\ref{vpd}, \ref{TheDimers} and \ref{vacBdy}. 
A dimer weight is assigned to the unique square face which is half-covered by the dimer as shown in Figure~\ref{TheDimers}. The statistical weights assigned to ``horizontal" and ``vertical" dimers are
\bea
\zeta_h(u)=a(u)=\rho\cos u,\qquad \zeta_v(u)=b(u)=\rho\sin u
\eea
Setting $g=\rho$, and allowing for the facts that (i) the $c_1$ face has two allowed configurations and (ii) no dimer covers the $c_2$ face, it follows that
\bea
c_1(u)=\zeta_h(u)^2+\zeta_v(u)^2=\rho^2(\cos^2 u+\sin^2 u)=\rho^2,\qquad c_2(u)=1
\eea
Additionally, fixing $\rho=\sqrt{2}$  at the isotropic point ($u=\frac{\lambda}{2}=\frac{\pi}{4}$) gives
\bea
a(\tfrac{\pi}{4})=1,\qquad b(\tfrac{\pi}{4})=1,\qquad c_1(\tfrac{\pi}{4})=2,\qquad c_2(\tfrac{\pi}{4})=1
\eea
It follows that, with this choice of gauge and normalization, the partition function at the isotropic point gives the correct counting of distinct dimer configurations. 


\def\leftzig{
\psline[linewidth=1.5pt,linestyle=solid,linecolor=red](0,0)(.5,.5)
\psline[linewidth=1.5pt,linestyle=solid,linecolor=red](0,1)(.5,.5)}
\def\rightzig{
\psline[linewidth=1.5pt,linestyle=solid,linecolor=red](.5,.5)(1,1)
\psline[linewidth=1.5pt,linestyle=solid,linecolor=red](.5,.5)(1,0)}
\def\leftzag{
\psline[linewidth=1.5pt,linestyle=solid,linecolor=red](0,0)(-.5,.5)(0,1)}
\def\rightzag{
\psline[linewidth=1.5pt,linestyle=solid,linecolor=red](1,0)(1.5,.5)(1,1)}

\psset{unit=1.4cm}
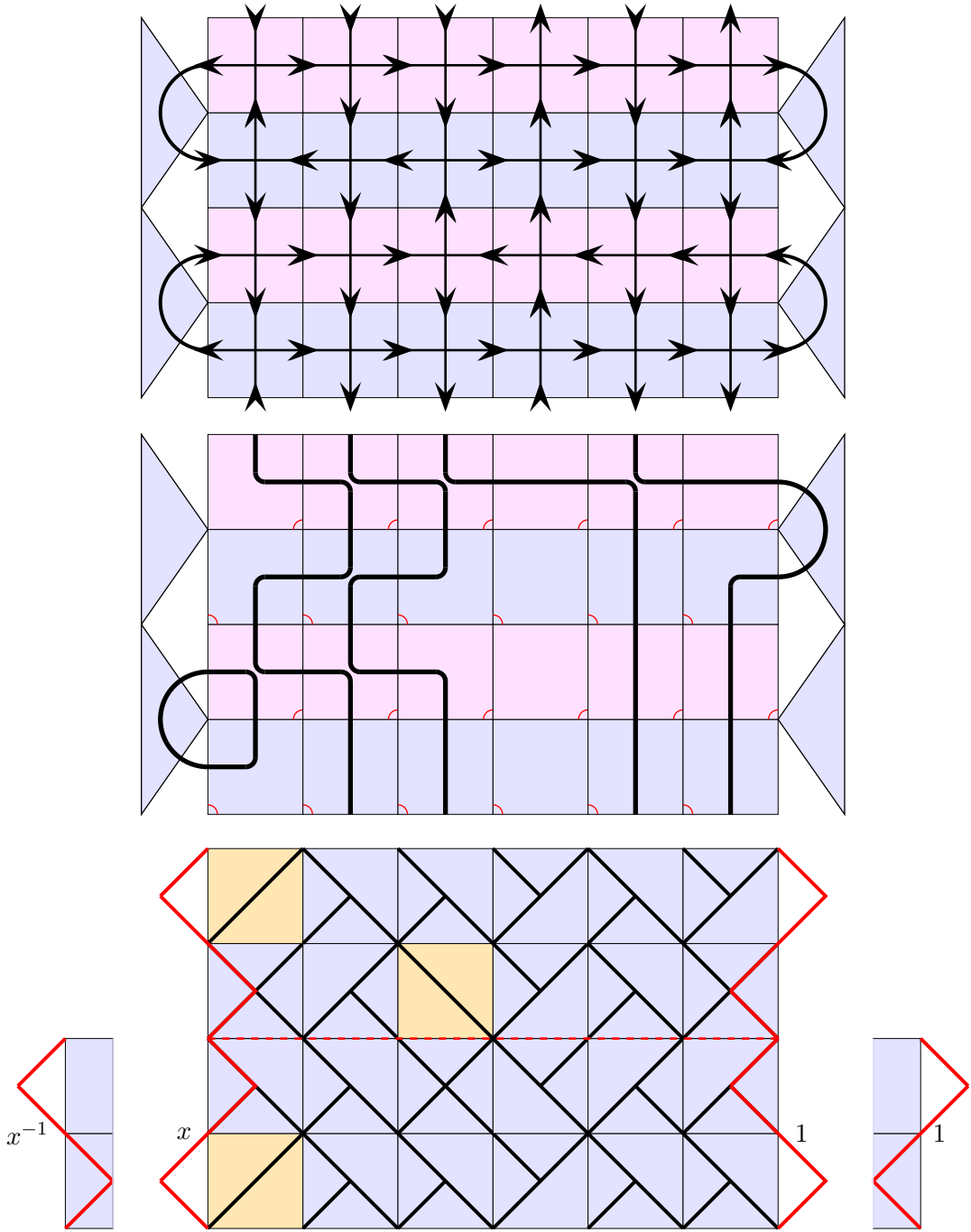
\begin{figure}
\begin{center}
\begin{pspicture}(0,0)(6,4)
\pspolygon[fillstyle=solid,fillcolor=lightlightblue,linewidth=.5pt](-.7,4)(-.7,2)(0,3)
\pspolygon[fillstyle=solid,fillcolor=lightlightblue,linewidth=.5pt](-.7,2)(-.7,0)(0,1)
\pspolygon[fillstyle=solid,fillcolor=lightlightblue,linewidth=.5pt](6.7,4)(6.7,2)(6,3)
\pspolygon[fillstyle=solid,fillcolor=lightlightblue,linewidth=.5pt](6.7,2)(6.7,0)(6,1)
\rput(0,0){\ve}
\rput(1,0){\vd}
\rput(2,0){\vd}
\rput(3,0){\va}
\rput(4,0){\vd}
\rput(5,0){\vd}
\rput(0,1){\vvd}
\rput(1,1){\vvd}
\rput(2,1){\vvf}
\rput(3,1){\vvc}
\rput(4,1){\vvb}
\rput(5,1){\vvb}
\rput(0,2){\vf}
\rput(1,2){\vb}
\rput(2,2){\ve}
\rput(3,2){\va}
\rput(4,2){\vd}
\rput(5,2){\vf}
\rput(0,3){\vve}
\rput(1,3){\vvd}
\rput(2,3){\vvd}
\rput(3,3){\vva}
\rput(4,3){\vvd}
\rput(5,3){\vva}
\psarc[linewidth=1.5pt](0,1){.5}{90}{270}
\psarc[linewidth=1.5pt](0,3){.5}{90}{270}
\psarc[linewidth=1.5pt](6,1){.5}{-90}{90}
\psarc[linewidth=1.5pt](6,3){.5}{-90}{90}
\end{pspicture}

\vspace{.2in}
\mbox{}\vspace{.8cm}\mbox{}\!\!
\begin{pspicture}(0,0)(6,4)
\pspolygon[fillstyle=solid,fillcolor=lightlightblue,linewidth=.5pt](-.7,4)(-.7,2)(0,3)
\pspolygon[fillstyle=solid,fillcolor=lightlightblue,linewidth=.5pt](-.7,2)(-.7,0)(0,1)
\pspolygon[fillstyle=solid,fillcolor=lightlightblue,linewidth=.5pt](6.7,4)(6.7,2)(6,3)
\pspolygon[fillstyle=solid,fillcolor=lightlightblue,linewidth=.5pt](6.7,2)(6.7,0)(6,1)
\rput(0,0){\pe}
\rput(1,0){\pd}
\rput(2,0){\pd}
\rput(3,0){\pa}
\rput(4,0){\pd}
\rput(5,0){\pd}
\rput(0,1){\qb}
\rput(1,1){\qb}
\rput(2,1){\qh}
\rput(3,1){\qa}
\rput(4,1){\qd}
\rput(5,1){\qd}
\rput(0,2){\pf}
\rput(1,2){\pb}
\rput(2,2){\pe}
\rput(3,2){\pa}
\rput(4,2){\pd}
\rput(5,2){\pf}
\rput(0,3){\qg}
\rput(1,3){\qb}
\rput(2,3){\qb}
\rput(3,3){\qc}
\rput(4,3){\qb}
\rput(5,3){\qc}
\psarc[linewidth=2pt](0,1){.5}{90}{270}
\psarc[linewidth=2pt](6,3){.5}{-90}{90}
\end{pspicture}

\vspace{-.13in}
\begin{pspicture}(0,0)(6,4)
\rput(0,0){\de}
\rput(1,0){\dd}
\rput(2,0){\dd}
\rput(3,0){\da}
\rput(4,0){\dd}
\rput(5,0){\dd}
\rput(0,1){\dd}
\rput(1,1){\dd}
\rput(2,1){\df}
\rput(3,1){\dc}
\rput(4,1){\db}
\rput(5,1){\db}
\rput(0,2){\df}
\rput(1,2){\db}
\rput(2,2){\dee}
\rput(3,2){\da}
\rput(4,2){\dd}
\rput(5,2){\df}
\rput(0,3){\de}
\rput(1,3){\dd}
\rput(2,3){\dd}
\rput(3,3){\da}
\rput(4,3){\dd}
\rput(5,3){\da}
\psframe[fillstyle=solid,fillcolor=lightlightblue,linewidth=0pt](-1.5,0)(-1.,1)
\psframe[fillstyle=solid,fillcolor=lightlightblue,linewidth=0pt](-1.5,1)(-1.,2)
\psframe[fillstyle=solid,fillcolor=lightlightblue,linewidth=0pt](7,0)(7.5,1)
\psframe[fillstyle=solid,fillcolor=lightlightblue,linewidth=0pt](7,1)(7.5,2)
\rput(-1,0){\rightzig}
\rput(-1,1){\rightzag}
\rput(-1,2){\rightzag}
\rput(-1,3){\rightzig}
\rput(-1.5,0){\leftzig}
\rput(-2.5,1){\rightzig}
\rput(5,0){\rightzag}
\rput(5,1){\rightzig}
\rput(5,2){\rightzig}
\rput(5,3){\rightzag}
\rput(6.5,0){\rightzig}
\rput(7.5,1){\leftzig}
\psline[linewidth=.5pt,linecolor=lightlightblue](-1.,0)(-1.,2)
\psline[linewidth=.5pt,linecolor=lightlightblue](7.,0)(7.,2)
\psline[linewidth=1pt,linecolor=red,linestyle=dashed,dash=3pt 3pt](0,2)(6,2)
\rput(-.25,1){$x$}
\rput(-1.9,1){$x^{-1}$}
\rput(6.25,1){$1$}
\rput(7.7,1){$1$}
\end{pspicture}
\end{center}
\caption{\label{vacBdy}Typical dimer configuration on a $6\times 4$ strip with vacuum boundary conditions in the vertex, particle and dimer representations. For the vertex representation, the boundary arrows can be in either one of the two possible directions (corresponding to a particle or vacancy in the particle representation). Particles move up and right on odd rows and up and left on even rows. The number of particles/down arrows inside the strip is conserved from double row to double row but not necessarily from single row to single row. For dimers, there are two different zigzag edges allowed independently on the left and right edges of each double row. The left boundary zigzags have weights $x,x^{-1}$ as shown. The right boundary zigzags have weight 1.} 
\end{figure}

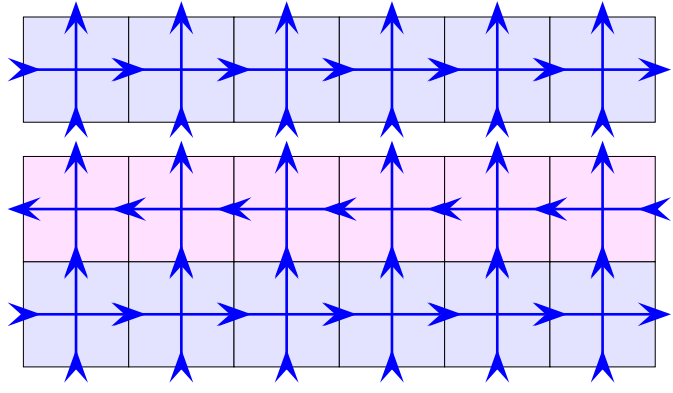
\begin{figure}
\begin{center}
\begin{pspicture}[shift=0](0,0)(6,1)
\rput(0,0){\var}
\rput(1,0){\var}
\rput(2,0){\var}
\rput(3,0){\var}
\rput(4,0){\var}
\rput(5,0){\var}
\end{pspicture}

\bigskip

\begin{pspicture}(0,0)(6,2)
\rput(0,0){\var}
\rput(1,0){\var}
\rput(2,0){\var}
\rput(3,0){\var}
\rput(4,0){\var}
\rput(5,0){\var}
\rput(0,1){\vcs}
\rput(1,1){\vcs}
\rput(2,1){\vcs}
\rput(3,1){\vcs}
\rput(4,1){\vcs}
\rput(5,1){\vcs}
\end{pspicture}
\end{center}
\caption{\label{RefStates}Reference states for the single and double row transfer matrices for mapping onto the particle representation. The reference arrows point up and to the right for the single row transfer matrices. For the double row transfer matrices, the reference arrows point up and right on odd rows and up and left on even rows.}
\end{figure}

In addition to the vertex and dimer representations, the six vertex free-fermion model admits a particle representation as shown in Figures \ref{vpd} and \ref{vacBdy}. A reference state on the strip is fixed as in Figure~\ref{RefStates}. An edge of a given vertex is a segment of a particle trajectory (and has particle occupation number $a_j=1$) if its arrow points in the opposite direction to that of the reference state. Otherwise, if the edge arrow points in the same direction as the reference state, the edge is not a segment of a particle trajectory (and the particle occupation is $a_j=0$). The segments of particle trajectories live on the medial lattice and are indicated with heavy lines in Figure~\ref{RefStates}. The number of particles is conserved and their trajectories are non-intersecting. 
The particle representation is the simplest of the three representations and is convenient for coding in Mathematica~\cite{Wolfram} and for manipulations in the diagrammatic planar algebra~\cite{Jones} so we usually work in the particle representation.
The ${\Bbb Z}_2$ arrow reversal symmetry of the vertex model implies a particle-hole duality in the particle representation. 

\subsection{Free-fermion and Temperley-Lieb algebras}
\label{SectTL}

\psset{unit=.75cm}
In this section, we consider the free-fermion model (\ref{FreeFermion}) with $\lambda=\frac{\pi}{2}$ and set $g=z:=e^{iu}$ and $\rho=1$. 

\subsubsection{Free-fermion algebra}
Regarding the elementary tiles as operators acting on an upper (zigzag) row particle configuration to produce a lower (zigzag) row particle configuration, we write them respectively as
\bea
E_j=n_j^{00},\ n_j^{11},\ f_j^\dagger f_{j+1},\ f_{j+1}^\dagger f_j,\ n_j^{10},\ n_j^{01},\qquad\quad 
n_j^{00}+n_j^{11}+n_j^{10}+n_j^{01}=I
\label{elemOps}
\eea
The four operators $n_j^{ab}$ are (diagonal) orthogonal projection operators which factorize into single-site orthogonal projectors corresponding to left and right  half (triangular) tiles
\bea
n_j^{ab}=n_j^an_{j+1}^b,\qquad n_j^a n_j^b=\delta_{ab}\,n_j^a,\qquad n_j^0+n_j^1=I,\qquad a,b=0,1
\eea
Here $n_j=n_j^1=f_j^\dagger f_j$ is the number operator counting single site occupancy at position $j$ and $n_j^0=f_j f_j^\dagger=1-f_j^\dagger f_j$ is the dual number operator counting the single-site vacancies at position $j$. The operators $f_j$ and $f_j^\dagger$ are single-site particle annihilation and creation operators respectively. 
It follows that all of the elementary tile operators can be written as combinations of bilinears in the fermion operators $f_j$ and $f_j^\dagger$. 
Diagrammatically, the particle hopping terms $f_j^\dagger f_{j+1}$ and $f_{j+1}^\dagger f_j$ factorize into left and right half (triangular) tiles 
\bea
f_j^\dagger f_{j+1}=\begin{pspicture}[shift=-.65](0,0)(1.5,1.5)
\pspolygon[linewidth=.5pt,fillstyle=solid,fillcolor=tile](.75,0)(1.5,.75)(.75,1.5)(0,.75)
\psline[linewidth=1.25pt](.375,.375)(1.125,1.125)
\psline[linewidth=.5pt,linestyle=dashed,dash=3pt 2pt](.75,0)(.75,1.5)
\rput[tr](.32,.375){\small $j$}
\rput[tl](1.125,.375){\small $j\!\!+\!\!1$}
\end{pspicture}\qquad\quad
f_{j+1}^\dagger f_j=\begin{pspicture}[shift=-.65](0,0)(1.5,1.5)
\pspolygon[linewidth=.5pt,fillstyle=solid,fillcolor=tile](.75,0)(1.5,.75)(.75,1.5)(0,.75)(.75,0)
\psline[linewidth=1.25pt](1.125,.375)(.375,1.125)
\psline[linewidth=.5pt,linestyle=dashed,dash=2pt 2pt](.75,0)(.75,1.5)
\rput[tr](.32,.375){\small $j$}
\rput[tl](1.125,.375){\small $j\!\!+\!\!1$}
\end{pspicture}
\eea
so that the fermion generators are represented by half (triangular) tiles
\bea
f_j^\dagger=\begin{pspicture}[shift=-.65](0,0)(.75,.75)
\pspolygon[linewidth=.5pt,fillstyle=solid,fillcolor=tile](.75,0)(.75,1.5)(0,.75)
\psline[linewidth=1.25pt](.375,.375)(.75,.75)
\rput[tr](.32,.375){\small $j$}
\end{pspicture}
\;=\;\begin{pspicture}[shift=-.65](.75,0)(1.5,1.5)
\pspolygon[linewidth=.5pt,fillstyle=solid,fillcolor=tile](.75,0)(1.5,.75)(.75,1.5)(.75,0)
\psline[linewidth=1.25pt](1.125,.375)(.75,.75)
\rput[tl](1.125,.375){\small $j$}
\end{pspicture},\qquad f_j=\begin{pspicture}[shift=-.65](0,0)(.75,.75)
\pspolygon[linewidth=.5pt,fillstyle=solid,fillcolor=tile](.75,0)(.75,1.5)(0,.75)
\psline[linewidth=1.25pt](.375,1.125)(.75,.75)
\rput[tr](.32,.375){\small $j$}
\end{pspicture}
\;=\;\begin{pspicture}[shift=-.65](.75,0)(1.5,1.5)
\pspolygon[linewidth=.5pt,fillstyle=solid,fillcolor=tile](.75,0)(1.5,.75)(.75,1.5)(.75,0)
\psline[linewidth=1.25pt](1.125,1.125)(.75,.75)
\rput[tl](1.125,.375){\small $j$}
\end{pspicture}\label{fermiTiles}
\eea

As defined here, the operators $f_j$ and $f_j^\dagger$ satisfy the mixed commutation relations
\bea
f_j^2=(f_j^\dagger)^2=0,\qquad \{f_j,f_j^\dagger\}=1,\qquad [f_j^\dagger,f_k]= [f_j^\dagger,f_k^\dagger]= [f_j,f_k]=0,\ \ j\ne k
\eea
so they are not, strictly speaking, fermion operators. However, it is straightforward~\cite{LSM61} to transform by a linear transformation to new operators that are strictly fermion operators. 

\subsubsection{Temperley-Lieb algebra}
The Temperley-Lieb algebra is realized~\cite{GST14, PasquierSaleur} by setting $x=e^{i\lambda}=i$ and defining the generators
\begin{subequations}
\begin{align}
e_j&=x \tilee+x^{-1} \tilef+\tilec+\tiled\label{TLgen}\\
&=xf_j^\dagger f_j (1-f_{j+1}^\dagger f_{j+1})+x^{-1}(1-f_j^\dagger f_j) f_{j+1}^\dagger f_{j+1} +f_j^\dagger f_{j+1}+f_{j+1}^\dagger f_j\\
&=xf_j^\dagger f_j+x^{-1}f_{j+1}^\dagger f_{j+1}+f_j^\dagger f_{j+1}+f_{j+1}^\dagger f_j\label{TLgenf}
\end{align}
\end{subequations}
The quartic (interacting) terms vanish, since $\beta=x+x^{-1}=0$, leaving bilinears in fermion operators. 
Using the planar algebra~\cite{Jones} of tiles, it readily follows that these operators yield a representation of the Temperley-Lieb algebra
\bea
e_j^2=\beta e_j=0,\qquad e_je_{j\pm 1}e_j=e_j,\qquad \beta=2\cos\lambda=x+x^{-1}=0
\eea
Equivalently this follows, purely from fermionic algebra, by writing the generators in terms of the fermionic operators $f_j$ and $f_j^\dagger$ as in (\ref{TLgenf}).

\section{Six Vertex Model on the Strip}
\label{secSixVertex}

The commuting double row transfer matrices of the six vertex model were constructed algebraically by Sklyanin~\cite{Sklyanin}. In this section, we develop a diagrammatic construction of the commuting double row transfer matrices of the six vertex model by generalizing the methods of \cite{BPO} and using planar algebras~\cite{Jones}.

\subsection{Local relations}

We describe the local relations satisfied by the six vertex face operators in the planar and linear algebra settings. 
Because it has local degrees of freedom, in the form of particle occupation numbers, the planar algebra of the six vertex model just involves local tensor contractions of the indices giving the particle numbers. By fixing the planar algebra operators to act in an arbitrary fixed direction, the local relations presented in this section are easily established concretely using matrix representations (for example in Mathematica~\cite{Wolfram}). Alternatively, a local relation can be established diagrammatically directly in the planar algebra setting. It then follows that the local relation holds for all matrix representations and for all choices of the direction of action.

\subsubsection{Face operators, symmetries and face weights}

As elements of a planar algebra, the face operators of the six vertex model in the particle representation decompose~\cite{BEPR2015}  into a sum of contributions from six elementary tiles
\psset{unit=.75cm}
\setlength{\unitlength}{.75cm}
\begin{align}
\mydiam{}{}{}{}{u,g}&=s_1(-u)\Bigg(\!\!\tilea \!+\!\tileb\!\!\Bigg)+s_0(u)\Bigg(\!\!\tilec\!+\!\tiled\!\!\Bigg)+g\tilee\!+g^{-1}\tilef
\label{faceDecomp31}
\end{align}
where $s_k(u)=\frac{\sin(u+k\lambda)}{\sin\lambda}$ and $g$ is a gauge factor. 
Multiplication of the tiles in the planar algebra is given~\cite{BEPR2015,PVO2017} by local tensor contraction of indices $a,b,c,d,\ldots=0,1$ specifying the particle occupation numbers on the centers of the tile edges. As orientated in (\ref{faceDecomp31}), the face operators are invariant under reflection about the horizontal diagonal and not invariant (for $g\ne 1$) under reflection about the vertical diagonal. 
Rotating the face operator by $90^\circ$ gives
\begin{align}
\mydiampink{}{}{}{}{u,g}&=s_1(-u)\Bigg(\!\!\tilea \!+\!\tilek\!\!\Bigg)+s_0(u)\Bigg(\!\!\tilec\!+\!\tiled\!\!\Bigg)+g\tileg\!+g^{-1}\tileh
\end{align}
Further rotations by $90^\circ$ give
\bea
\mydiamtop{}{}{}{}{u,g}=\mydiam{}{}{}{}{u,\!g^{-1}},\qquad \mydiampinkleft{}{}{}{}{u,g}=\mydiampink{}{}{}{}{u,\!g^{-1}}
\eea
Only the occupation numbers $a,b,c,d=0,1$ of the edges are important. The colors of the face operators (indicating their relative orientation) and the internal particle trajectories are just for easy visual identification so that
\bea
\tileb\equiv\tilek\equiv\tilecd
\eea

Usually, we work in the fixed gauge $g=z:=e^{iu}$ with $x=e^{i\lambda}$ and set
\bea
\mydiam{}{}{}{}{u}=\mydiam{}{}{}{}{u,z}
\eea
Using this gauge, gives 
\bea
\mydiam{}{}{}{}{u}=s_1(-u)\;\mydiam{}{}{}{}{0}+s_0(u)\;\mydiam{}{}{}{}{\lambda}
\eea
where the generators of the planar Temperley-Lieb algebra are
\begin{subequations}
\begin{align}
\mydiam{}{}{}{}{0}&=\tilea +\tileb+\tilee+\tilef\\
\mydiam{}{}{}{}{\lambda}&=\tilec+\!\tiled+x\tilee+x^{-1}\tilef
\end{align}
\end{subequations}
Acting vertically, the first operator acts as the identity and the second, acting at position $j$, acts as the Temperley-Lieb generator $e_j$.

More conventionally, the bulk face weights of the six vertex model are\\[-28pt]
\psset{unit=.85cm}
\begin{align}
\WGblue0000z2g&=
\begin{pspicture}[shift=-.4](-.4,0)(1.3,2)
\pspolygon[linewidth=.25pt,fillstyle=solid,fillcolor=lightlightblue](0,0)(1,0)(1,1)(0,1)(0,0)
\rput(-.25,.5){\small $0$}
\rput(.5,-.25){\small $0$}
\rput(1.25,.5){\small $0$}
\rput(.5,1.25){\small $0$}
\psarc[linewidth=.5pt,linecolor=red](0,0){.125}{0}{90}
\end{pspicture}=
\begin{pspicture}[shift=-.4](-.4,0)(1.3,2)
\pspolygon[linewidth=.25pt,fillstyle=solid,fillcolor=lightpurple](0,0)(1,0)(1,1)(0,1)(0,0)
\rput(-.25,.5){\small $0$}
\rput(.5,-.25){\small $0$}
\rput(1.25,.5){\small $0$}
\rput(.5,1.25){\small $0$}
\psarc[linewidth=.5pt,linecolor=red](1,0){.125}{90}{180}
\end{pspicture}=s_1(-u)\nonumber\\
\WGblue0101z2g&=
\begin{pspicture}[shift=-.4](-.4,0)(1.3,2)
\pspolygon[linewidth=.25pt,fillstyle=solid,fillcolor=lightlightblue](0,0)(1,0)(1,1)(0,1)(0,0)
\rput(-.25,.5){\small $0$}
\rput(.5,-.25){\small $1$}
\rput(1.25,.5){\small $0$}
\rput(.5,1.25){\small $1$}
\multirput(0.5,0)(.1,0){1}{\psline[linewidth=1pt](0,0)(0,1.)}
\psarc[linewidth=.5pt,linecolor=red](0,0){.125}{0}{90}
\end{pspicture}=
\begin{pspicture}[shift=-.4](-.4,0)(1.3,2)
\pspolygon[linewidth=.25pt,fillstyle=solid,fillcolor=lightpurple](0,0)(1,0)(1,1)(0,1)(0,0)
\rput(-.25,.5){\small $0$}
\rput(.5,-.25){\small $1$}
\rput(1.25,.5){\small $0$}
\rput(.5,1.25){\small $1$}
\multirput(0.5,0)(.1,0){1}{\psline[linewidth=1pt](0,0)(0,1.)}
\psarc[linewidth=.5pt,linecolor=red](1,0){.125}{90}{180}
\end{pspicture}=s_0(u)\nonumber\\
\WGblue0110z2g&=
\begin{pspicture}[shift=-.4](-.4,0)(1.3,2)
\pspolygon[linewidth=.25pt,fillstyle=solid,fillcolor=lightlightblue](0,0)(1,0)(1,1)(0,1)(0,0)
\rput(-.25,.5){\small $0$}
\rput(.5,-.25){\small $1$}
\rput(1.25,.5){\small $1$}
\rput(.5,1.25){\small $0$}
\psbezier[linewidth=1pt,linecolor=black](.5,0)(.5,.4)(.6,.5)(1,.5)
\psarc[linewidth=.5pt,linecolor=red](0,0){.125}{0}{90}
\end{pspicture}=
\begin{pspicture}[shift=-.4](-.4,0)(1.3,2)
\pspolygon[linewidth=.25pt,fillstyle=solid,fillcolor=lightpurple](0,0)(1,0)(1,1)(0,1)(0,0)
\rput(-.25,.5){\small $0$}
\rput(.5,-.25){\small $0$}
\rput(1.25,.5){\small $1$}
\rput(.5,1.25){\small $1$}
\psbezier[linewidth=1pt,linecolor=black](1,.5)(.6,.5)(.5,.6)(.5,1)
\psarc[linewidth=.5pt,linecolor=red](1,0){.125}{90}{180}
\end{pspicture}=g^{-1}\label{bulkFaceWts}\\
\WGblue1001z2g&=
\begin{pspicture}[shift=-.4](-.4,0)(1.3,2)
\pspolygon[linewidth=.25pt,fillstyle=solid,fillcolor=lightlightblue](0,0)(1,0)(1,1)(0,1)(0,0)
\rput(-.25,.5){\small $1$}
\rput(.5,-.25){\small $0$}
\rput(1.25,.5){\small $0$}
\rput(.5,1.25){\small $1$}
\psbezier[linewidth=1pt,linecolor=black](0,.5)(.4,.5)(.5,.6)(.5,1)
\psarc[linewidth=.5pt,linecolor=red](0,0){.125}{0}{90}
\end{pspicture}=
\begin{pspicture}[shift=-.4](-.4,0)(1.3,2)
\pspolygon[linewidth=.25pt,fillstyle=solid,fillcolor=lightpurple](0,0)(1,0)(1,1)(0,1)(0,0)
\rput(-.25,.5){\small $1$}
\rput(.5,-.25){\small $1$}
\rput(1.25,.5){\small $0$}
\rput(.5,1.25){\small $0$}
\psbezier[linewidth=1pt,linecolor=black](.5,0)(.5,.4)(.4,.5)(0,.5)
\psarc[linewidth=.5pt,linecolor=red](1,0){.125}{90}{180}
\end{pspicture}=g\nonumber\\
\WGblue1010z2g&=
\begin{pspicture}[shift=-.4](-.4,0)(1.3,2)
\pspolygon[linewidth=.25pt,fillstyle=solid,fillcolor=lightlightblue](0,0)(1,0)(1,1)(0,1)(0,0)
\rput(-.25,.5){\small $1$}
\rput(.5,-.25){\small $0$}
\rput(1.25,.5){\small $1$}
\rput(.5,1.25){\small $0$}
\multirput(0,0)(0,0){1}{\psline[linewidth=1pt](0,.5)(1,.5)}
\psarc[linewidth=.5pt,linecolor=red](0,0){.125}{0}{90}
\end{pspicture}=
\begin{pspicture}[shift=-.4](-.4,0)(1.3,2)
\pspolygon[linewidth=.25pt,fillstyle=solid,fillcolor=lightpurple](0,0)(1,0)(1,1)(0,1)(0,0)
\rput(-.25,.5){\small $1$}
\rput(.5,-.25){\small $0$}
\rput(1.25,.5){\small $1$}
\rput(.5,1.25){\small $0$}
\multirput(0,0)(0,0){1}{\psline[linewidth=1pt](0,.5)(1,.5)}
\psarc[linewidth=.5pt,linecolor=red](1,0){.125}{90}{180}
\end{pspicture}=s_0(u)\nonumber\\
\WGblue1111z2g&=
\begin{pspicture}[shift=-.4](-.4,0)(1.3,2)
\pspolygon[linewidth=.25pt,fillstyle=solid,fillcolor=lightlightblue](0,0)(1,0)(1,1)(0,1)(0,0)
\rput(-.25,.5){\small $1$}
\rput(.5,-.25){\small $1$}
\rput(1.25,.5){\small $1$}
\rput(.5,1.25){\small $1$}
\psbezier[linewidth=1pt,linecolor=black](0,.5)(.4,.5)(.5,.6)(.5,1)
\psbezier[linewidth=1pt,linecolor=black](.5,0)(.5,.4)(.6,.5)(1,.5)
\psarc[linewidth=.5pt,linecolor=red](0,0){.125}{0}{90}
\end{pspicture}=
\begin{pspicture}[shift=-.4](-.4,0)(1.3,2)
\pspolygon[linewidth=.25pt,fillstyle=solid,fillcolor=lightpurple](0,0)(1,0)(1,1)(0,1)(0,0)
\rput(-.25,.5){\small $1$}
\rput(.5,-.25){\small $1$}
\rput(1.25,.5){\small $1$}
\rput(.5,1.25){\small $1$}
\psbezier[linewidth=1pt,linecolor=black](.5,0)(.5,.4)(.4,.5)(0,.5)
\psbezier[linewidth=1pt,linecolor=black](1,.5)(.6,.5)(.5,.6)(.5,1)
\psarc[linewidth=.5pt,linecolor=red](1,0){.125}{90}{180}
\end{pspicture}=s_1(-u)\nonumber\\[-6pt] \nonumber
\end{align}
The set of six allowed (blue) faces is not invariant under rotations through $90^\circ$. There is therefore no crossing symmetry. Instead, we distinguish the set of six rotated faces (pink) by the position of the corner marked by the (red) arc. In the blue faces, the particles move up and to the right and, in the pink faces, they move up and to the left. A face weight is unchanged under a rotation if the face configuration and the marked corner are rotated together. Again, the colour of the faces is just for easy visual identification.

\begin{subequations}
The six vertex face weights can be organized into an $\check R$-matrix. Explicitly, choosing the particular basis $\{(0,0),(0,1),(1,0),(1,1)\}$ gives
\bea
\WGblue abcdz2g=X(u,g)_{ab}{}^{dc},\qquad X(u,g)=\begin{pmatrix} s_1(-u)&0&0&0\\ 0&g^{-1}&s_0(u)&0\\0&s_0(u)&g&0\\ 0&0&0&s_1(-u)
\end{pmatrix}\\
\WGblue abcdz2g=\tilde X(u,g)_{da}{}^{cb},\qquad \tilde X(u,g)=\begin{pmatrix} s_1(-u)&0&0&g^{-1}\\ 0&0&s_0(u)&0\\0&s_0(u)&0&0\\ g&0&0&s_1(-u)
\end{pmatrix}
\eea
\end{subequations}
Let us define
\bea
X_j(u,g)=I\otimes I\otimes\cdots I\otimes X(u,g)\otimes I\cdots \otimes I\otimes I
\eea
acting on $({\Bbb C}^2)^{\otimes N}$ where $X(u,g)$ acts in the slots $j$ and $j+1$ and similarly for $\tilde X(u,g)$. Setting 
\bea
X_j(u)=X_j(u,z)=s_1(-u)\,I + s_0(u)\,e_j
\label{faceOps2}
\eea
the generators of the linear Temperley-Lieb algebra are then
\bea
X_j(0)=I,\qquad X_j(\lambda)=e_j, \quad j=1,2,\ldots,N-1
\eea
satisfying
\bea
e_j^2=\beta e_j,\qquad e_je_{j\pm 1}e_j=e_j,\qquad j=1,2,\ldots,N-1,\qquad \beta=x+x^{-1}
\eea
This corresponds to the linear vertical action of the planar algebra. 
\psset{unit=.6cm}

\subsubsection{Inversion relations}

The elementary face weights satisfy two distinct inversion relations. 
In the planar algebra, they are
\begin{subequations}
\bea
\mbox{Inv1}:\quad\raisebox{-.7cm}{
\psset{unit=.8cm}
\begin{pspicture}(-1.5,-1)(2.5,1)
\rput(0,0){\diamodd{u,g}}
\psarc[linewidth=.5pt,linecolor=red](-.9,0){.15}{-45}{45}
\rput(2,0){\diamodd{-u,\frac{1}{g}}}
\psarc[linewidth=.5pt,linecolor=red](1.1,0){.15}{-45}{45}
\pscircle[fillstyle=solid,fillcolor=black](1.6,-.5){.05}
\pscircle[fillstyle=solid,fillcolor=black](.6,-.5){.05}
\pscircle[fillstyle=solid,fillcolor=black](1.6,.5){.05}
\pscircle[fillstyle=solid,fillcolor=black](.6,.5){.05}
\psline[linestyle=dashed,linewidth=.5pt](.6,-.5)(1.6,-.5)
\psline[linestyle=dashed,linewidth=.5pt](.6,.5)(1.6,.5)
\end{pspicture}} \quad
\;=\eta_1(u)\,
\psset{unit=.8cm}
\raisebox{-.7cm}{\diamoddleft{0}},\qquad \eta_1(u)=s_1(u)s_1(-u)\\[4pt]
\mbox{Inv2}:\quad\raisebox{-.7cm}{
\psset{unit=.8cm}
\begin{pspicture}(-1.5,-1)(2.5,1)
\rput(0,0){\diamoddy{2\lambda\!-\!u,g}}
\psarc[linewidth=.5pt,linecolor=red](.1,-1){.15}{45}{135}
\rput(2,0){\diamodd{u,g}}
\psarc[linewidth=.5pt,linecolor=red](2.1,-1){.15}{45}{135}
\pscircle[fillstyle=solid,fillcolor=black](1.6,-.5){.05}
\pscircle[fillstyle=solid,fillcolor=black](.6,-.5){.05}
\pscircle[fillstyle=solid,fillcolor=black](1.6,.5){.05}
\pscircle[fillstyle=solid,fillcolor=black](.6,.5){.05}
\psline[linestyle=dashed,linewidth=.5pt](.6,-.5)(1.6,-.5)
\psline[linestyle=dashed,linewidth=.5pt](.6,.5)(1.6,.5)
\end{pspicture}} \quad
\;=\eta_2(u) \,
\psset{unit=.8cm}
\raisebox{-.7cm}{\diamoddleft{0}},\qquad \eta_2(u)=s_0(u)s_2(-u)
\eea
\label{InRels}
\end{subequations}
In the linear algebra acting from left to right, these become
\begin{subequations}
\begin{align}
&X_j(u,g)X_j(-u,1/g)=s_1(u)s_1(-u)\,I\\[2pt]
&\tilde X_j(2\lambda-u,g) \tilde X_j(u,g)=s_0(u)s_2(-u)\,I
\end{align}
\end{subequations}
Up to the scalar on the right side, the face $\tilde X_j(2\lambda-u,g)$ (shown in yellow) is the inverse of the face $\tilde X_j(u,g)$. 
We also observe the commutation relations
\bea
[X_j(u),X_j(v)]=0,\qquad [\tilde X_j(u,g),\tilde X_j(v,g)]=0
\eea

\def\gYBELHS #1#2#3#4#5{
\begin{pspicture}[shift=-1](-.2,-.2)(2.4,2)
\pspolygon[fillstyle=solid,fillcolor=lightlightblue](0.5,0)(1.5,0)(2,0.866)(1.5,1.732)(0.5,1.732)(-1.5,1.732)(-2,0.866)(-1.5,0)(0.5,0)
\pspolygon[fillstyle=solid,fillcolor=lightlightblue](0,0.866)(0.5,1.732)(-.5,1.732)(-1.,0.866)(-.5,0)(0.5,0)
\psline(1.5,0)(1,0.866)(0,0.866)
\psline(1,0.866)(1.5,1.732)
\psarc[linecolor=red](-1.5,0){0.1}{0}{120}
\psarc[linecolor=red](-1,0.866){0.1}{60}{180}
\psarc[linecolor=red](0,0.866){0.1}{60}{180}
\psarc[linecolor=red](0.5,0){0.1}{0}{120}
\psarc[linecolor=red](1,0.866){0.1}{60}{180}
\psarc[linecolor=red](1.5,0){0.1}{60}{120}
\psarc[linecolor=red](-.5,0){0.1}{0}{120}
\rput(-1.2,0.433){\scriptsize $#1$}
\rput(-1.2,1.299){\scriptsize $#2$}
\rput(.8,0.433){\scriptsize $#3$}
\rput(.8,1.299){\scriptsize $#4$}
\rput(1.5,0.866){\scriptsize $#5$}
\psline[linewidth=1.8pt,linecolor=red,linestyle=solid](1,0.866)(-2.1,0.866)
\psline[linewidth=1.8pt,linecolor=red,linestyle=solid](1,0.866)(1.55,1.8)
\psline[linewidth=1.8pt,linecolor=red,linestyle=solid](1,0.866)(1.55,-.1)
\psline[linewidth=1.pt,linestyle=dotted](-.5,1.256)(0.,1.256)
\psline[linewidth=1.pt,linestyle=dotted](-.5,0.38)(0.,0.38)
\rput(-0.5,1.){\color{red}{\scriptsize $cut 1$}}
\rput(1.5,1.299){\color{red}{\scriptsize $cut 2$}}
\rput(1.5,0.433){\color{red}{\scriptsize $cut 3$}}
\multiput(-1.5,0.866)(1,0){3}{\pscircle[fillstyle=solid,fillcolor=black](0,0){0.045}}
\pscircle[fillstyle=solid,fillcolor=black](1.24,1.26){0.045}
\pscircle[fillstyle=solid,fillcolor=black](1.24,0.44){0.045}
\rput[br](-1.85,0.38){\scriptsize $a$}
\rput[br](0,-0.2){\scriptsize $b$}
\rput[bl](1.85,0.38){\scriptsize $c$}
\rput[bl](1.85,1.256){\scriptsize $d$}
\rput[bl](0,1.85){\scriptsize $e$}
\rput[br](-1.8,1.256){\scriptsize $f$}
\end{pspicture}
}


\def\gYBERHS #1#2#3#4#5{
\begin{pspicture}[shift=-1](-.2,-.2)(2.2,2)
\pspolygon[fillstyle=solid,fillcolor=lightlightblue](0,0.866)(0.5,0)(3.5,0)(4,0.866)(3.5,1.732)(0.5,1.732)(0,0.866)
\pspolygon[fillstyle=solid,fillcolor=lightlightblue](2,0.866)(1.5,1.732)(2.5,1.732)(3,0.866)(2.5,0)(1.5,0)(2,0.866)
\psline(0.5,0)(1,0.866)(0.5,1.732)
\psline(1,0.866)(2,0.866)
\rput(0.5,0.866){\scriptsize $#1$}
\rput(1.2,0.433){\scriptsize $#2$}
\rput(1.2,1.299){\scriptsize $#3$}
\rput(3.2,0.433){\scriptsize $#4$}
\rput(3.2,1.299){\scriptsize $#5$}
\psarc[linecolor=red](1,0.866){0.1}{0}{120}
\psarc[linecolor=red](2,0.866){0.1}{0}{120}
\psarc[linecolor=red](1.5,0){0.1}{60}{180}
\psarc[linecolor=red](2.5,0){0.1}{60}{180}
\psarc[linecolor=red](0.5,0.){0.1}{60}{120}
\psarc[linecolor=red](3,0.866){0.1}{0}{120}
\psarc[linecolor=red](3.5,0){0.1}{60}{180}
\psline[linewidth=1.8pt,linecolor=red,linestyle=solid](1,0.866)(4.1,0.866)
\psline[linewidth=1.8pt,linecolor=red,linestyle=solid](1,0.866)(0.43,1.85)
\psline[linewidth=1.8pt,linecolor=red,linestyle=solid](1,0.866)(0.43,-.1)
\psline[linewidth=1.pt,linestyle=dotted](2,1.256)(2.5,1.256)
\psline[linewidth=1.pt,linestyle=dotted](2,0.38)(2.5,0.38)
\rput(2.5,1.){\color{red}{\scriptsize $cut 1$}}
\rput(0.55,1.3){\color{red}{\scriptsize $cut 3$}}
\rput(0.55,0.5){\color{red}{\scriptsize $cut 2$}}
\multiput(1.5,0.866)(1,0){3}{\pscircle[fillstyle=solid,fillcolor=black](0,0){0.045}}
\pscircle[fillstyle=solid,fillcolor=black](0.76,1.26){0.045}
\pscircle[fillstyle=solid,fillcolor=black](0.76,0.44){0.045}
\rput[br](0.2,0.38){\scriptsize $a$}
\rput[br](2,-0.2){\scriptsize $b$}
\rput[bl](3.85,0.38){\scriptsize $c$}
\rput[bl](3.85,1.256){\scriptsize $d$}
\rput[bl](2,1.85){\scriptsize $e$}
\rput[br](0.2,1.256){\scriptsize $f$}
\end{pspicture}
}

\def\YBELHS #1#2#3{
\begin{pspicture}[shift=-1](-.2,-.2)(2.4,2)
\pspolygon[fillstyle=solid,fillcolor=lightlightblue,linewidth=.5pt](0,0.866)(0.5,0)(1.5,0)(2,0.866)(1.5,1.732)(0.5,1.732)(0,0.866)
\psline[linewidth=.5pt](1.5,0)(1,0.866)(0,0.866)
\psline[linewidth=.5pt](1,0.866)(1.5,1.732)
\psarc[linecolor=red](0.5,0){0.1}{0}{120}
\psarc[linecolor=red](1,0.866){0.1}{60}{180}
\psarc[linecolor=red](1.5,0){0.1}{60}{120}
\rput(0.75,0.433){\scriptsize $#1$}
\rput(0.75,1.299){\scriptsize $#2$}
\rput(1.5,0.866){\scriptsize $#3$}
\pscircle[fillstyle=solid,fillcolor=black](0.5,0.866){0.045}
\pscircle[fillstyle=solid,fillcolor=black](1.24,1.26){0.045}
\pscircle[fillstyle=solid,fillcolor=black](1.24,0.44){0.045}
\rput[br](0.15,0.38){\scriptsize $a$}
\rput[br](1.,-0.2){\scriptsize $b$}
\rput[bl](1.85,0.38){\scriptsize $c$}
\rput[bl](1.85,1.256){\scriptsize $d$}
\rput[bl](1.,1.85){\scriptsize $e$}
\rput[br](0.15,1.256){\scriptsize $f$}
\end{pspicture}
}


\def\YBERHS #1#2#3{
\begin{pspicture}[shift=-1](-.2,-.2)(2.2,2)
\pspolygon[fillstyle=solid,fillcolor=lightlightblue,linewidth=.5pt](0,0.866)(0.5,0)(1.5,0)(2,0.866)(1.5,1.732)(0.5,1.732)(0,0.866)
\psline[linewidth=.5pt](0.5,0)(1,0.866)(0.5,1.732)
\psline[linewidth=.5pt](1,0.866)(2,0.866)
\rput(0.5,0.866){\scriptsize $#1$}
\rput(1.25,0.433){\scriptsize $#2$}
\rput(1.25,1.299){\scriptsize $#3$}
\psarc[linecolor=red](1,0.866){0.1}{0}{120}
\psarc[linecolor=red](1.5,0){0.1}{60}{180}
\psarc[linecolor=red](0.5,0.){0.1}{60}{120}
\pscircle[fillstyle=solid,fillcolor=black](1.5,0.866){0.045}
\pscircle[fillstyle=solid,fillcolor=black](0.76,1.26){0.045}
\pscircle[fillstyle=solid,fillcolor=black](0.76,0.44){0.045}
\rput[br](0.15,0.38){\scriptsize $a$}
\rput[br](1,-0.2){\scriptsize $b$}
\rput[bl](1.85,0.38){\scriptsize $c$}
\rput[bl](1.85,1.256){\scriptsize $d$}
\rput[bl](1.,1.85){\scriptsize $e$}
\rput[br](0.15,1.256){\scriptsize $f$}
\end{pspicture}
}

\subsubsection{Yang-Baxter equations}

The fundamental Yang-Baxter Equation (YBE)~\cite{BaxBook} in the planar and linear algebra is
\begin{subequations}
\label{YBE}
\psset{unit=.7cm}
\begin{align}
\raisebox{-1.3cm}{
\begin{pspicture}(0,0)(3,3.9)
\rput(1,1){\diam {}{}{}{}u}
\rput(1,3){\diam {}{}{}{}v}
\rput(2,2){\diam {}{}{}{}{u\!+\!v}}
\rput(.5,1.5){\pscircle[fillstyle=solid,fillcolor=black](.05,-.015){.07}}
\rput(.5,2.55){\pscircle[fillstyle=solid,fillcolor=black](.05,-.015){.07}}
\rput(1.5,1.5){\pscircle[fillstyle=solid,fillcolor=black](.05,-.015){.07}}
\rput(1.5,2.5){\pscircle[fillstyle=solid,fillcolor=black](.05,-.015){.07}}
\psline[linestyle=dashed](.55,1.5)(.55,2.5)
\end{pspicture}}\quad
&=\ 
\raisebox{-1.3cm}{
\begin{pspicture}(0,0)(3,3.9)
\rput(2,3){\diam {}{}{}{}u}
\rput(2,1){\diam {}{}{}{}v}
\rput(1,2){\diam {}{}{}{}{u\!+\!v}}
\rput(1.5,1.5){\pscircle[fillstyle=solid,fillcolor=black](.05,-.015){.07}}
\rput(1.5,2.55){\pscircle[fillstyle=solid,fillcolor=black](.05,-.015){.07}}
\rput(2.5,1.5){\pscircle[fillstyle=solid,fillcolor=black](.05,-.015){.07}}
\rput(2.5,2.5){\pscircle[fillstyle=solid,fillcolor=black](.05,-.015){.07}}
\psline[linestyle=dashed](2.55,1.5)(2.55,2.5)
\end{pspicture}}\\[4pt]
X_j(u)X_{j+1}(u+v)X_j(v)&=X_{j+1}(v)X_j(u+v)X_{j+1}(u)
\end{align}
\end{subequations}
Distorting the faces into rhombi leads to the alternative representation of the YBE as the following diagrammatic equality holding for all values of the indices $a,b,c,d,e,f=0,1$, of the two partition functions
\begin{equation}
\psset{unit=1.1cm}
\label{eq:YBEH}
\YBELHS uv{v+u}\!\!\! = \YBERHS {v+u}vu
\end{equation}

To establish commuting transfer matrices with Kac boundary conditions, we need three independent YBEs. In the planar algebra, these are\\
\bea
\psset{unit=.8cm}
\mbox{YBE1}:\quad\raisebox{-1.5cm}{
\begin{pspicture}(0,0)(3,3.8)
\rput(1,1){\diam {}{}{}{}{v\!-\!\xi}}
\rput(1,3){\diampink{}{}{}{}{u\!+\!\xi}}
\rput(2,2){\diam{}{}{}{}{u\!+\!v}}
\rput(.5,1.5){\pscircle[fillstyle=solid,fillcolor=black](.05,-.015){.07}}
\rput(.5,2.5){\pscircle[fillstyle=solid,fillcolor=black](.05,-.015){.07}}
\rput(1.5,1.5){\pscircle[fillstyle=solid,fillcolor=black](.05,-.015){.07}}
\rput(1.5,2.5){\pscircle[fillstyle=solid,fillcolor=black](.05,-.015){.07}}
\psline[linestyle=dashed](.55,1.5)(.55,2.5)
\end{pspicture}}\quad
=
\raisebox{-1cm}{
\begin{pspicture}(0,0)(3.8,2.8)
\rput(1,2.4){\rbfacepink {u\!+\!\xi}}
\rput(1,1){\lbface {v\!-\!\xi}}
\rput(2.4,1.4){\diam {}{}{}{}{u\!+\!v}}
\psarc[linewidth=1pt,linecolor=red](2.45,0.4){.15}{45}{135}
\rput(.75,1.4){\pscircle[fillstyle=solid,fillcolor=black](.05,-.015){.07}}
\rput(1.44,0.7){\pscircle[fillstyle=solid,fillcolor=black](.05,-.015){.07}}
\rput(1.94,0.88){\pscircle[fillstyle=solid,fillcolor=black](.05,-.015){.07}}
\rput(1.94,1.94){\pscircle[fillstyle=solid,fillcolor=black](.05,-.015){.07}}
\rput(1.44,2.1){\pscircle[fillstyle=solid,fillcolor=black](.05,-.015){.07}}
\psline[linestyle=dashed](1.44,0.65)(1.94,0.83)
\psline[linestyle=dashed](1.44,2.1)(1.94,1.94)
\end{pspicture}}
 =
\raisebox{-1cm}{
\begin{pspicture}(0,0)(3.8,2.8)
\rput(3,1){\rbfacepink {u\!+\!\xi}}
\rput(3,2.4){\lbface {v\!-\!\xi}}
\rput(1,1.4){\diam {}{}{}{}{u\!+\!v}}
\psarc[linewidth=1pt,linecolor=red](1.1,0.4){.15}{45}{135}
\rput(2.75,1.4){\pscircle[fillstyle=solid,fillcolor=black](.05,-.015){.07}}
\rput(2.,0.7){\pscircle[fillstyle=solid,fillcolor=black](.05,-.015){.07}}
\rput(1.48,0.88){\pscircle[fillstyle=solid,fillcolor=black](.05,-.015){.07}}
\rput(1.5,1.94){\pscircle[fillstyle=solid,fillcolor=black](.05,-.015){.07}}
\rput(2.,2.1){\pscircle[fillstyle=solid,fillcolor=black](.05,-.015){.07}}
\psline[linestyle=dashed](2,0.7)(1.48,0.88)
\psline[linestyle=dashed](2,2.1)(1.5,1.94)
\end{pspicture}}
\quad
=
\raisebox{-1.5cm}{
\begin{pspicture}(0,0)(3,3.8)
\rput(2,3){\diam {}{}{}{}{v\!-\!\xi}}
\rput(2,1){\diampink {}{}{}{}{u\!+\!\xi}}
\rput(1,2){\diam {}{}{}{}{u\!+\!v}}
\rput(1.5,1.5){\pscircle[fillstyle=solid,fillcolor=black](.05,-.015){.07}}
\rput(1.5,2.5){\pscircle[fillstyle=solid,fillcolor=black](.05,-.015){.07}}
\rput(2.5,1.5){\pscircle[fillstyle=solid,fillcolor=black](.05,-.015){.07}}
\rput(2.5,2.5){\pscircle[fillstyle=solid,fillcolor=black](.05,-.015){.07}}
\psline[linestyle=dashed](2.55,1.5)(2.55,2.5)
\end{pspicture}}\qquad
\eea\\
\bea
\psset{unit=.8cm}
\mbox{YBE2}:\quad\raisebox{-1.5cm}{
\begin{pspicture}(0,0)(3,3.8)
\rput(1,1){\diam {}{}{}{}{u\!-\!\xi}}
\rput(1,3){\ldiam{}{}{}{}{v\!-\!\xi}}
\rput(2,2){\ldiam{}{}{}{}{v\!-\!u}}
\rput(.5,1.5){\pscircle[fillstyle=solid,fillcolor=black](.05,-.015){.07}}
\rput(.5,2.5){\pscircle[fillstyle=solid,fillcolor=black](.05,-.015){.07}}
\rput(1.5,1.5){\pscircle[fillstyle=solid,fillcolor=black](.05,-.015){.07}}
\rput(1.5,2.5){\pscircle[fillstyle=solid,fillcolor=black](.05,-.015){.07}}
\psline[linestyle=dashed](.55,1.5)(.55,2.5)
\end{pspicture}}\quad
=
\raisebox{-1cm}{
\begin{pspicture}(0,0)(3.8,2.8)
\rput(1,2.4){\lbface{v\!-\!\xi}}
\rput(1,1){\lbface{u\!-\!\xi}}
\rput(2.4,1.4){\ldiam{}{}{}{}{v\!-\!u}}
\psarc[linewidth=1pt,linecolor=red](1.5,1.4){.15}{-45}{45}
\rput(.75,1.4){\pscircle[fillstyle=solid,fillcolor=black](.05,-.015){.07}}
\rput(1.44,0.7){\pscircle[fillstyle=solid,fillcolor=black](.05,-.015){.07}}
\rput(1.94,0.88){\pscircle[fillstyle=solid,fillcolor=black](.05,-.015){.07}}
\rput(1.94,1.94){\pscircle[fillstyle=solid,fillcolor=black](.05,-.015){.07}}
\rput(1.44,2.1){\pscircle[fillstyle=solid,fillcolor=black](.05,-.015){.07}}
\psline[linestyle=dashed](1.44,0.65)(1.94,0.83)
\psline[linestyle=dashed](1.44,2.1)(1.94,1.94)
\end{pspicture}}
=
\raisebox{-1cm}{
\begin{pspicture}(0,0)(3.8,2.8)
\rput(3,1){\lbface {v\!-\!\xi}}
\rput(3,2.4){\lbface {u\!-\!\xi}}
\rput(1,1.4){\ldiam {}{}{}{}{v\!-\!u}}
\psarc[linewidth=1pt,linecolor=red](.1,1.4){.15}{-45}{45}
\rput(2.75,1.4){\pscircle[fillstyle=solid,fillcolor=black](.05,-.015){.07}}
\rput(2.,0.7){\pscircle[fillstyle=solid,fillcolor=black](.05,-.015){.07}}
\rput(1.48,0.88){\pscircle[fillstyle=solid,fillcolor=black](.05,-.015){.07}}
\rput(1.5,1.94){\pscircle[fillstyle=solid,fillcolor=black](.05,-.015){.07}}
\rput(2.,2.1){\pscircle[fillstyle=solid,fillcolor=black](.05,-.015){.07}}
\psline[linestyle=dashed](2,0.7)(1.48,0.88)
\psline[linestyle=dashed](2,2.1)(1.5,1.94)
\end{pspicture}}
\quad =
\raisebox{-1.5cm}{
\begin{pspicture}(0,0)(3,3.8)
\rput(2,3){\diam {}{}{}{}{u\!-\!\xi}}
\rput(2,1){\ldiam {}{}{}{}{v\!-\!\xi}}
\rput(1,2){\ldiam {}{}{}{}{v\!-\!u}}
\rput(1.5,1.5){\pscircle[fillstyle=solid,fillcolor=black](.05,-.015){.07}}
\rput(1.5,2.5){\pscircle[fillstyle=solid,fillcolor=black](.05,-.015){.07}}
\rput(2.5,1.5){\pscircle[fillstyle=solid,fillcolor=black](.05,-.015){.07}}
\rput(2.5,2.5){\pscircle[fillstyle=solid,fillcolor=black](.05,-.015){.07}}
\psline[linestyle=dashed](2.55,1.5)(2.55,2.5)
\end{pspicture}}\qquad
\eea\\
\bea
\psset{unit=.8cm}
\mbox{YBE3}:\quad\raisebox{-1.5cm}{
\begin{pspicture}(0,0)(3,3.8)
\rput(1,1){\rdiampink {}{}{}{}{u\!+\!\xi}}
\rput(1,3){\diampink{}{}{}{}{v\!+\!\xi}}
\rput(2,2){\rdiam{}{}{}{}{u\!-\!v}}
\rput(.5,1.5){\pscircle[fillstyle=solid,fillcolor=black](.05,-.015){.07}}
\rput(.5,2.5){\pscircle[fillstyle=solid,fillcolor=black](.05,-.015){.07}}
\rput(1.5,1.5){\pscircle[fillstyle=solid,fillcolor=black](.05,-.015){.07}}
\rput(1.5,2.5){\pscircle[fillstyle=solid,fillcolor=black](.05,-.015){.07}}
\psline[linestyle=dashed](.55,1.5)(.55,2.5)
\psline[linestyle=dashed](.55,1.5)(.55,2.5)
\end{pspicture}}\quad
=
\raisebox{-1cm}{
\begin{pspicture}(0,0)(3.8,2.8)
\rput(1,2.4){\rbfacepink{v\!+\!\xi}}
\rput(1,1){\rbfacepink{u\!+\!\xi}}
\rput(2.4,1.4){\rdiam {}{}{}{}{u\!-\!v}}
\rput(.75,1.4){\pscircle[fillstyle=solid,fillcolor=black](.05,-.015){.07}}
\rput(1.44,0.7){\pscircle[fillstyle=solid,fillcolor=black](.05,-.015){.07}}
\rput(1.94,0.88){\pscircle[fillstyle=solid,fillcolor=black](.05,-.015){.07}}
\rput(1.94,1.94){\pscircle[fillstyle=solid,fillcolor=black](.05,-.015){.07}}
\rput(1.44,2.1){\pscircle[fillstyle=solid,fillcolor=black](.05,-.015){.07}}
\psline[linestyle=dashed](1.44,0.65)(1.94,0.83)
\psline[linestyle=dashed](1.44,2.1)(1.94,1.94)
\end{pspicture}}
 =
\raisebox{-1cm}{
\begin{pspicture}(0,0)(3.8,2.8)
\rput(3,1){\rbfacepink {v\!+\!\xi}}
\rput(3,2.4){\rbfacepink{u\!+\!\xi}}
\rput(1,1.4){\rdiam {}{}{}{}{u\!-\!v}}
\rput(2.75,1.4){\pscircle[fillstyle=solid,fillcolor=black](.05,-.015){.07}}
\rput(2.,0.7){\pscircle[fillstyle=solid,fillcolor=black](.05,-.015){.07}}
\rput(1.48,0.88){\pscircle[fillstyle=solid,fillcolor=black](.05,-.015){.07}}
\rput(1.5,1.94){\pscircle[fillstyle=solid,fillcolor=black](.05,-.015){.07}}
\rput(2.,2.1){\pscircle[fillstyle=solid,fillcolor=black](.05,-.015){.07}}
\psline[linestyle=dashed](2,0.7)(1.48,0.88)
\psline[linestyle=dashed](2,2.1)(1.5,1.94)
\end{pspicture}}\quad
=
\raisebox{-1.5cm}{
\begin{pspicture}(0,0)(3,3.8)
\rput(2,3){\rdiampink {}{}{}{}{u\!+\!\xi}}
\rput(2,1){\diampink {}{}{}{}{v\!+\!\xi}}
\rput(1,2){\rdiam {}{}{}{}{u\!-\!v}}
\rput(1.5,1.5){\pscircle[fillstyle=solid,fillcolor=black](.05,-.015){.07}}
\rput(1.5,2.5){\pscircle[fillstyle=solid,fillcolor=black](.05,-.015){.07}}
\rput(2.5,1.5){\pscircle[fillstyle=solid,fillcolor=black](.05,-.015){.07}}
\rput(2.5,2.5){\pscircle[fillstyle=solid,fillcolor=black](.05,-.015){.07}}
\psline[linestyle=dashed](2.55,1.5)(2.55,2.5)
\end{pspicture}}\qquad
\eea\\
Here $\xi$ is an arbitrary boundary field.

\subsubsection{Boundary Yang-Baxter equations}
\psset{unit=.8cm}

In the presence of a boundary, there are addtional local relations in the form of boundary Yang-Baxter or reflection equations~\cite{Cherednik,Sklyanin,BPO}. 
The nonzero left and right boundary triangle weights and the corresponding planar operators are independent of $u$ and given by
\begin{subequations}
\label{BdyCond}
\bea
&K^L\Big({b\atop a}\Big)=x^{1-2a}\delta(a,b),\qquad K^R\Big({b\atop a}\Big)=\delta(a,b)&\\[8pt]
\psset{unit=.8cm}
&\begin{pspicture}[shift=-.9](-1,0)(.2,2)
\pspolygon[fillstyle=solid,fillcolor=lightlightblue,linewidth=.5pt](-1,2)(-1,0)(0,1)
\rput(-.6,1){$u$}
\end{pspicture}= x\;
\begin{pspicture}[shift=-.9](-1,0)(.2,2)
\pspolygon[fillstyle=solid,fillcolor=tile,linewidth=.5pt](-1,2)(-1,0)(0,1)
\end{pspicture}+x^{-1}\;
\begin{pspicture}[shift=-.9](-1,0)(.2,2)
\pspolygon[fillstyle=solid,fillcolor=tile,linewidth=.5pt](-1,2)(-1,0)(0,1)
\psarc[linewidth=1pt](0,1){.65}{135}{225}
\end{pspicture},\qquad
\begin{pspicture}[shift=-.9](-1,0)(.2,2)
\pspolygon[fillstyle=solid,fillcolor=lightlightblue,linewidth=.5pt](-1,1)(0,0)(0,2)
\rput(-.4,1){$u$}
\end{pspicture}\,=\;
\begin{pspicture}[shift=-.9](-1,0)(.2,2)
\pspolygon[fillstyle=solid,fillcolor=tile,linewidth=.5pt](0,2)(0,0)(-1,1)
\end{pspicture}+\;
\begin{pspicture}[shift=-.9](-1,0)(.2,2)
\pspolygon[fillstyle=solid,fillcolor=tile,linewidth=.5pt](0,2)(0,0)(-1,1)
\psarc[linewidth=1pt](-1,1){.65}{-45}{45}
\end{pspicture}&
\eea
\end{subequations}
\begin{subequations}
The general Right Boundary Yang-Baxter Equation (RBYBE) is
\bea
X_j(u-v)K^R_{j+1}(u)X_j(u+v)K^R_{j+1}(v)=
K^R_{j+1}(v)X_j(u+v)K^R_{j+1}(u)X_j(u-v)\qquad\qquad\quad\\[-8pt]
\psset{unit=.9cm}
\mbox{RBYBE:}\quad\setlength{\unitlength}{.9cm}
\raisebox{-.5cm}{
\raisebox{-2cm}{\begin{pspicture}(2,4.4)
\rput(1,1){\diamodd{u-v}}
\rput(1,3){\diamodd{u+v}}
\psarc[linewidth=1pt,linecolor=red](1.1,2){.15}{45}{135}
\psarc[linewidth=1pt,linecolor=red](1.1,0){.15}{45}{135}
\rput(1,1){\righttri {}{}{}}
\rput(1,3){\righttri {}{}{}}
\rput[bl](1.6,3.9){\small $v$}
\rput[bl](1.6,1.9){\small $u$}
\rput(.5,1.5){\pscircle[fillstyle=solid,fillcolor=black](.05,-.015){.07}}
\rput(.5,2.5){\pscircle[fillstyle=solid,fillcolor=black](.05,-.015){.07}}
\rput(1.5,1.5){\pscircle[fillstyle=solid,fillcolor=black](.05,-.015){.07}}
\rput(1.5,2.5){\pscircle[fillstyle=solid,fillcolor=black](.05,-.015){.07}}
\rput(1.5,3.5){\pscircle[fillstyle=solid,fillcolor=black](.05,-.015){.07}}
\psline[linestyle=dashed](.55,1.5)(.55,2.5)
\end{pspicture}}\qquad
\raisebox{.5cm}{\mbox{\!\!\!\!=\ }}
\qquad\ 
\raisebox{-2.cm}{\begin{pspicture}(2,4.7)
\rput(1,3){\diamodd{u+v}}
\psarc[linewidth=1pt,linecolor=red](1.1,2){.15}{45}{135}
\rput(0,2){\diamodd{u-v}}
\psarc[linewidth=1pt,linecolor=red](-.9,2){.15}{-45}{45}
\rput(1,1){\righttri {}{}{}}
\rput(1.,3.){\righttri {}{}{}}
\rput[bl](1.6,3.9){\small $v$}
\rput[bl](1.6,1.9){\small $u$}
\rput(.5,1.5){\pscircle[fillstyle=solid,fillcolor=black](.05,-.015){.07}}
\rput(.5,2.5){\pscircle[fillstyle=solid,fillcolor=black](.05,-.015){.07}}
\rput(1.5,1.5){\pscircle[fillstyle=solid,fillcolor=black](.05,-.015){.07}}
\rput(1.5,2.5){\pscircle[fillstyle=solid,fillcolor=black](.05,-.015){.07}}
\rput(1.5,3.5){\pscircle[fillstyle=solid,fillcolor=black](.05,-.015){.07}}
\psline[linestyle=dashed](.55,1.5)(1.55,1.5)
\end{pspicture}}\qquad
\raisebox{.5cm}{\mbox{\!\!\!\!=\ }}
\qquad
\raisebox{-1.2cm}{\begin{pspicture}(2,4.7)
\rput(1,2){\diamodd{u+v}}
\psarc[linewidth=1pt,linecolor=red](1.1,1){.15}{45}{135}
\rput(0,3){\diamodd{u-v}}
\psarc[linewidth=1pt,linecolor=red](1.1,3){.15}{135}{225}
\rput(1.,0){\righttri {}{}{}}
\rput(1,2){\righttri {}{}{}}
\rput[bl](1.6,2.9){\small $u$}
\rput[bl](1.6,.9){\small $v$}
\rput(.5,2.5){\pscircle[fillstyle=solid,fillcolor=black](.05,-.015){.07}}
\rput(.5,3.5){\pscircle[fillstyle=solid,fillcolor=black](.05,-.015){.07}}
\rput(1.5,2.5){\pscircle[fillstyle=solid,fillcolor=black](.05,-.015){.07}}
\rput(1.5,3.5){\pscircle[fillstyle=solid,fillcolor=black](.05,-.015){.07}}
\rput(1.5,1.5){\pscircle[fillstyle=solid,fillcolor=black](.05,-.015){.07}}
\rput(.5,2.5){\pscircle[fillstyle=solid,fillcolor=black](.05,-.015){.07}}
\rput(1.5,1.5){\pscircle[fillstyle=solid,fillcolor=black](.05,-.015){.07}}
\rput(1.5,2.5){\pscircle[fillstyle=solid,fillcolor=black](.05,-.015){.07}}
\rput(1.5,3.5){\pscircle[fillstyle=solid,fillcolor=black](.05,-.015){.07}}
\psline[linestyle=dashed](.55,3.5)(1.55,3.5)
\end{pspicture}} \quad
\raisebox{.5cm}{\mbox{=\!\!\!\!}}
\quad
\raisebox{-2.cm}{\begin{pspicture}(2,4.7)
\rput(1,2){\diamodd{u+v}}
\psarc[linewidth=1pt,linecolor=red](1.1,1){.15}{45}{135}
\rput(1,4){\diamodd{u-v}}
\psarc[linewidth=1pt,linecolor=red](1.1,3){.15}{45}{135}
\rput(1.,0){\righttri{}{}{}}
\rput(1,2){\righttri {}{}{}}
\rput[bl](1.6,2.9){\small $u$}
\rput[bl](1.6,0.9){\small $v$}
\rput(.5,2.5){\pscircle[fillstyle=solid,fillcolor=black](.05,-.015){.07}}
\rput(.5,3.5){\pscircle[fillstyle=solid,fillcolor=black](.05,-.015){.07}}
\rput(1.5,2.5){\pscircle[fillstyle=solid,fillcolor=black](.05,-.015){.07}}
\rput(1.5,3.5){\pscircle[fillstyle=solid,fillcolor=black](.05,-.015){.07}}
\rput(1.5,1.5){\pscircle[fillstyle=solid,fillcolor=black](.05,-.015){.07}}
\rput(.5,2.5){\pscircle[fillstyle=solid,fillcolor=black](.05,-.015){.07}}
\rput(1.5,1.5){\pscircle[fillstyle=solid,fillcolor=black](.05,-.015){.07}}
\rput(1.5,2.5){\pscircle[fillstyle=solid,fillcolor=black](.05,-.015){.07}}
\rput(1.5,3.5){\pscircle[fillstyle=solid,fillcolor=black](.05,-.015){.07}}
\psline[linestyle=dashed](.55,2.5)(.55,3.5)
\end{pspicture}}}\qquad\qquad
\eea
where the relevant position is $j=N-1$. After removing the right boundary triangles $K^R_{j+1}(u)=I$, this reduces to the commutation relation $[X_j(u-v),X_j(u+v)]=0$.
\end{subequations}
\begin{subequations}

With $z:=e^{iu}$, $w:=e^{iv}$, the general Left Boundary Yang-Baxter Equation (LBYBE) is
\be
{X}_{j+1}(v\!-\!u,\frac{z}{w})K^L_{j}(u)\tilde X_{j+1}(2\lambda\!-\!u\!-\!v,z w)K^L_{j}(v)=
K^L_{j}(v)\tilde X_{j+1}(2\lambda\!-\!u\!-\!v,z w)K^L_{j}(u){X}_{j+1}(v\!-\!u,\frac{z}{w})
\ee
\bea
\psset{unit=.9cm}
\setlength{\unitlength}{.9cm}
\mbox{LBYBE:}\quad\raisebox{-2.45cm}{\raisebox{.8cm}{\begin{pspicture}(2,3.8)
\rput(1,2){\sdiam{}{}{}{}{2\lambda\!-\!u\!-\!v}}
\rput(1,0){\diamodd{v-u}}
\psarc[linewidth=.5pt,linecolor=red](1.08,1){.15}{225}{315}
\rput(0.,0){\lefttri{}{}{}}
\rput(0,2){\lefttri {}{}{}}
\rput[bl](0.4,2.9){\small $v$}
\rput[bl](0.4,0.9){\small $u$}
\rput(.5,.5){\pscircle[fillstyle=solid,fillcolor=black](.05,-.015){.07}}
\rput(.5,1.5){\pscircle[fillstyle=solid,fillcolor=black](.05,-.015){.07}}
\rput(.5,2.5){\pscircle[fillstyle=solid,fillcolor=black](.05,-.015){.07}}
\rput(1.5,0.5){\pscircle[fillstyle=solid,fillcolor=black](.05,-.015){.07}}
\rput(1.5,1.5){\pscircle[fillstyle=solid,fillcolor=black](.05,-.015){.07}}
\psline[linestyle=dashed](1.55,0.5)(1.55,1.5)
\end{pspicture}}\quad
\raisebox{2.5cm}{\mbox{=}}
\quad
\raisebox{0.cm}{\begin{pspicture}(2,3.8)
\rput(1,3){\sdiam{}{}{}{}{2\lambda\!-\!u\!-\!v}}
\rput(2,2){\diamodd{v-u}}
\psarc[linewidth=.5pt,linecolor=red](1.08,2){.15}{-45}{45}
\rput(0,1){\lefttri {}{}{}}
\rput(0.,3.){\lefttri {}{}{}}
\rput[bl](0.4,3.9){\small $v$}
\rput[bl](0.4,1.9){\small $u$}
\rput(.5,1.5){\pscircle[fillstyle=solid,fillcolor=black](.05,-.015){.07}}
\rput(.5,2.5){\pscircle[fillstyle=solid,fillcolor=black](.05,-.015){.07}}
\rput(.5,3.5){\pscircle[fillstyle=solid,fillcolor=black](.05,-.015){.07}}
\rput(1.5,1.5){\pscircle[fillstyle=solid,fillcolor=black](.05,-.015){.07}}
\rput(1.5,2.5){\pscircle[fillstyle=solid,fillcolor=black](.05,-.015){.07}}
\psline[linestyle=dashed](.55,1.5)(1.55,1.5)
\end{pspicture}}\qquad
\raisebox{2.5cm}{\mbox{\ \ \ =\!\!\!\!\!}}
\qquad
\raisebox{.9cm}{\begin{pspicture}(2,3.8)
\rput(2,3){\diamodd{v-u}}
\psarc[linewidth=.5pt,linecolor=red](3.08,3){.15}{135}{225}
\rput(1,2){\sdiam{}{}{}{}{2\lambda\!-\!u\!-\!v}}
\rput(0,0){\lefttri {}{}{}}
\rput(0,2){\lefttri {}{}{}}
\rput[bl](0.4,2.9){\small $u$}
\rput[bl](0.4,0.9){\small $v$}
\rput(.5,2.5){\pscircle[fillstyle=solid,fillcolor=black](.05,-.015){.07}}
\rput(.5,3.5){\pscircle[fillstyle=solid,fillcolor=black](.05,-.015){.07}}
\rput(1.5,2.5){\pscircle[fillstyle=solid,fillcolor=black](.05,-.015){.07}}
\rput(1.5,3.5){\pscircle[fillstyle=solid,fillcolor=black](.05,-.015){.07}}
\rput(.5,1.5){\pscircle[fillstyle=solid,fillcolor=black](.05,-.015){.07}}
\psline[linestyle=dashed](0.55,3.5)(1.55,3.5)
\end{pspicture}}\qquad
\raisebox{2.5cm}{\mbox{\ \ \ \ =\hspace{-8pt}}}
\qquad
\raisebox{-.2cm}{\begin{pspicture}(2,3.8)
\rput(1,2){\sdiam{}{}{}{}{2\lambda\!-\!u\!-\!v}}
\rput(1,4){\diamodd{v-u}}
\psarc[linewidth=.5pt,linecolor=red](1.075,5){.15}{225}{315}
\rput(0.,0){\lefttri{}{}{}}
\rput(0,2){\lefttri {}{}{}}
\rput[bl](0.4,2.9){\small $u$}
\rput[bl](0.4,0.9){\small $v$}
\rput(.5,2.5){\pscircle[fillstyle=solid,fillcolor=black](.05,-.015){.07}}
\rput(.5,3.5){\pscircle[fillstyle=solid,fillcolor=black](.05,-.015){.07}}
\rput(1.5,2.5){\pscircle[fillstyle=solid,fillcolor=black](.05,-.015){.07}}
\rput(1.5,3.5){\pscircle[fillstyle=solid,fillcolor=black](.05,-.015){.07}}
\rput(.5,1.5){\pscircle[fillstyle=solid,fillcolor=black](.05,-.015){.07}}
\rput(1.5,2.5){\pscircle[fillstyle=solid,fillcolor=black](.05,-.015){.07}}
\rput(1.5,3.5){\pscircle[fillstyle=solid,fillcolor=black](.05,-.015){.07}}
\psline[linestyle=dashed](1.55,2.5)(1.55,3.5)
\end{pspicture}}}\qquad\qquad
\eea
where the gauge factors have been omitted in the diagrams and the relevant position is $j=0$. 
For the dimer model under consideration, the boundary triangles are independent of the spectral parameters.
\end{subequations}

\subsection{Commuting double row transfer matrices}

The general double row transfer matrices are defined diagrammatically by
\bea
\vec D(u)=\;
\psset{unit=1.2cm}
\begin{pspicture}[shift=-1.1](-.7,.75)(6.7,3.2)
\multiput(0,1)(0,2){2}{\psline[linewidth=.25pt,linestyle=dashed](-.7,0)(6.7,0)}
\facegrid{(0,1)}{(6,2)}
\facegridpink{(0,2)}{(6,3)}
\pspolygon[fillstyle=solid,fillcolor=lightlightblue](0,2)(-.7,1)(-.7,3)
\pspolygon[fillstyle=solid,fillcolor=lightlightblue](6,2)(6.7,1)(6.7,3)
\rput(-.4,2.){\small $u$}
\rput(0.5,1.5){\small $u$}
\rput(4.5,1.5){\small $u$}
\rput(5.5,1.5){\small $u-\xi$}
\rput(0.5,2.5){\small $u$}
\rput(4.5,2.5){\small $u$}
\rput(5.5,2.5){\small $u+\xi$}
\rput(6.4,2.){\small $u$}
\rput(2.5,1.5){\small $\dots$}
\rput(2.5,2.5){\small $\dots$}
\psarc[linecolor=red,linewidth=.5pt](0,1){.15}{0}{90}
\psarc[linecolor=red,linewidth=.5pt](4,1){.15}{0}{90}
\psarc[linecolor=red,linewidth=.5pt](5,1){.15}{0}{90}
\psarc[linecolor=red,linewidth=.5pt](1,2){.15}{90}{180}
\psarc[linecolor=red,linewidth=.5pt](5,2){.15}{90}{180}
\psarc[linecolor=red,linewidth=.5pt](6,2){.15}{90}{180}
\psline[linecolor=red,linewidth=1pt,linestyle=dashed](5,.7)(5,3.3)
\end{pspicture}\quad
\label{D}
\eea
where $\xi$ is an arbitrary boundary field. There are a total of $N$ columns in the bulk and $w=0,1$ columns in the boundary. We are primarily interested in the following two cases: (i) $w=0$ in which case there is no boundary column and the system is homogeneous and (ii) $w=1$ for which the boundary consists of the right-most column with $\xi=\frac{\lambda}{2}$.  The specialization $\xi=\frac{\lambda}{2}$ has nice properties compared to other nonzero values of $\xi$. In particular, the inversion identity can be solved exactly for $\xi=\frac{\lambda}{2}$.

\subsubsection{Sectors and quantum numbers}
In the six vertex arrow (or spin) representation, the total magnetization 
\bea
S_z=\sum_{j=1}^{\calN} \sigma_j=-\calN,-\calN+2,\ldots, \calN-2, \calN,\qquad \calN=N+w
\eea
is conserved under the action of the transfer matrix. By the $\mathbb{Z}_2$ up-down symmetry, 
the spectrum for the sectors $S_z=\pm m$ coincide for $m>0$. 
More generally, the number of down spins is $d=\half(\calN-S_z)$. The number of up spins is thus $\calN-d=\half(\calN+S_z)$ and 
the counting of states in the $S_z$ sector is given by the binomial $\genfrac{(}{)}{0pt}{}{\calN}{d}$ 
with $S_z=\calN$ mod 2. 
In the particle representation, a particle configuration along a row of the double row transfer matrix takes the form
\bea
a=\{a_1,a_2,\ldots,a_{\calN-1},a_{\calN}\},\qquad a_j=0,1\ \mbox{for $j=1,2,\ldots, \calN$}
\eea
The total number of particles $d=\sum_{j=1}^\calN a_j$ coincides with the number of down arrows and is also conserved. 
The transfer matrix and vector space of states thus decompose as
\bea
\vec D(u)=\mathop{\bigoplus}_{d=0}^\calN \vec D_d(u),\qquad \mathop{\mbox{dim}} {\cal V}^{(\calN)}=\sum_{d=0}^\calN \mathop{\mbox{dim}}{\cal V}^{(\calN)}_d=\sum_{d=0}^\calN
\genfrac{(}{)}{0pt}{}{\calN}{d}=2^\calN=\mbox{dim}\,(\mathbb{C}^2)^{\otimes \calN}\label{Tdecomp}
\eea 
For comparing the spectra sector-by-sector with critical dense polymers~\cite{PRV}, it is useful to define
\bea
\ell=|\calN-2d|=|S_z|=\begin{cases}
0,2,4,\ldots,\calN,&\mbox{$\calN$ even}\\
1,3,5,\ldots,\calN,&\mbox{$\calN$ odd}\end{cases}\label{matching}
\eea
In the context of critical dense polymers, $\ell$ is the number of defects.

\subsubsection{Diagrammatic proof of commutation}
Setting $\eta_1=\eta_1(u-v), \eta_2=\eta_2(u+v)$ as in (\ref{InRels}), the commutation of the double row transfer matrices is established diagrammatically\\
\begin{subequations}
\begin{flalign}
&\raisebox{-.0cm}{$\vec D(u)\vec D(v)\;=$\hspace{-6pt}}
\raisebox{-2.1cm}{
\psset{unit=1.1cm}
\begin{pspicture}(-1.5,0)(6.5,3.75)
\multiput(0,0)(0,1){5}{\psline[linewidth=.25pt,linestyle=dashed](-1,0)(6,0)}
\rput(-1.05,0){\lefttri {}{}{u}}
\rput(4.95,0){\righttri {}{}u}
\rput(-1.05,2){\lefttri {}{}{v}}
\rput(4.95,2){\righttri {}{}v}
\rput(0,0){\oface u}
\rput(1,0){\oface u}
\rput(2,0){\oface {}}
\rput(3,0){\oface u}
\rput(4,0){\oface {u-\xi}}
\rput(0,1){\eface u}
\rput(1,1){\eface u}
\rput(2,1){\eface {}}
\rput(3,1){\eface u}
\rput(4,1){\eface {u+\xi}}
\rput(0,2){\oface v}
\rput(1,2){\oface v}
\rput(2,2){\oface {}}
\rput(3,2){\oface v}
\rput(4,2){\oface {v-\xi}}
\rput(0,3){\eface v}
\rput(1,3){\eface v}
\rput(2,3){\eface {}}
\rput(3,3){\eface v}
\rput(4,3){\eface {v+\xi}}
\rput[t](2.5,2.4){\small $\dots$}
\rput[b](2.5,1.4){\small $\cdots$}
\rput[t](2.5,3.4){\small $\cdots$}
\rput[b](2.5,0.4){\small $\dots$}
\psline[linecolor=red,linewidth=1pt,linestyle=dashed](4,-.3)(4,4.3)
\end{pspicture}}\qquad\qquad\qquad\qquad\qquad\\[1cm]
&\raisebox{0cm}{ $\overset{\text{Inv2}}={\disp\frac{1}{\eta_2}}$\ }
\raisebox{-2.1cm}{
\psset{unit=1.1cm}
\begin{pspicture}(-1,0)(6.5,3.5)
\multiput(0,0)(0,1){5}{\psline[linewidth=.25pt,linestyle=dashed](-1,0)(9,0)}
\rput(-1.058,0){\lefttri {}{}{u}}
\rput(7.95,0){\righttri {}{}u}
\rput(-1.058,2){\lefttri {}{}{v}}
\rput(7.95,2){\righttri {}{}v}
\rput(3,0){\oface {u}}
\rput(4,0){\oface {u}}
\rput(5,0){\oface {}}
\rput(6,0){\oface u}
\rput(7,0){\oface {u-\xi}}
\rput(3,1){\eface {u}}
\rput(4,1){\eface {u}}
\rput(5,1){\eface {}}
\rput(6,1){\eface u}
\rput(7,1){\eface {u+\xi}}
\rput(3,2){\oface {v}}
\rput(4,2){\oface {v}}
\rput(5,2){\oface {}}
\rput(6,2){\oface v}
\rput(7,2){\oface {v-\xi}}
\rput(3,3){\eface {v}}
\rput(4,3){\eface {v}}
\rput(5,3){\eface {}}
\rput(6,3){\eface v}
\rput(7,3){\eface {v+\xi}}
\rput[t](5.5,2.4){\small $\dots$}
\rput[b](5.5,1.4){\small $\cdots$}
\rput[t](5.5,3.4){\small $\cdots$}
\rput[b](5.5,0.4){\small $\dots$}
\rput(-.05,2){\sdiam {}{}{}{}{2\lambda\!-\!u\!-\!v}}
\rput(1.95,2){\diam {}{}{}{}{u+v}}
\psline[linecolor=red,linewidth=1pt,linestyle=dashed](7,-.3)(7,4.3)
\end{pspicture}}\\[1cm]
&\raisebox{0cm}{$\overset{\text{YBE1}}=\;{\disp\frac{1}{\eta_2}}$\ }\raisebox{-2.1cm}{
\psset{unit=1.1cm}
\begin{pspicture}(-1.,0)(6.5,3.5)
\multiput(0,0)(0,1){5}{\psline[linewidth=.25pt,linestyle=dashed](-1,0)(8,0)}
\rput(-1.058,0){\lefttri {}{}{u}}
\rput(6.99,0){\righttri {}{}u}
\rput(-1.058,2){\lefttri {}{}{v}}
\rput(6.99,2){\righttri {}{}v}
\rput(-.07,2){\sdiam {}{}{}{}{2\lambda\!-\!u\!-\!v}}
\rput(6.98,2){\diam {}{}{}{}{u+v}}
\rput(1,0){\oface u}
\rput(2,0){\oface u}
\rput(3,0){\oface {}}
\rput(4,0){\oface u}
\rput(5,0){\oface {u-\xi}}
\rput(1,1){\oface v}
\rput(2,1){\oface v}
\rput(3,1){\oface {}}
\rput(4,1){\oface v}
\rput(5,1){\oface {v-\xi}}
\rput(1,2){\eface u}
\rput(2,2){\eface u}
\rput(3,2){\eface {}}
\rput(4,2){\eface u}
\rput(5,2){\eface {u+\xi}}
\rput(1,3){\eface v}
\rput(2,3){\eface v}
\rput(3,3){\eface {}}
\rput(4,3){\eface v}
\rput(5,3){\eface {v+\xi}}
\rput[t](3.5,2.4){\small $\cdots$}
\rput[b](3.5,1.4){\small $\cdots$}
\rput[t](3.5,3.4){\small $\cdots$}
\rput[b](3.5,0.4){\small $\cdots$}
\psline[linecolor=red,linewidth=1pt,linestyle=dashed](5,-.3)(5,4.3)
\end{pspicture}}\\[1cm]
&\raisebox{0cm}{$\disp \overset{\text{Inv1}}=\disp\frac{1}{\eta_1\eta_2}$\ }\raisebox{-2.1cm}{\qquad\quad
\begin{pspicture}(-1.7,0)(6.5,6.3)
\psset{unit=1.1cm}
\multiput(0,0)(0,1){5}{\psline[linewidth=.25pt,linestyle=dashed](-2,0)(10,0)}
\rput(-2.058,0){\lefttri {}{}{u}}
\rput(8.95,0){\righttri {}{}u}
\rput(-2.058,2){\lefttri {}{}{v}}
\rput(8.95,2){\righttri {}{}v}
\rput(0,0){\oface u}
\rput(5,0){\oface {}}
\rput(6,0){\oface u}
\rput(7,0){\oface {u-\xi}}
\rput(0,1){\oface v}
\rput(5,1){\oface {}}
\rput(6,1){\oface v}
\rput(7,1){\oface {v-\xi}}
\rput(0,2){\eface u}
\rput(5,2){\eface {}}
\rput(6,2){\eface u}
\rput(7,2){\eface {u+\xi}}
\rput(0,3){\eface v}
\rput(5,3){\eface {}}
\rput(6,3){\eface v}
\rput(7,3){\eface {v+\xi}}
\rput[t](5.5,2.4){\small $\cdots$}
\rput[b](5.5,1.4){\small $\cdots$}
\rput[t](5.5,3.4){\small $\cdots$}
\rput[b](5.5,0.4){\small $\cdots$}
\rput(1.95,1){\ldiam {}{}{}{}{v-u}}
\rput(3.95,1){\ldiam {}{}{}{}{u-v}}
\rput(-1.07,2){\sdiam {}{}{}{}{2\lambda\!-\!u\!-\!v}}
\rput(8.96,2){\diam {}{}{}{}{u+v}}
\psline[linecolor=red,linewidth=1pt,linestyle=dashed](7,-.3)(7,4.3)
\end{pspicture}}\\[1cm]
&\raisebox{0cm}{$\disp\overset{\text{YBE2}}=\frac{1}{\eta_1\eta_2}$\ \;}\raisebox{-2.1cm}{\qquad\quad
\begin{pspicture}(-1.7,0)(6.5,6.3)
\psset{unit=1.1cm}
\multiput(0,0)(0,1){5}{\psline[linewidth=.25pt,linestyle=dashed](-2,0)(9,0)}
\rput(-2.1,0){\lefttri {}{}{u}}
\rput(7.91,0){\righttri {}{}u}
\rput(-2.1,2){\lefttri {}{}{v}}
\rput(7.91,2){\righttri {}{}v}
\rput(-1.1,2){\sdiam {}{}{}{}{{2\lambda\!-\!u\!-\!v}}}
\rput(7.92,2){\diam {}{}{}{}{u+v}}
\rput(-.09,1){\ldiam {}{}{}{}{v-u}}
\rput(6.92,1){\ldiam {}{}{}{}{u-v}}
\rput(1,0){\oface v}
\rput(2,0){\oface v}
\rput(3,0){\oface {}}
\rput(4,0){\oface v}
\rput(5,0){\oface {v-\xi}}
\rput(1,1){\oface u}
\rput(2,1){\oface u}
\rput(3,1){\oface {}}
\rput(4,1){\oface u}
\rput(5,1){\oface {u-\xi}}
\rput(1,2){\eface u}
\rput(2,2){\eface u}
\rput(3,2){\eface {}}
\rput(4,2){\eface u}
\rput(5,2){\eface {v+\xi}}
\rput(1,3){\eface v}
\rput(2,3){\eface v}
\rput(3,3){\eface {}}
\rput(4,3){\eface v}
\rput(5,3){\eface {v+\xi}}
\rput[t](3.5,2.4){\small $\cdots$}
\rput[b](3.5,1.4){\small $\cdots$}
\rput[t](3.5,3.4){\small $\cdots$}
\rput[b](3.5,0.4){\small $\cdots$}
\psline[linecolor=red,linewidth=1pt,linestyle=dashed](5,-.3)(5,4.3)
\end{pspicture}}\\[1cm]
&\raisebox{0cm}{$\disp\overset{\text{BYBE}}=\frac{1}{\eta_1\eta_2}$\ }
\raisebox{-2.1cm}{
\psset{unit=1.1cm}
\begin{pspicture}(-2,0)(6.5,3.5)
\multiput(0,0)(0,1){5}{\psline[linewidth=.25pt,linestyle=dashed](-2,0)(9,0)}
\rput(-2.1,0){\lefttri {}{}{v}}
\rput(7.91,0){\righttri {}{}v}
\rput(-2.1,2){\lefttri {}{}{u}}
\rput(7.91,2){\righttri {}{}u}
\rput(-1.1,2){\sdiam {}{}{}{}{{2\lambda\!-\!u\!-\!v}}}
\rput(7.92,2){\diam {}{}{}{}{u+v}}
\rput(-.09,3){\rdiam {}{}{}{}{v-u}}
\rput(6.92,3){\rdiam {}{}{}{}{u-v}}
\rput(1,0){\oface v}
\rput(2,0){\oface v}
\rput(3,0){\oface {}}
\rput(4,0){\oface v}
\rput(5,0){\oface {v-\xi}}
\rput(1,1){\oface u}
\rput(2,1){\oface u}
\rput(3,1){\oface {}}
\rput(4,1){\oface u}
\rput(5,1){\oface {u-\xi}}
\rput(1,2){\eface u}
\rput(2,2){\eface u}
\rput(3,2){\eface {}}
\rput(4,2){\eface u}
\rput(5,2){\eface {u+\xi}}
\rput(1,3){\eface v}
\rput(2,3){\eface v}
\rput(3,3){\eface {}}
\rput(4,3){\eface v}
\rput(5,3){\eface {v+\xi}}
\rput[t](3.5,2.4){\small $\cdots$}
\rput[b](3.5,1.4){\small $\cdots$}
\rput[t](3.5,3.4){\small $\cdots$}
\rput[b](3.5,0.4){\small $\cdots$}
\psline[linecolor=red,linewidth=1pt,linestyle=dashed](5,-.3)(5,4.3)
\end{pspicture}}\\[1cm]
&\raisebox{0cm}{$\disp\overset{\text{YBE3}}=
\frac{1}{\eta_1\eta_2}$\ }
\raisebox{-2.1cm}{
\psset{unit=1.1cm}
\begin{pspicture}(-2.,0)(6.5,3.5)
\multiput(0,0)(0,1){5}{\psline[linewidth=.25pt,linestyle=dashed](-2,0)(10,0)}
\rput(-2.1,0){\lefttri {}{}{v}}
\rput(8.93,0){\righttri {}{}v}
\rput(-2.1,2){\lefttri {}{}{u}}
\rput(8.93,2){\righttri {}{}u}
\rput(-1.1,2){\sdiam {}{}{}{}{{2\lambda\!-\!u\!-\!v}}}
\rput(8.95,2){\diam {}{}{}{}{u+v}}
\rput(5.96,3){\rdiam {}{}{}{}{v-u}}
\rput(7.96,3){\rdiam {}{}{}{}{u-v}}
\rput(0,0){\oface v}
\rput(1,0){\oface v}
\rput(2,0){\oface v}
\rput(3,0){\oface {}}
\rput(4,0){\oface {v-\xi}}
\rput(0,1){\oface u}
\rput(1,1){\oface u}
\rput(2,1){\oface u}
\rput(3,1){\oface {}}
\rput(4,1){\oface {u-\xi}}
\rput(0,2){\eface v}
\rput(1,2){\eface v}
\rput(2,2){\eface v}
\rput(3,2){\eface {}}
\rput(4,2){\eface {v+\xi}}
\rput(0,3){\eface u}
\rput(1,3){\eface u}
\rput(2,3){\eface u}
\rput(3,3){\eface {}}
\rput(4,3){\eface {u+\xi}}
\rput[t](3.5,2.4){\small $\cdots$}
\rput[b](3.5,1.4){\small $\cdots$}
\rput[t](3.5,3.4){\small $\cdots$}
\rput[b](3.5,0.4){\small $\cdots$}
\psline[linecolor=red,linewidth=1pt,linestyle=dashed](4,-.3)(4,4.3)
\end{pspicture}}\\[1cm]
&\raisebox{0cm}{$\overset{\text{Inv1}}={\disp\frac{1}{\eta_2}}$\ }\raisebox{-2.1cm}{
\psset{unit=1.1cm}
\begin{pspicture}(-2,0)(6.5,3.5)
\multiput(0,0)(0,1){5}{\psline[linewidth=.25pt,linestyle=dashed](-2,0)(7,0)}
\rput(-2.1,0){\lefttri {}{}{v}}
\rput(5.93,0){\righttri {}{}v}
\rput(-2.1,2){\lefttri {}{}{u}}
\rput(5.93,2){\righttri {}{}u}
\rput(-1.1,2){\sdiam {}{}{}{}{{2\lambda\!-\!u\!-\!v}}}
\rput(5.95,2){\diam {}{}{}{}{u+v}}
\rput(0,0){\oface v}
\rput(1,0){\oface v}
\rput(2,0){\oface v}
\rput(3,0){\oface {}}
\rput(4,0){\oface {v-\xi}}
\rput(0,1){\oface u}
\rput(1,1){\oface u}
\rput(2,1){\oface u}
\rput(3,1){\oface {}}
\rput(4,1){\oface {u-\xi}}
\rput(0,2){\eface v}
\rput(1,2){\eface v}
\rput(2,2){\eface v}
\rput(3,2){\eface {}}
\rput(4,2){\eface {v+\xi}}
\rput(0,3){\eface u}
\rput(1,3){\eface u}
\rput(2,3){\eface u}
\rput(3,3){\eface {}}
\rput(4,3){\eface {u+\xi}}
\rput[t](3.5,2.4){\small $\cdots$}
\rput[b](3.5,1.4){\small $\cdots$}
\rput[t](3.5,3.4){\small $\cdots$}
\rput[b](3.5,0.4){\small $\cdots$}
\psline[linecolor=red,linewidth=1pt,linestyle=dashed](4,-.3)(4,4.3)
\end{pspicture}}\\[1cm]
&\raisebox{0cm}{$\overset{\text{YBE1}}={\disp\frac{1}{\eta_2}}$\ }\raisebox{-2.1cm}{
\psset{unit=1.1cm}
\begin{pspicture}(-1.,0)(6.5,3.5)
\multiput(0,0)(0,1){5}{\psline[linewidth=.25pt,linestyle=dashed](-1,0)(9,0)}
\rput(-1.058,0){\lefttri {}{}{v}}
\rput(7.95,0){\righttri {}{}v}
\rput(-1.058,2){\lefttri {}{}{u}}
\rput(7.95,2){\righttri {}{}u}
\rput(3,0){\oface {v}}
\rput(4,0){\oface {v}}
\rput(5,0){\oface {}}
\rput(6,0){\oface v}
\rput(7,0){\oface {v-\xi}}
\rput(3,1){\eface {v}}
\rput(4,1){\eface {v}}
\rput(5,1){\eface {}}
\rput(6,1){\eface v}
\rput(7,1){\eface {v+\xi}}
\rput(3,2){\oface {u}}
\rput(4,2){\oface {u}}
\rput(5,2){\oface {}}
\rput(6,2){\oface u}
\rput(7,2){\oface {u-\xi}}
\rput(3,3){\eface {u}}
\rput(4,3){\eface {u}}
\rput(5,3){\eface {}}
\rput(6,3){\eface u}
\rput(7,3){\eface {u+\xi}}
\rput[t](5.5,2.4){\small $\dots$}
\rput[b](5.5,1.4){\small $\cdots$}
\rput[t](5.5,3.4){\small $\cdots$}
\rput[b](5.5,0.4){\small $\dots$}
\rput(-.05,2){\sdiam {}{}{}{}{2\lambda\!-\!u\!-\!v}}
\rput(1.95,2){\diam {}{}{}{}{u+v}}
\psline[linecolor=red,linewidth=1pt,linestyle=dashed](7,-.3)(7,4.3)
\end{pspicture}}\\[1cm]
&\raisebox{0cm}{$\overset{\text{Inv2}}=\;$}
\raisebox{-2.1cm}{
\psset{unit=1.1cm}
\begin{pspicture}(-1.,0)(6.5,3.5)
\multiput(0,0)(0,1){5}{\psline[linewidth=.25pt,linestyle=dashed](-1,0)(6,0)}
\rput(-1.058,0){\lefttri {}{}{v}}
\rput(4.95,0){\righttri {}{}v}
\rput(-1.058,2){\lefttri {}{}{u}}
\rput(4.95,2){\righttri {}{}u}
\rput(0,0){\oface v}
\rput(1,0){\oface v}
\rput(2,0){\oface {}}
\rput(3,0){\oface v}
\rput(4,0){\oface {v-\xi}}
\rput(0,1){\eface v}
\rput(1,1){\eface v}
\rput(2,1){\eface {}}
\rput(3,1){\eface v}
\rput(4,1){\eface {v+\xi}}
\rput(0,2){\oface u}
\rput(1,2){\oface u}
\rput(2,2){\oface {}}
\rput(3,2){\oface u}
\rput(4,2){\oface {u-\xi}}
\rput(0,3){\eface u}
\rput(1,3){\eface u}
\rput(2,3){\eface {}}
\rput(3,3){\eface u}
\rput(4,3){\eface {u+\xi}}
\rput[t](2.5,2.4){\small $\dots$}
\rput[b](2.5,1.4){\small $\cdots$}
\rput[t](2.5,3.4){\small $\cdots$}
\rput[b](2.5,0.4){\small $\dots$}
\psline[linecolor=red,linewidth=1pt,linestyle=dashed](4,-.3)(4,4.3)
\end{pspicture}}\raisebox{0cm}{$\!\!\!=\;\vec D(v)\vec D(u)$}
\end{flalign}
\end{subequations}

\goodbreak
\section{Solution of Dimers on a Strip and Finite-Size Spectra}
\label{SecStrip}

In this section, we specialise to the six vertex model at the free-fermion point with $\lambda=\frac{\pi}{2}$ and $x=i$ corresponding to dimers.

\subsection{Inversion identities on the strip}

In Appendix~A, we show that the double row transfer matrices (\ref{D}) satisfy the inversion identities
\begin{align}
w=0:\ \  &\vec D(u)\vec D(u\!+\!\lambda)= -\tan^2 2u\Big[\cos^{2N}\!u-\sin^{2N}\!u\Big]^2\! \Ib\label{InvIda}\\
w=1:\ \ &\vec D(u)\vec D(u\!+\!\lambda)=-\tan^2 2u\Big[\sin(u\!+\!\xi)\sin(u\!-\!\xi)\cos^{2N}\!u-\cos(u\!+\!\xi)\cos(u\!-\!\xi)\sin^{2N}\!u\Big]^2\! \Ib
\label{InvIdb}
\end{align}
The first inversion identity is obtained from the second by dividing both sides by $\cos^4\xi$ and taking the braid limit $\xi\to i\infty$. It is useful to introduce the normalized transfer matrices
\bea
\vec d(u)=\begin{cases}
\disp\frac{\vec D(u)}{\sin 2u},&w=0\\[6pt]
\disp\frac{\vec D(u)}{\sin^2(\xi+\lambda)\sin 2u},&w=1
\end{cases}\label{d}
\eea
satisfying the initial condition and crossing symmetry
\bea
\vec d(0)=\Ib,\qquad  \vec d(\lambda-u)= \vec d(u)
\eea

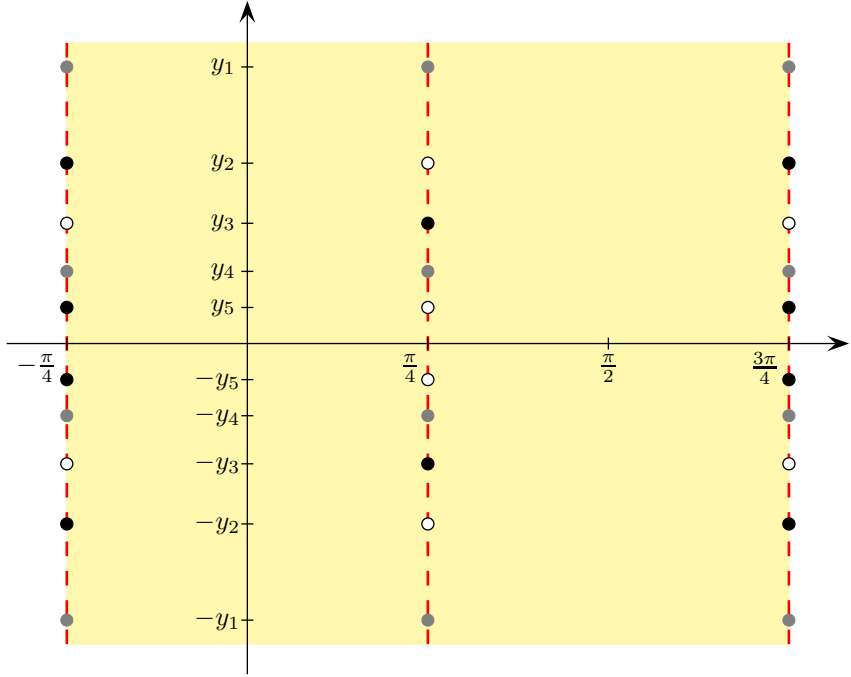
\begin{figure}
\psset{unit=.8cm}
\setlength{\unitlength}{.8cm}
\begin{center}
\begin{pspicture}[shift=-5.2](-.25,.7)(14,11.5)
\psframe[linecolor=yellow!40!white,linewidth=0pt,fillstyle=solid,fillcolor=yellow!40!white](1,1)(13,11)
\psline[linecolor=black,linewidth=.5pt,arrowsize=6pt]{->}(4,.5)(4,11.7)
\psline[linecolor=black,linewidth=.5pt,arrowsize=6pt]{->}(0,6)(14,6)
\psline[linecolor=red,linewidth=1pt,linestyle=dashed,dash=.25 .25](1,1)(1,11)
\psline[linecolor=red,linewidth=1pt,linestyle=dashed,dash=.25 .25](7,1)(7,11)
\psline[linecolor=red,linewidth=1pt,linestyle=dashed,dash=.25 .25](13,1)(13,11)
\psline[linecolor=black,linewidth=.5pt](1,5.9)(1,6.1)
\psline[linecolor=black,linewidth=.5pt](7,5.9)(7,6.1)
\psline[linecolor=black,linewidth=.5pt](10,5.9)(10,6.1)
\psline[linecolor=black,linewidth=.5pt](13,5.9)(13,6.1)
\rput(.5,5.6){\small $-\frac{\pi}{4}$}
\rput(6.7,5.6){\small $\frac{\pi}{4}$}
\rput(10,5.6){\small $\frac{\pi}{2}$}
\rput(12.6,5.6){\small $\frac{3\pi}{4}$}
\psline[linecolor=black,linewidth=.5pt](3.9,6.6)(4.1,6.6)
\psline[linecolor=black,linewidth=.5pt](3.9,7.2)(4.1,7.2)
\psline[linecolor=black,linewidth=.5pt](3.9,8.0)(4.1,8.0)
\psline[linecolor=black,linewidth=.5pt](3.9,9.0)(4.1,9.0)
\psline[linecolor=black,linewidth=.5pt](3.9,10.6)(4.1,10.6)
\psline[linecolor=black,linewidth=.5pt](3.9,5.4)(4.1,5.4)
\psline[linecolor=black,linewidth=.5pt](3.9,4.8)(4.1,4.8)
\psline[linecolor=black,linewidth=.5pt](3.9,4.0)(4.1,4.0)
\psline[linecolor=black,linewidth=.5pt](3.9,3.0)(4.1,3.0)
\psline[linecolor=black,linewidth=.5pt](3.9,1.4)(4.1,1.4)
\rput(3.6,6.6){\small $y_5$}
\rput(3.6,7.2){\small $y_4$}
\rput(3.6,8.0){\small $y_3$}
\rput(3.6,9.0){\small $y_2$}
\rput(3.6,10.6){\small $y_1$}
\rput(3.5,5.4){\small $-y_5$}
\rput(3.5,4.8){\small $-y_4$}
\rput(3.5,4.0){\small $-y_3$}
\rput(3.5,3.0){\small $-y_2$}
\rput(3.5,1.4){\small $-y_1$}
\psarc[linecolor=black,linewidth=.5pt,fillstyle=solid,fillcolor=black](1,6.6){.1}{0}{360}
\psarc[linecolor=gray,linewidth=0pt,fillstyle=solid,fillcolor=gray](1,7.2){.1}{0}{360}
\psarc[linecolor=black,linewidth=.5pt,fillstyle=solid,fillcolor=white](1,8.0){.1}{0}{360}
\psarc[linecolor=black,linewidth=.5pt,fillstyle=solid,fillcolor=black](1,9.0){.1}{0}{360}
\psarc[linecolor=gray,linewidth=0pt,fillstyle=solid,fillcolor=gray](1,10.6){.1}{0}{360}
\psarc[linecolor=black,linewidth=.5pt,fillstyle=solid,fillcolor=white](7,6.6){.1}{0}{360}
\psarc[linecolor=gray,linewidth=0pt,fillstyle=solid,fillcolor=gray](7,7.2){.1}{0}{360}
\psarc[linecolor=black,linewidth=.5pt,fillstyle=solid,fillcolor=black](7,8.0){.1}{0}{360}
\psarc[linecolor=black,linewidth=.5pt,fillstyle=solid,fillcolor=white](7,9.0){.1}{0}{360}
\psarc[linecolor=gray,linewidth=0pt,fillstyle=solid,fillcolor=gray](7,10.6){.1}{0}{360}
\psarc[linecolor=black,linewidth=.5pt,fillstyle=solid,fillcolor=black](13,6.6){.1}{0}{360}
\psarc[linecolor=gray,linewidth=0pt,fillstyle=solid,fillcolor=gray](13,7.2){.1}{0}{360}
\psarc[linecolor=black,linewidth=.5pt,fillstyle=solid,fillcolor=white](13,8.0){.1}{0}{360}
\psarc[linecolor=black,linewidth=.5pt,fillstyle=solid,fillcolor=black](13,9.0){.1}{0}{360}
\psarc[linecolor=gray,linewidth=0pt,fillstyle=solid,fillcolor=gray](13,10.6){.1}{0}{360}
\psarc[linecolor=black,linewidth=.5pt,fillstyle=solid,fillcolor=black](1,5.4){.1}{0}{360}
\psarc[linecolor=gray,linewidth=0pt,fillstyle=solid,fillcolor=gray](1,4.8){.1}{0}{360}
\psarc[linecolor=black,linewidth=.5pt,fillstyle=solid,fillcolor=white](1,4.0){.1}{0}{360}
\psarc[linecolor=black,linewidth=.5pt,fillstyle=solid,fillcolor=black](1,3.0){.1}{0}{360}
\psarc[linecolor=gray,linewidth=0pt,fillstyle=solid,fillcolor=gray](1,1.4){.1}{0}{360}
\psarc[linecolor=black,linewidth=.5pt,fillstyle=solid,fillcolor=white](7,5.4){.1}{0}{360}
\psarc[linecolor=gray,linewidth=0pt,fillstyle=solid,fillcolor=gray](7,4.8){.1}{0}{360}
\psarc[linecolor=black,linewidth=.5pt,fillstyle=solid,fillcolor=black](7,4.0){.1}{0}{360}
\psarc[linecolor=black,linewidth=.5pt,fillstyle=solid,fillcolor=white](7,3.0){.1}{0}{360}
\psarc[linecolor=gray,linewidth=0pt,fillstyle=solid,fillcolor=gray](7,1.4){.1}{0}{360}
\psarc[linecolor=black,linewidth=.5pt,fillstyle=solid,fillcolor=black](13,5.4){.1}{0}{360}
\psarc[linecolor=gray,linewidth=0pt,fillstyle=solid,fillcolor=gray](13,4.8){.1}{0}{360}
\psarc[linecolor=black,linewidth=.5pt,fillstyle=solid,fillcolor=white](13,4.0){.1}{0}{360}
\psarc[linecolor=black,linewidth=.5pt,fillstyle=solid,fillcolor=black](13,3.0){.1}{0}{360}
\psarc[linecolor=gray,linewidth=0pt,fillstyle=solid,fillcolor=gray](13,1.4){.1}{0}{360}
\end{pspicture}
\end{center}
\caption{\label{uplane}A typical pattern of zeros in the complex $u$-plane associated to a transfer matrix eigenvalue. Single zeros are shown by grey disks, double zeros are shown by black disks and the absence of zeros is shown by white disks. The upper and lower half-planes are related under the ${\Bbb Z}_2$ complex conjugation symmetry.}
\end{figure}

Remarkably, the normalized double row transfer matrices (\ref{d}) satisfy precisely the same inversion identities as critical dense polymers~\cite{PR2007,PRVKac}. Specifically, we find
\begin{subequations}
\label{InvIdStrip}
\begin{align}
&w=0:\quad &\vec d(u)&=\frac{\vec D(u)}{\sin 2u},&\qquad \vec d(u)\vec d(u+\lambda)\,&=\,\Big(\frac{\cos^{2N}\!u-\sin^{2N}\!u}{\cos^2\!u-\sin^2\!u}
  \Big)^{\!2}\Ib
\label{dd0}\\
&w=1,\ \xi=\frac{\lambda}{2}:\quad& \vec d(u)&=\frac{2\vec D(u)}{\sin 2u},&\qquad \vec d(u)\vec d(u+\lambda)\,&=\,\big({\cos^{2N}\!u+\sin^{2N}\!u}
  \big)^{\!2}\Ib
\label{dd1}
\end{align}
\end{subequations}
Using standard inversion identity techniques~\cite{Felderhof,BaxBook,OPW1996}, the last two functional equations can be solved, for arbitrary finite sizes $N$, for the eigenvalues $d(u)$ of $\vec d(u)$.  
The calculation of the eigenvalues by solving the functional equations (\ref{dd0}) and (\ref{dd1}) follows exactly the same path as in \cite{PRVKac}. 
So let us just summarize the salient facts. 
The eigenvalues $d(u)$ are Laurent polynomials in $z=e^{iu}$. Consequently, they are determined by their complex zeros in the analyticity strip $-\frac{\pi}{4}\le \Re u\le \frac{3\pi}{4}$. Following~\cite{PRVKac}, these zeros occur as \mbox{1-strings} in the center of the analyticity strip or as ``2-strings" with one zero on the boundary $\Re u=-\frac{\pi}{4}$ of the analyticity strip and its periodic image on the other boundary $\Re u=\frac{3\pi}{4}$. The ordinates of the 1- and 2-strings are quantized and given by
\bea
y_j\,=\,-\frac{i}{2}\ln\tan\frac{E_j\pi}{2N},\qquad 
E_j=\begin{cases}j,&\mbox{$N+w$ even}\\ j-\half,&\mbox{$N+w$ odd}\end{cases}\qquad j\in{\Bbb Z}
\label{yj}
\eea
At each allowed ordinate, there is either two 1-strings, two 2-strings or one 1-string and one 2-string. The fact that double zeros occur has its origins in the relation between critical dense polymers and symplectic fermions~\cite{Symplectic}. Due to complex conjugation symmetry, the pattern of zeros in the upper and lower half-planes is the same. We can therefore focus solely on the lower half-plane. 
A typical pattern of zeros is shown in Figure~\ref{uplane}.

A pattern of zeros is completely determined by specifying the location of the 1-strings. A 1-string at position $j$ is a local elementary excitation with associated conformal energy $E_j$. In the ground state, with energy $E_0$, there are no 1-strings. Counting the doubled 1-strings as two separate 1-strings, the conformal excitation energy above the ground state is given by
\bea
E=E_0+\sum_j E_j,\qquad \mbox{$j=$ position of 1-strings}
\eea

The lowest state energy is $E_0=-\frac{c}{24}+\Delta_s$ where $c$ is the central charge and $\Delta_s$ is the conformal weight associated with the particular sector labelled by 
\bea
s=|S_z|+1,\qquad \mbox{$\calN+s$ odd}
\eea
The lowest states in each sector exactly coincide with those of critical dense polymers for arbitrary finite sizes. The zero patterns for these lowest states are encoded as double column diagrams in Figure~\ref{groundstates}.
On the strip, the only difference between dimers and critical dense polymers with $(r,s)=(1,1)$ boundary conditions resides in the degeneracy of energy levels and the counting of states. The finite-size corrections based on Euler-Maclaurin calculations therefore also coincide, yielding $c_\text{eff}=1$. 
As justified in Section~5, we conclude that the central charge is $c=-2$ and the conformal weights are $\Delta_s=\big((2-s)^2-1\big)/8$ with $s=1,2,3,\ldots$. 

\psset{unit=.5cm}
\setlength{\unitlength}{.6cm}
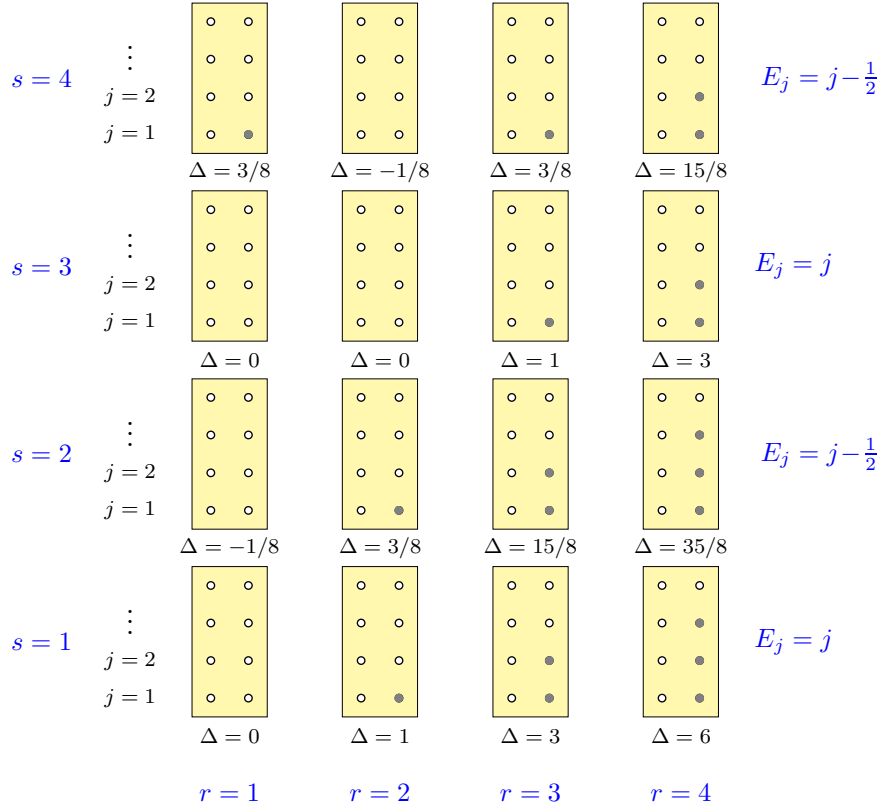
\begin{figure}[tb]
\begin{center}
\begin{pspicture}[shift=-5](-.25,-2.25)(.6,5)
\multirput(0,0)(0,5){4}{\rput(0,.5){\scriptsize $j=1$}
\rput(0,1.5){\scriptsize $j=2$}
\rput(0,2.75){$\vdots$}}
\end{pspicture}
\hspace{4pt}
\begin{pspicture}[shift=-5](-.25,-2.25)(15,19)
\multirput(0,0)(0,5){4}{\multirput(0,0)(4,0){4}{\psframe[linewidth=0pt,fillstyle=solid,fillcolor=yellow!40!white](0,0)(2,4)}}
\multirput(0,0)(4,0){4}{
\multirput(0,0)(0,5){4}{\multirput(0,0)(0,1){4}{\multirput(0,0)(1,0){2}{\psarc[linecolor=black,linewidth=.5pt,fillstyle=solid,fillcolor=white](0.5,0.5){.1}{0}{360}}}}}
\multirput(0,0)(0,5){2}{
\multirput(4,0)(4,0){3}{\psarc[linecolor= gray,linewidth=.5pt,fillstyle=solid,fillcolor=gray](1.5,.5){.1}{0}{360}}
\multirput(8,1)(4,0){2}{\psarc[linecolor= gray,linewidth=.5pt,fillstyle=solid,fillcolor=gray](1.5,.5){.1}{0}{360}}
\multirput(12,2)(4,0){1}{\psarc[linecolor= gray,linewidth=.5pt,fillstyle=solid,fillcolor=gray](1.5,.5){.1}{0}{360}}
}
\multirput(0,10)(0,5){2}{
\multirput(8,0)(4,0){2}{\psarc[linecolor= gray,linewidth=.5pt,fillstyle=solid,fillcolor=gray](1.5,.5){.1}{0}{360}}
\multirput(12,1)(4,0){1}{\psarc[linecolor= gray,linewidth=.5pt,fillstyle=solid,fillcolor=gray](1.5,.5){.1}{0}{360}}
}
\rput(0,15){\psarc[linecolor= gray,linewidth=.5pt,fillstyle=solid,fillcolor=gray](1.5,.5){.1}{0}{360}}
\rput(1,-.5){\scriptsize $\Delta=0$}
\rput(5,-.5){\scriptsize $\Delta=1$}
\rput(9,-.5){\scriptsize $\Delta=3$}
\rput(13,-.5){\scriptsize $\Delta=6$}
\rput(1,4.5){\scriptsize $\Delta=-1/8$}
\rput(5,4.5){\scriptsize $\Delta=3/8$}
\rput(9,4.5){\scriptsize $\Delta=15/8$}
\rput(13,4.5){\scriptsize $\Delta=35/8$}
\rput(1,9.5){\scriptsize $\Delta=0$}
\rput(5,9.5){\scriptsize $\Delta=0$}
\rput(9,9.5){\scriptsize $\Delta=1$}
\rput(13,9.5){\scriptsize $\Delta=3$}
\rput(1,14.5){\scriptsize $\Delta=3/8$}
\rput(5,14.5){\scriptsize $\Delta=-1/8$}
\rput(9,14.5){\scriptsize $\Delta=3/8$}
\rput(13,14.5){\scriptsize $\Delta=15/8$}
\rput(-4,2){\small \color{blue} $s=1$}
\rput(-4,7){\small \color{blue} $s=2$}
\rput(-4,12){\small \color{blue} $s=3$}
\rput(-4,17){\small \color{blue} $s=4$}
\rput(1,-2){\small \color{blue} $r=1$}
\rput(5,-2){\small \color{blue} $r=2$}
\rput(9,-2){\small \color{blue} $r=3$}
\rput(13,-2){\small \color{blue} $r=4$}
\rput(16,2){\small \color{blue} $E_j=j$}
\rput(16.7,7){\small \color{blue} $E_j=j\!-\!\half$}
\rput(16,12){\small \color{blue} $E_j=j$}
\rput(16.7,17){\small \color{blue} $E_j=j\!-\!\half$}
\end{pspicture}
\vspace{-.1in}
\end{center}
\caption{Lowest or groundstate double-column configurations arranged by sectors in a Kac table for $r,s=1,2,3,4$ for critical dense polymers. 
The continuation of the pattern for larger values of $r$ and $s$ is clear. Only the first column with $r=1$ relates to dimers. 
The solid grey dots represent single 1-strings in the center of the analyticity strip. 
There are no double zeros in the center of the analyticity strip for these groundstates. The vacuum sector with $\Delta=0$ lies at $(r,s)=(1,1)$.
}
\label{groundstates}
\end{figure}

It follows that the finitized characters take the form
\bea
\chit_s^{(N)}(q)=q^{-c/24+\Delta_s} \sum_E q^E,\qquad \mbox{$E=$ eigenvalue excitation energy}
\eea
This is a truncated set of conformal eigenenergies of the infinite system. The finitized characters are the spectrum generating functions for the finite set of conformal energies. 
The parameter $q$ is the modular nome and arises through the finite-size calculation as
\bea
q=\exp(-2\pi {N'\over N}\sin 2u)
\eea
where $N'/N$ is the fixed lattice aspect ratio. 
The remaining problem is thus to classify the allowed patterns of zeros and their degeneracies. This is a combinatorial problem and, since not all patterns of zeros occur, it entails certain selection rules. We determine the classification of zero patterns empirically based on examining the patterns of zeros for modest sizes $N$. For critical dense polymers on the strip, the empirical selection rules obtained were ultimately shown to be correct~\cite{MorinDuchesneSelection}. 

\subsection{Combinatorial analysis of patterns of zeros}

\renewcommand{\gauss}[2]{\left[\!\!\begin{array}{c} {#1}\\ {#2} \end{array}\!\!\right]}
\nc{\sgauss}[2]{\Big[\!\!\begin{array}{c} {#1}\\[-3pt] {#2} \end{array}\!\!\Big]}
\renewcommand{\sbin}[2]{\Big\{\!\begin{array}{c} {#1}\\ {#2} 
\end{array}\!\Big\}}
\renewcommand{\sbinlr}[2]{\Big\langle\!\!\begin{array}{c} {#1}\\ {#2} 
\end{array}\!\!\Big\rangle}
\renewcommand{\bino}[2]{\left(\!\!\begin{array}{c} {#1}\\ {#2} \end{array}\!\!\right)}
\def\cat#1#2#3{C_{#1,#2}(#3)}
\def\catt#1#2#3{C_{#1,#2}'(#3)}

Combinatorially, the key building blocks are $q$-Narayana numbers (or equivalently skew $q$-binomials) enumerated by double-column diagrams with dominance.

The information in a zero pattern is simply encoded in a double-column diagram. A double-column configuration $S=(L,R)$
is called {\em admissible} if $L\preceq R$ with respect to the partial ordering
\be
  L\ \preceq\ R\ \ \ \ \ {\rm if}\ \ \ \ \ L_j\ \leq\ R_j,\ \ j=1,2,\ldots,m
\label{order}
\ee
which presupposes that
\be
  0\ \leq\ m\ \leq\ n\ \leq\ M
\label{mnM}
\ee
Admissibility is characterized diagrammatically as in the following example
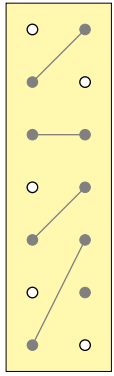
\psset{unit=.7cm}
\setlength{\unitlength}{.7cm}
\be
\begin{pspicture}[shift=-3.45](-.25,-.25)(2,7.3)
\psframe[linewidth=0pt,fillstyle=solid,fillcolor=yellow!40!white](0,0)(2,7)
\psarc[linecolor=black,linewidth=.5pt,fillstyle=solid,fillcolor=white](0.5,6.5){.1}{0}{360}
\psarc[linecolor=gray,linewidth=0pt,fillstyle=solid,fillcolor=gray](0.5,5.5){.1}{0}{360}
\psarc[linecolor=gray,linewidth=0pt,fillstyle=solid,fillcolor=gray](0.5,4.5){.1}{0}{360}
\psarc[linecolor=black,linewidth=.5pt,fillstyle=solid,fillcolor=white](0.5,3.5){.1}{0}{360}
\psarc[linecolor=gray,linewidth=0pt,fillstyle=solid,fillcolor=gray](0.5,2.5){.1}{0}{360}
\psarc[linecolor=black,linewidth=.5pt,fillstyle=solid,fillcolor=white](0.5,1.5){.1}{0}{360}
\psarc[linecolor=gray,linewidth=0pt,fillstyle=solid,fillcolor=gray](0.5,0.5){.1}{0}{360}
\psarc[linecolor=gray,linewidth=0pt,fillstyle=solid,fillcolor=gray](1.5,6.5){.1}{0}{360}
\psarc[linecolor=black,linewidth=.5pt,fillstyle=solid,fillcolor=white](1.5,5.5){.1}{0}{360}
\psarc[linecolor=gray,linewidth=0pt,fillstyle=solid,fillcolor=gray](1.5,4.5){.1}{0}{360}
\psarc[linecolor=gray,linewidth=0pt,fillstyle=solid,fillcolor=gray](1.5,3.5){.1}{0}{360}
\psarc[linecolor=gray,linewidth=0pt,fillstyle=solid,fillcolor=gray](1.5,2.5){.1}{0}{360}
\psarc[linecolor=gray,linewidth=0pt,fillstyle=solid,fillcolor=gray](1.5,1.5){.1}{0}{360}
\psarc[linecolor=black,linewidth=.5pt,fillstyle=solid,fillcolor=white](1.5,0.5){.1}{0}{360}
\psline[linecolor=gray,linewidth=.5pt](0.5,5.5)(1.5,6.5)
\psline[linecolor=gray,linewidth=.5pt](0.5,4.5)(1.5,4.5)
\psline[linecolor=gray,linewidth=.5pt](0.5,2.5)(1.5,3.5)
\psline[linecolor=gray,linewidth=.5pt](0.5,0.5)(1.5,2.5)
\end{pspicture}
\label{adm}
\ee
One draws line segments between the occupied sites of greatest height
in the two columns, then between the occupied sites of second-to-greatest
height and so on. The double-column configuration is now admissible  
if $m\le n$ and it does {\em not\/} involve line segments with a {\em strictly negative slope}.
Thus, in an admissible double-column configuration, there are either no line segments ($m=0$)
or each line segment appears with a non-negative slope. Such admissible diagrams are said to satisfy dominance. 
At each position or height $j$, there is zero, one or two occupied sites corresponding to zero, one or two 1-strings in the lower half-plane.

Combinatorially, the (generalized) $q$-Narayana numbers $\sbinlr{M}{m,n}_{\!q}$ are defined 
as the sum of the monomials associated to all admissible double-column configurations of height $M$ with exactly $m$ and $n$ occupied sites in 
the left and right columns respectively
\be
\sbinlr{M}{m,n}_{\!q}=\sum_{S:\,|L|=m, |R|=n} q^{E(S)}
\ee
These are the basic building blocks to describe the allowed patterns of zeros in each sector. 
Physically, these are the generating functions for the spectrum encoded in a double column diagram with conformal energies $E_j=j$. The monomials $q^{E(S)}$ need to be scaled by the factor $q^{-\frac{1}{2}(m+n)}$ in sectors with $E_j=j-\half$. 
The $q$-Narayana numbers admit the closed-form expressions
\begin{align}
  \sbinlr{M}{m,n}_{\!q}&=q^{\hf m(m+1)+\hf n(n+1)}\sbin{M}{m,n}_q \label{qNara}\\[10pt]
  &=q^{\hf m(m+1)+\hf n(n+1)}
   \bigg(\gauss{M}{m}_{{\!}q}{\gauss{M}{n}}_{{\!}q}-q^{n-m+1} {\gauss{M}{m-1}}_{\!q}{\gauss{M}{n+1}}_{{\!}q}
   \bigg)\quad\label{sbin}
    \end{align}
where $\sgauss{M}{m}_{\!q}$
is a $q$-binomial (Gaussian polynomial) and $\sbin{M}{m,n}_q$ are skew $q$-binomials, as in Appendix~\ref{SkewqBinom}.  The (generalized) $q$-Narayana numbers coincide with
$q$-Narayana numbers~\cite{Narayana,Branden} when $m=n$.

\subsection{Empirical selection rules}
\nc{\qqcat}[3]{C_{#1,#2}(#3)}
\nc{\qqcatt}[3]{C_{#1,#2}'(#3)}

In this Section we consider the empirical classification of patterns of zeros for the cases $w=0,1$.
Empirically, using Mathematica~\cite{Wolfram} to examine the spectra out to $\calN=N+w=8$, we find that the finitized characters are classified in terms of  patterns of zeros, double column diagrams and $q$-Narayana numbers by
\bea
\chit_s^{(N)}(q)=\begin{cases}
\disp q^{-c/24+\Delta_1}\,\sum_{m,n=0}^{\floor{\frac{\calN-1}{2}}} A_{m,n}^{(s)}\, \sbinlr{\floor{\frac{\calN-1}{2}}}{m,n}_{\!q},&\mbox{$\Delta_1=0$, $s$ odd}\\[14pt]
\disp q^{-c/24+\Delta_2}\,\sum_{m,n=0}^{\floor{\frac{\calN-1}{2}}} B_{m,n}^{(s)}\, q^{-\frac{1}{2}(m+n)}\,\sbinlr{\floor{\frac{\calN-1}{2}}}{m,n}_{\!q},&\mbox{$\Delta_2=-\frac{1}{8}$, $s$ even}
\end{cases}
\label{NaraDecomp}
\eea
The $\floor{\frac{\calN+1}{2}}\times \floor{\frac{\calN+1}{2}}$ matrices $A^{(s)}$ and $B^{(s)}$ are special Toeplitz matrices with a simple structure as indicated in the following examples
\begin{align}
&\calN=8\!:\ A^{(1)}\!=\!\tmat{2&2&2&2\\ 0&2&2&2\\ 0&0&2&2\\ 0&0&0&2},\  A^{(3)}\!=\!\tmat{1&2&2&2\\ 0&1&2&2\\ 0&0&1&2\\ 0&0&0&1},\  A^{(5)}\!=\!\tmat{0&1&2&2\\ 0&0&1&2\\ 0&0&0&1\\ 0&0&0&0},\ \dots,\  A^{(9)}\!=\!\tmat{0&0&0&1\\ 0&0&0&0\\ 0&0&0&0\\ 0&0&0&0}\\
&\calN=7\!:\ \ B^{(2)}\!=\!\tmat{1&1&1&1\\ 0&1&1&1\\ 0&0&1&1\\ 0&0&0&1},\  \ B^{(4)}\!=\!\tmat{0&1&1&1\\ 0&0&1&1\\ 0&0&0&1\\ 0&0&0&0},\  \ B^{(6)}\!=\!\tmat{0&0&1&1\\ 0&0&0&1\\ 0&0&0&0\\ 0&0&0&0},\ \ B^{(8)}\!=\!\tmat{0&0&0&1\\ 0&0&0&0\\ 0&0&0&0\\ 0&0&0&0}
\end{align}

Following~\cite{PRV}, for $r\ge 1$, we define the following (generalized) $q$-analogs of Catalan numbers 
\bea
\begin{array}{ll}
\disp\cat Mrq=\sum_{m=0}^{M-r+1}\sbinlr{M}{m,m\!+\!r\!-\!1}_{\!q}=q^{\frac{r(r-1)}{2}}\frac{(1-q^r)}{(1-q^{M+1})}\gauss{2M+2}{M+1-r}_{\!q}\quad &\mbox{$s$ odd}\\[16pt]
\disp\catt Mrq=q^{-\frac{r-1}{2}}\sum_{m=0}^{M-r+1}q^{-m} \sbinlr{M}{m,m\!+\!r\!-\!1}_{\!q}
=q^{\frac{(r-1)^2}{2}}\frac{(1-q^{2r})}{(1-q^{M+r+1})}\gauss{2M+1}{M+1-r}_{\!q}\quad &\mbox{$s$ even}
\end{array}\label{qCat}
\eea
These $q$-Catalan polynomials are simply related to finitized irreducible characters~\cite{PRVKac}
\bea
\mch_{r,1}^{(M)}(q)=q^{-{c\over 24}} \cat Mrq,\qquad \mch_{r,2}^{(M)}(q)=q^{-{c\over 24}-{1\over 8}} \catt Mrq
\label{catfin}
\eea
Using the result
\bea
\lim_{M\to\infty} \gauss{M}{m}_{\!q}={1\over (q)_m},\qquad (q)_m=\prod_{k=1}^m (1-q^k)
\label{qbinlimit}
\eea
it follows from (\ref{qCat}) that, in the thermodynamic limit, the irreducible characters are
\begin{align}
\mch_{r,1}(q)&=\lim_{M\to\infty} \mch_{r,1}^{(M)}(q)=q^{-{c\over 24}+\frac{r(r-1)}{2}}\,\frac{1-q^r}{(q)_\infty}\\ 
\mch_{r,2}(q)&=\lim_{M\to\infty} \mch_{r,2}^{(M)}(q)=q^{-{c\over 24}-{1\over 8}+\frac{(r-1)^2}{2}}\,\frac{1-q^{2r}}{(q)_\infty}
\label{catfinlimit}
\end{align}
Summing over diagonals in (\ref{NaraDecomp}) with $n-m=r-1\ge 0$ gives the decomposition into finitized irreducible characters
\bea
\chit_s^{(N)}(q)=\begin{cases}
\disp \sum_{r=1}^{\floor{\frac{\calN+1}{2}}} A_{1,r}^{(s)}\; \mch_{r,1}^{(M)}(q),&\mbox{$s$ odd}\\[14pt]
\disp \sum_{r=1}^{\floor{\frac{\calN+1}{2}}} B_{1,r}^{(s)}\; \mch_{r,2}^{(M)}(q),&\mbox{$s$ even}
\end{cases}\qquad\qquad M= \floor{\tfrac{\calN-1}{2}}
\label{qCatDecomp}
\eea

For $w=0,1$ and $s$ odd or even, the finitized characters can be written in terms of $q$-binomials
\bea
\chit_s^{(N)}(q)=q^{-c/24+\Delta_s}\, \frac{1+q^{(s-1)/2}}{1+q^{\calN/2}} \gauss{\calN}{\frac{1}{2}(\calN+s-1)}_q,\qquad \calN=N+w
\label{finChar}
\eea
where
\bea
\Delta_s=\frac{(2-s)^2-1}{8}
\eea
Setting $q=1$ gives the correct counting of states $\chit_s^{(N)}(1)=\binom{\calN}{\frac{1}{2}(\calN+s-1)}$. 
Observing that $|q|<1$ and using the result (\ref{qbinlimit}),
it follows that, in the thermodynamic limit,
\bea
\chit_s(q)=\lim_{N\to\infty} \chit_s^{(N)}(q)=\frac{q^{-c/24+\Delta_s}}{(q)_\infty} \,(1+q^{(s-1)/2})=\frac{q^{-c/24}}{(q)_\infty}(q^{\Delta_s}+q^{\Delta_{s+2}})
\label{chars}
\eea
Notice that, for $s=1$, all states are doubly degenerate and that, for $s$ even, $q^{(s-1)/2}$ is a half-integer power of $q$.

\section{Jordan Decompositions and Irreducible Modules}
\label{secJordan}

\subsection{Isotropic double row transfer matrices}

It is easy to verify that, at the isotropic point $u=\frac{\lambda}{2}$, the double row transfer matrix $\vec D(u)$ is not Hermitian. Nevertheless, it has real eigenvalues. For $\calN$ odd, we find no Jordan blocks. But, for $\calN$ even, we find the Jordan decomposition produces nontrivial Jordan blocks of rank 2. Explicitly, for $w=0$ and $\calN=N$ even,
\begin{align}
\calN=2,\  S_z=0:&\qquad \smat{1&1\\0&1}\\
\calN=4,\  S_z=0:&\qquad \half\oplus \smat{\frac{3}{2}-\sqrt{2}&1\\ 0& \frac{3}{2}-\sqrt{2}}\oplus \smat{\frac{3}{2}+\sqrt{2}&1\\ 0 &\frac{3}{2}+\sqrt{2}}\oplus\half
\end{align} 
with similar results for $w=1$, $N$ odd and $\xi=\tfrac{\lambda}{2}$.

\subsection{Quantum Hamiltonians}

The quantum Hamiltonians are given by the logarithmic derivative
\bea
{\cal H}_w=-\half\,\frac{d}{du} \log \vec D(u)\Big|_{u=0},\qquad w=0,1
\eea

As pointed out in \cite{PVO2017}, the Hamiltonian of dimers on the strip with $w=0$ precisely coincides with the $U_q(sl(2))$-invariant XX Hamiltonian of the free-fermion six vertex model
\begin{align}
{\cal H}_{w=0}&=-\sum_{j=1}^{N-1}e_j=-\half\sum_{j=1}^{N-1} (\sigma_j^x\sigma_{j+1}^x+\sigma_j^y\sigma_{j+1}^y)-\half i(\sigma_1^z-\sigma_N^z)\\
&=-\sum_{j=1}^{N-1}(f_j^\dagger f_{j+1}+f_{j+1}^\dagger f_j)-i( f_1^\dagger f_1- f_N^\dagger f_N)
\end{align}
where $\sigma_j^{x,y,z}$ are the usual Pauli matrices and $f_j=\half(\sigma_j^x-i\sigma_j^y)$, $f_j^\dagger=\half(\sigma_j^x+i\sigma_j^y)$. This Hamiltonian is manifestly not Hermitian. 
Nevertheless, the eigenvalues of the Hamiltonian are real~\cite{MDRRSA2015}. For $\calN=N$ odd, we find no non-trivial Jordan blocks but, for $\calN$ even, we find rank 2 Jordan blocks. 
Separating into $S_z=0,\pm 2,\pm 4,\ldots$ sectors and identifying the equivalent $\pm S_z$ sectors, the Jordan canonical forms for $\calN=2$ and $\calN=4$ are respectively
\begin{align}
\calN=2:&\quad \big[\mbox{\scriptsize $\begin{pmatrix}0&1\\ 0&0\end{pmatrix}$}\big]\oplus 2[0]\label{N2}\\
\calN=4:&\quad \big[\mbox{\scriptsize $\begin{pmatrix}\sqrt{2}&1\\ 0&\sqrt{2}\end{pmatrix}$}\oplus 0
\oplus 0\oplus\mbox{\scriptsize $\begin{pmatrix}-\sqrt{2}&1\\ 0&-\sqrt{2}\end{pmatrix}$}\big]
\oplus2\big[\sqrt{2}\oplus\mbox{\scriptsize $\begin{pmatrix}0&1\\ 0&0\end{pmatrix}$}\oplus (-\sqrt{2})\big] \oplus 2[0]\label{N4}
\end{align}
Such Jordan blocks for the quantum group invariant XX Hamiltonian were observed in \cite{GHNS2015}. 
More generally, we find empirically that there are $\binom{N-2}{d-1}$ rank 2 Jordan blocks in sectors with $d$ down arrows where $1\le d\le N-1$,
in accordance with \cite{MorinDuchesneSelection}. 
By comparison, the transfer matrices and Hamiltonians of the $(r,s)$ sectors of critical dense polymers are diagonalizable~\cite{PR2007} and do not exhibit Jordan blocks.

The Hamiltonian of dimers on the strip with $w=1$ agrees with the $\rho=2$ Hamiltonian of critical dense polymers~\cite{PRVKac}
\bea
{\cal H}_{w=1}&=-\sum_{j=1}^{N-1}e_j+\frac{1}{s_0(\xi)s_2(\xi)}\;e_N
\eea
This can be written in terms of fermion operators using (\ref{TLgenf}). Specializing to $\xi=\frac{\lambda}{2}=\frac{\pi}{4}$, we find no Jordan blocks for $\calN=N+1$ odd but rank 2 Jordan blocks for $\calN$ even
\begin{align}
\calN=2:&\quad \big[\mbox{\scriptsize $\begin{pmatrix}0&1\\ 0&0\end{pmatrix}$}\big]\oplus 2[0]\label{NN2}\\
\calN=4:&\quad \big[\mbox{\scriptsize $\begin{pmatrix}\sqrt{3}&1\\ 0&\sqrt{3}\end{pmatrix}$}\oplus 0
\oplus 0\oplus\mbox{\scriptsize $\begin{pmatrix}-\sqrt{3}&1\\ 0&-\sqrt{3}\end{pmatrix}$}\big]
\oplus2\big[\sqrt{3}\oplus\mbox{\scriptsize $\begin{pmatrix}0&1\\ 0&0\end{pmatrix}$}\oplus (-\sqrt{3})\big] \oplus 2[0]\label{NN4}
\end{align}
Although the eigenvalues are different, the patterns of the appearance of Jordan blocks is the same as for $w=0$. This is easily seen for $\calN=2,4$ by comparing (\ref{N2}), (\ref{N4}) with (\ref{NN2}), (\ref{NN4}).

\subsection{Representation theory}

In the continuum scaling limit, the $w=0$ and $w=1$ Hamiltonians give rise to the Virasoro dilatation operator $L_0$. 
For $s$ even ($\mathcal{N}$ odd), we do not find any non-trivial Jordan blocks. For $s$ odd ($\mathcal{N}$ even), on the other hand, we {\em do} find non-trivial Jordan blocks, all of rank $2$. Since the indications are that these Jordan blocks persist in the scaling limit, we see that the dimer model admits higher-rank 
representations of the Virasoro algebra.
We thus conclude that, as a CFT, our dimer model is {\em logarithmic}.

\psset{unit=1.3cm}
\begin{figure}[htb]
\begin{center}
\begin{pspicture}(0,0)(6,1.5)
\rput(-1,.5){$s=1$:}
\rput(-.4,1.3){$\Delta=$}
\rput(0,1.3){$0$}
\rput(1,1.3){$1$}
\rput(2,1.3){$3$}
\rput(3,1.3){$6$}
\rput(4,1.3){$10$}
\rput(5,1.3){$15$}
\multirput(0,0)(2,0){3}{\psline[linewidth=1.5pt,arrowsize=8pt]{->}(0,1)(0,.1)}
\multirput(0,0)(2,0){3}{\psline[linecolor=gray,linestyle=dashed,linewidth=1pt,arrowsize=8pt]{->}(.9,0)(.1,0)}
\multirput(1,0)(2,0){3}{\psline[linecolor=gray,linewidth=1pt,arrowsize=8pt]{<-}(.9,0)(.1,0)}
\multirput(0,1)(2,0){3}{\psline[linecolor=gray,linewidth=1pt,arrowsize=8pt]{<-}(.9,0)(.1,0)}
\multirput(1,1)(2,0){3}{\psline[linecolor=gray,linestyle=dashed,linewidth=1pt,arrowsize=8pt]{->}(.9,0)(.1,0)}
\multirput(0,0)(1,0){6}{\psline[linecolor=gray,linewidth=1pt,arrowsize=8pt]{->}(.95,.95)(.07,.07)}
\multirput(0,0)(2,0){3}{\pscircle[fillstyle=solid,fillcolor=black](0,0){.1}}
\multirput(1,0)(2,0){3}{\pscircle[fillstyle=solid,fillcolor=gray](0,0){.1}}
\multirput(0,1)(2,0){3}{\pscircle[fillstyle=solid,fillcolor=white](0,0){.1}}
\multirput(1,1)(2,0){3}{\pscircle[fillstyle=solid,fillcolor=gray](0,0){.1}}
\multirput(0,0)(0,.5){3}{\rput(6.2,0){$\cdots$}}
\end{pspicture}
\\
\vspace{12pt}
\begin{pspicture}(0,0)(6,1.3)
\rput(-1,.5){$s=3$:}
\multirput(1,0)(2,0){3}{\psline[linewidth=1.5pt,arrowsize=8pt]{->}(0,1)(0,.1)}
\multirput(0,0)(2,0){3}{\psline[linecolor=gray,linewidth=1pt,arrowsize=8pt]{<-}(.9,0)(.1,0)}
\multirput(1,0)(2,0){3}{\psline[linecolor=gray,linestyle=dashed,linewidth=1pt,arrowsize=8pt]{->}(.9,0)(.1,0)}
\multirput(2,1)(2,0){2}{\psline[linecolor=gray,linestyle=dashed,linewidth=1pt,arrowsize=8pt]{->}(.9,0)(.1,0)}
\multirput(1,1)(2,0){3}{\psline[linecolor=gray,linewidth=1pt,arrowsize=8pt]{<-}(.9,0)(.1,0)}
\multirput(0,0)(1,0){6}{\psline[linecolor=gray,linewidth=1pt,arrowsize=8pt]{->}(.95,.95)(.07,.07)}
\multirput(0,0)(2,0){3}{\pscircle[fillstyle=solid,fillcolor=gray](0,0){.1}}
\multirput(1,0)(2,0){3}{\pscircle[fillstyle=solid,fillcolor=black](0,0){.1}}
\multirput(1,1)(2,0){3}{\pscircle[fillstyle=solid,fillcolor=white](0,0){.1}}
\multirput(2,1)(2,0){2}{\pscircle[fillstyle=solid,fillcolor=gray](0,0){.1}}
\multirput(0,0)(0,.5){3}{\rput(6.2,0){$\cdots$}}
\end{pspicture}
\\
\vspace{12pt}
\begin{pspicture}(0,-.6)(6,1.3)
\rput(-1,.5){$s=5$:}
\multirput(2,0)(2,0){2}{\psline[linewidth=1.5pt,arrowsize=8pt]{->}(0,1)(0,.1)}
\multirput(2,0)(2,0){2}{\psline[linecolor=gray,linestyle=dashed,linewidth=1pt,arrowsize=8pt]{->}(.9,0)(.1,0)}
\multirput(1,0)(2,0){3}{\psline[linecolor=gray,linewidth=1pt,arrowsize=8pt]{<-}(.9,0)(.1,0)}
\multirput(2,1)(2,0){2}{\psline[linecolor=gray,linewidth=1pt,arrowsize=8pt]{<-}(.9,0)(.1,0)}
\multirput(3,1)(2,0){2}{\psline[linecolor=gray,linestyle=dashed,linewidth=1pt,arrowsize=8pt]{->}(.9,0)(.1,0)}
\multirput(1,0)(1,0){5}{\psline[linecolor=gray,linewidth=1pt,arrowsize=8pt]{->}(.95,.95)(.07,.07)}
\multirput(2,0)(2,0){2}{\pscircle[fillstyle=solid,fillcolor=black](0,0){.1}}
\multirput(1,0)(2,0){3}{\pscircle[fillstyle=solid,fillcolor=gray](0,0){.1}}
\multirput(2,1)(2,0){2}{\pscircle[fillstyle=solid,fillcolor=white](0,0){.1}}
\multirput(3,1)(2,0){2}{\pscircle[fillstyle=solid,fillcolor=gray](0,0){.1}}
\multirput(0,0)(0,.5){3}{\rput(6.2,0){$\cdots$}}
\rput(-.4,-.3){$\Delta=$}
\rput(0,-.3){$0$}
\rput(1,-.3){$1$}
\rput(2,-.3){$3$}
\rput(3,-.3){$6$}
\rput(4,-.3){$10$}
\rput(5,-.3){$15$}
\rput(3.,-.8){\bf $\vdots$}
\end{pspicture}
\end{center}
\caption{\label{Loewy1}Loewy diagrams of the Virasoro modules $\mathcal{D}_s$ for $s=1,3,5,\ldots$. 
The nodes represent irreducible
sub-quotients, or equivalently, sub-singular vectors generating these sub-quotients.
The black arrows indicate the 
off-diagonal action of $L_0$ in the rank 2 Jordan blocks. 
The gray arrows indicate the conjectured action
by the Virasoro algebra linking the irreducible sub-quotients.}
\end{figure}
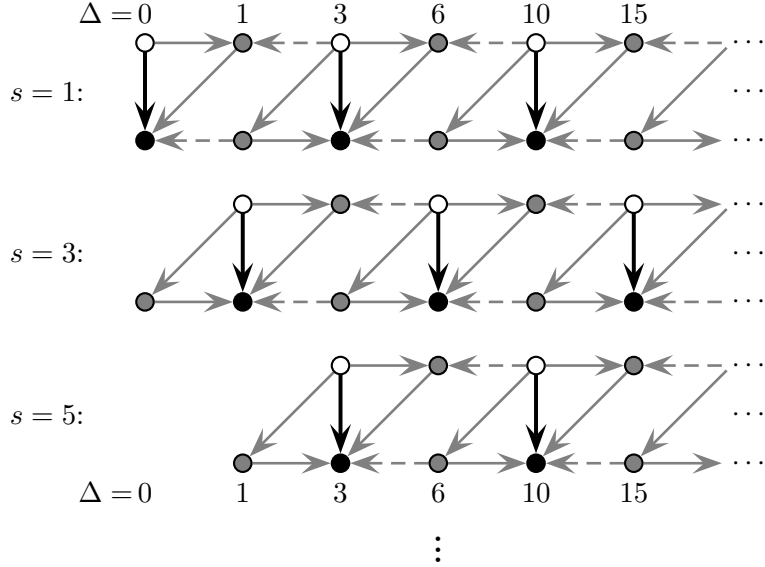

\begin{figure}[htb]
\begin{center}
\begin{pspicture}(0,-.2)(6,1.3) 
\multirput(1,0)(2,0){3}{\psline[linewidth=1.5pt,arrowsize=8pt]{->}(0,1)(0,.1)}
\multirput(0,0)(2,0){3}{\psline[linecolor=gray,linewidth=1pt,arrowsize=8pt]{<-}(.9,0)(.1,0)}
\multirput(1,0)(2,0){3}{\psline[linecolor=gray,linestyle=dashed,linewidth=1pt,arrowsize=8pt]{->}(.9,0)(.1,0)}
\multirput(2,1)(2,0){2}{\psline[linecolor=gray,linestyle=dashed,linewidth=1pt,arrowsize=8pt]{->}(.9,0)(.1,0)}
\multirput(1,1)(2,0){3}{\psline[linecolor=gray,linewidth=1pt,arrowsize=8pt]{<-}(.9,0)(.1,0)}
\multirput(0,0)(1,0){6}{\psline[linecolor=gray,linewidth=1pt,arrowsize=8pt]{->}(.95,.95)(.07,.07)}
\multirput(0,0)(2,0){3}{\pscircle[fillstyle=solid,fillcolor=gray](0,0){.1}}
\multirput(1,0)(2,0){3}{\pscircle[fillstyle=solid,fillcolor=black](0,0){.1}}
\multirput(1,1)(2,0){3}{\pscircle[fillstyle=solid,fillcolor=white](0,0){.1}}
\multirput(2,1)(2,0){2}{\pscircle[fillstyle=solid,fillcolor=gray](0,0){.1}}
\multirput(0,0)(0,.5){3}{\rput(6.2,0){$\cdots$}}
\rput(0,-.3){$\Delta_s$}
\rput(1,-.3){$\Delta_{s+2}$}
\rput(2,-.3){$\Delta_{s+4}$}
\rput(3,-.3){$\Delta_{s+6}$}
\rput(4,-.3){$\Delta_{s+8}$}
\rput(5,-.3){$\Delta_{s+10}$}
\end{pspicture}
\end{center}
\caption{\label{Loewys}Loewy diagram for $s$ odd with $s>1$.}
\end{figure}
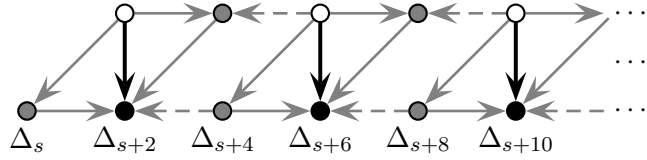

Restricting to $s$ odd and analysing the patterns of appearance of the Jordan blocks yields crucial insight into the structures of 
the ensuing Virasoro modules, here denoted by $\mathcal{D}_s$, $s=1,3,5,\ldots$.
Although not uniquely determined, see comment below, the Loewy diagrams in Figures~\ref{Loewy1} and~\ref{Loewys} are 
compatible with this analysis and with the results for the finitized characters.
Thus, in the Loewy diagram indicated
for $\mathcal{D}_s$, the black dots represent singular vectors which generate the socle $C^{\bullet}$, i.e., the maximal completely reducible submodule of $\mathcal{D}_s$.
The gray dots represent sub-singular vectors that are singular in the quotient 
$\mathcal{D}_s/C^{\bullet}$; they  generate the socle $C^{\textcolor{gray}{\bullet}}$ 
of $\mathcal{D}_s/C^{\bullet}$. Finally, the white dots represent sub-singular vectors that are singular 
in $(\mathcal{D}_s/C^{\bullet})/C^{\textcolor{gray}{\bullet}}$;
they generate the head, i.e., the maximal completely reducible quotient of $\mathcal{D}_s$.

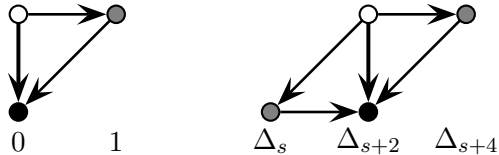
\begin{figure}[htb]
\begin{center}
\begin{pspicture}(1,-.2)(2,1.3)
\multirput(1,0)(2,0){1}{\psline[linewidth=1.5pt,arrowsize=8pt]{->}(0,1)(0,.1)}
\multirput(1,1)(2,0){1}{\psline[linewidth=1pt,arrowsize=8pt]{<-}(.9,0)(.1,0)}
\multirput(1,0)(2,0){1}{\psline[linewidth=1pt,arrowsize=8pt]{->}(.95,.95)(.07,.07)}
\multirput(1,0)(2,0){1}{\pscircle[fillstyle=solid,fillcolor=black](0,0){.1}}
\multirput(1,1)(2,0){1}{\pscircle[fillstyle=solid,fillcolor=white](0,0){.1}}
\multirput(2,1)(2,0){1}{\pscircle[fillstyle=solid,fillcolor=gray](0,0){.1}}
\rput(1,-.3){$0$}
\rput(2,-.3){$1$}
\end{pspicture}
\qquad\qquad\quad
\begin{pspicture}(0,-.2)(2,1.3)
\multirput(1,0)(2,0){1}{\psline[linewidth=1.5pt,arrowsize=8pt]{->}(0,1)(0,.1)}
\multirput(0,0)(2,0){1}{\psline[linewidth=1pt,arrowsize=8pt]{<-}(.9,0)(.1,0)}
\multirput(1,1)(2,0){1}{\psline[linewidth=1pt,arrowsize=8pt]{<-}(.9,0)(.1,0)}
\multirput(0,0)(1,0){2}{\psline[linewidth=1pt,arrowsize=8pt]{->}(.95,.95)(.07,.07)}
\multiput(0,0)(2,0){1}{\pscircle[fillstyle=solid,fillcolor=gray](0,0){.1}}
\multirput(1,0)(2,0){1}{\pscircle[fillstyle=solid,fillcolor=black](0,0){.1}}
\multirput(1,1)(2,0){1}{\pscircle[fillstyle=solid,fillcolor=white](0,0){.1}}
\multirput(2,1)(2,0){1}{\pscircle[fillstyle=solid,fillcolor=gray](0,0){.1}}
\rput(0,-.3){$\Delta_s$}
\rput(1,-.3){$\Delta_{s+2}$}
\rput(2,-.3){$\Delta_{s+4}$}
\end{pspicture}
\end{center}
\caption{\label{Staggered}Staggered modules for $c=-2$, corresponding respectively to $s=1$ and $s=3,5,\ldots$.}
\end{figure}

We note that our analysis so far is incapable of determining whether the left-pointing horizontal 
(dashed)
arrows, in particular, are actually present in the diagrams for $\mathcal{D}_s$ in Figures~\ref{Loewy1} and~\ref{Loewys}. As drawn, the diagrams describe {\em indecomposable} modules, whereas the
diagrams obtained by removing the left-pointing horizontal arrows would describe {\em decomposable} modules.
The latter option is supported by the analysis in \cite{GST14}.
However, independent of the presence of these arrows, the familiar $c=-2$ staggered modules \cite{GK96} in Figure~\ref{Staggered}
appear as {\em submodules} of our dimer modules $\mathcal{D}_s$. This resembles the way the Virasoro Kac modules \cite{Ras11} appear as submodules of the Feigin-Fuchs modules \cite{FF} arising in the dimer model \cite{MDRR2015}. It therefore seems natural to expect that modules similar to our dimer modules exist for other logarithmic minimal models as well.

Following the conjectured module structures (with or without the left-pointing horizontal arrows), the finitized characters $\chit_s^{(N)}(q)$ decompose into finitized characters of the irreducible sub-quotients of conformal weights 
$\Delta\in\{0,1,3,6,10,15,\ldots\}$. Parameterising these weights as
\bea
 \Delta_{r,1}=\Delta_{2r+1}=\frac{(2r-1)^2-1}{8},\quad r=1,2,3,\ldots
\eea
the decompositions are given by (\ref{qCatDecomp}), where the finitized irreducible characters are denoted by $\mch_{r,1}^{(M)}(q)$ with $M= \floor{\tfrac{\calN-1}{2}}$. 
We can refine these decompositions by indicating the appearances of the Jordan blocks.
For simplicity, we do this for the full Virasoro characters. Similar expressions for the finitized characters follow readily. We thus write
\bea
\chit_s(q)=\begin{cases}
\disp \sum_{r\in1+2\mathbb{Z}_{\ge 0}}\hat{2}\,\mch_{r,1}(q)+\sum_{r\in2+2\mathbb{Z}_{\ge 0}}2\,\mch_{r,1}(q),\quad &s=1
\\[18pt]
\disp \mch_{\frac{s-1}{2},1}(q)+\sum_{r\in\frac{s+1}{2}+2\mathbb{Z}_{\ge 0}}\hat{2}\,\mch_{r,1}(q)+\sum_{r\in\frac{s+3}{2}+2\mathbb{Z}_{\ge 0}}2\,\mch_{r,1}(q),\quad &s=3,5,\ldots
\end{cases}
\eea
where a hat on a multiplicity, $\hat{2}$, indicates that Jordan blocks are formed between the matching vectors in the two copies of the corresponding irreducible sub-quotient.

\section{Conclusion}
\label{secConclusion}

Although the dimer model
was first solved many years ago, there remain a number of unanswered questions concerning 
its CFT description.
In a previous paper~\cite{PVO2017}, the dimer model on a cylinder, with $45\degree$ rotated dimers, was solved exactly. Moreover, the modular invariant conformal partition function was obtained from finite-size corrections and shown to precisely agree with the modular invariant partition function of critical dense polymers. 

In this paper, we have solved exactly the dimer model on a strip with $U_q(sl(2))$ invariant boundary conditions by viewing it as a free-fermion six vertex model and using Yang-Baxter techniques. 
The key to solving the lattice model is to show that the commuting double row transfer matrices satisfy special functional equations in the form of inversion identities. Due to the common underlying Temperley-Lieb algebra, these inversion identities coincide with those of critical dense polymers. This implies, essentially through the Temperley-Lieb equivalence, that the two models have the same eigenvalues,
although the eigenvalue degeneracies and the counting of states differ. 
Indeed, the lowest eigenvalues in each sector (labelled by $(r,s)=(1,s)$) coincide leading to the same finite-size corrections. 
This leads us to conclude that, for the $U_q(sl(2))$ invariant boundary conditions, the two models
share the same central charge and a common infinite set of conformal weights
\bea
c=-2,\qquad \Delta_s=\Delta_{1,s}=\frac{(2-s)^2-1}{8},\quad s=1,2,3,\ldots
\eea
The common negative conformal weight
\bea
\Delta_2=\Delta_{1,2}=-\frac{1}{8}
\eea
implies that both CFTs are nonunitary.
However, despite these similarities, combinatorial analysis of the patterns of zeros of the transfer matrix eigenvalues of dimers on the strip leads to finitized and conformal characters (\ref{finChar}) and (\ref{chars}) that are distinct from the Kac characters of critical dense polymers. So it appears that the dimer model, with these boundary conditions,
lies in
a different ``universality class" to that of critical dense polymers. 

Moreover, we have shown that for dimers on the strip, with $U_q(sl(2))$ invariant boundary conditions, the Jordan canonical forms of the isotropic double row transfer matrices and quantum Hamiltonians exhibit nontrivial Jordan blocks. Assuming, as argued, that these Jordan blocks persist in the scaling limit, this implies that the dilatation operator $L_0$ admits higher-rank Virasoro representations. All this 
points to a CFT description of the dimer model that is not rational. Indeed, we argue that, with $U_q(sl(2))$ invariant boundary conditions, the model
is described by a logarithmic CFT with central charge $c=-2$, minimal conformal weight $\Delta_\text{min}=-\frac{1}{8}$  and effective central charge $c_\text{eff}=1$. 

Notwithstanding our conclusion, some additional comments are in order. 
First, we have chosen to apply boundary conditions (\ref{BdyCond}) with $x=i$ compatible with $U_q(sl(2))$ invariance. 
Our boundary weights (\ref{BdyCond}) are thus complex and lead to non-positive Boltzmann weights for the lattice model, a non-Hermitian XX Hamiltonian and a nonunitary CFT.
So, on this basis, it could be argued that our dimer model is ``unphysical". In fact, our dimer model is just the free-fermion point of the $U_q(sl(2))$-invariant XXZ Hamiltonian~\cite{PasquierSaleur} which is a {\em standard\/} and very well-studied quantum chain. While the $U_q(sl(2))$-invariant XX Hamiltonian is non-Hermitian, its eigenvalues are all real. We have also demonstrated that it exhibits proper conformal properties in the continuum scaling limit. We thus argue that our dimer model on the strip and the $U_q(sl(2))$-invariant XX Hamiltonian should not be regarded as ``unphysical" inasmuch as they relate to a seemingly well-defined, albeit logarithmic, CFT. 
The situation is analagous to the two-dimensional RSOS(2,5) lattice model studied by Forrester and Baxter~\cite{FB1985}. 
Indeed, it could similarly be argued that this model is ``unphysical" in the sense that its Boltzmann weights are non-positive, its associated quantum Hamiltonian ${\cal H}_{2,5}$ is non-Hermitian and its associated CFT, which is the Yang-Lee minimal model ${\cal M}(2,5)$, is nonunitary. 
Nevertheless, the free energy, correlation length, local height probabilities and critical exponents of RSOS(2,5) all seem to make good sense. Indeed, for the Ising model in a complex magnetic field, the RSOS(2,5) theory describes~\cite{FisherYL,CardyYL} the critical behaviour associated with the closure of the Yang-Lee zeros on the unit circle. This Yang-Lee edge singularity exemplifies the Yang-Lee universality class. Although ${\cal H}_{2,5}$ is non-Hermitian, 
its
eigenvalues are all real and it provides a good example of non-Hermitian quantum mechanics~\cite{BenderBoettcher,Moiseyev,Bender} which is now accepted to have many ``physical" applications. Similarly, although it is nonunitary, the minimal model ${\cal M}(2,5)$ is
a perfectly well-defined rational CFT~\cite{BPZ,BajnokPDeeb}.

\subsection*{Acknowledgments}

JR was supported by the Australian Research Council under the Discovery Project scheme, project number DP160101376. 
AVO is supported by a Melbourne International Research Scholarship and a Melbourne International Fee Remission Scholarship. 
Parts of this work were carried out while PAP was visiting the APCTP, Pohang, Korea as an ICTP Visiting Scholar. 
The authors thank Alexi Morin-Duchesne and the referee for helpful comments.

\goodbreak

\appendix

\section{Proof of Inversion Identities on the Strip}
\label{StripInvProof}

In this appendix, we prove the inversion identity (\ref{InvIda}) for dimers on the strip
\bea
\vec D(u)\vec D(u+\lambda)=-\tan^2 2u
\left(\cos^{2N}\!u-\sin^{2N}\!u\right)^2 \Ib,\quad&\mbox{$w=0$}
\label{AInvIdStrip}
\eea
where $\vec D(u)$ is the double row transfer matrix (\ref{D}) with $w=0$. The inversion identity (\ref{InvIdb}) is proved similarly.
Throughout this section, we work in the Temperley-Lieb representation with the gauge $g=z:=e^{iu}$.

For a column at position $j$ with fixed $a_j,b_j=0,1$, let us define the four $16\times 16$ matrices\\[5pt]
\bea
\psset{unit=1cm}
R\Big(\begin{matrix}a_j\\ b_j\end{matrix}\Big)=\begin{pspicture}(-.4,2)(1.3,4)
\rput(0,0){\oface u}
\rput(0,1){\eface u}
\rput(0,2){\oface{u\!+\!\lambda}}
\rput(0,3){\eface{u\!+\!\lambda}}
\rput[t](.5,-.1){\small $b_j$}
\rput[b](.5,4.1){\small $a_j$}
\rput[r](-.1,.5){\small $f$}
\rput[r](-.1,1.5){\small $e$}
\rput[r](-.1,2.5){\small $d$}
\rput[r](-.1,3.5){\small $c$}
\rput[l](1.1,.5){\small $f'$}
\rput[l](1.1,1.5){\small $e'$}
\rput[l](1.1,2.5){\small $d'$}
\rput[l](1.1,3.5){\small $c'$}
\end{pspicture}
\eea \\[2cm]
The matrix elements of the product of double row transfer matrices, with upper and lower particle state configurations $\vec a=\{a_1,a_2,\dots,a_N\}$ and $\vec b=\{b_1,b_2,\dots,b_N\}$, are then given by
\begin{align}
\big[\vec D(u)\vec D(u+\lambda)\big]_{\svec b,\svec a}= \langle \mbox{left}|\, \prod_{j=1}^N R\Big(\begin{matrix}a_j\\ b_j\end{matrix}\Big)\,|\mbox{right}\rangle,\qquad a_j,b_j=0,1\label{lDr}
\end{align}
where the left and right boundary vectors are
\begin{align}
&\langle \mbox{left}|=(-1,1,1,-1,0,0,0,0,0,0,0,0,0,0,0,0)\in\calV_6\\
&|\mbox{right}\rangle=(1,1,1,1,0,0,0,0,0,0,0,0,0,0,0,0)^T\in\calV_6
\end{align}

\def\tildeV{\tilde{\calV}}
Setting
\bea
s=\sin u,\; c=\cos u,\; z=e^{i u}, \; x=i
\eea
and ordering the sixteen intermediate basis states as
\begin{align}
\scriptsize
\begin{pmatrix}
 0 \\
 0 \\
 0 \\
 0 \\
\end{pmatrix}\!\!,
\begin{pmatrix}
 0 \\
 0 \\
 1 \\
 1 \\
\end{pmatrix}\!\!,
\begin{pmatrix}
 1 \\
 1 \\
 0 \\
 0 \\
\end{pmatrix}\!\!,
\begin{pmatrix}
 1 \\
 1 \\
 1 \\
 1 \\
\end{pmatrix}\!\!,
\begin{pmatrix}
 0 \\
 1 \\
 1 \\
 0 \\
\end{pmatrix}\!\!,
\begin{pmatrix}
 1 \\
 0 \\
 0 \\
 1 \\
\end{pmatrix}\!\!;\quad
\begin{pmatrix}
 0 \\
 0 \\
 0 \\
 1 \\
\end{pmatrix}\!\!,
\begin{pmatrix}
 0 \\
 0 \\
 1 \\
 0 \\
\end{pmatrix}\!\!,
\begin{pmatrix}
 0 \\
 1 \\
 0 \\
 0 \\
\end{pmatrix}\!\!,
\begin{pmatrix}
 0 \\
 1 \\
 0 \\
 1 \\
\end{pmatrix}\!\!,
\begin{pmatrix}
 0 \\
 1 \\
 1 \\
 1 \\
\end{pmatrix}\!\!,
\begin{pmatrix}
 1 \\
 0 \\
 0 \\
 0 \\
\end{pmatrix}\!\!,
\begin{pmatrix}
 1 \\
 0 \\
 1 \\
 0 \\
\end{pmatrix}\!\!,
\begin{pmatrix}
 1 \\
 0 \\
 1 \\
 1 \\
\end{pmatrix}\!\!,
\begin{pmatrix}
 1 \\
 1 \\
 0 \\
 1 \\
\end{pmatrix}\!\!,
\begin{pmatrix} 1 \\
 1 \\
 1 \\
 0 \\
\end{pmatrix}
\end{align}
the four {\small $R\Big(\begin{matrix}a_j\\ b_j\end{matrix}$\Big)} matrices are given explicitly by
\setlength{\arraycolsep}{1.4pt}
\setcounter{MaxMatrixCols}{20}
\begin{align}
R\Big(\begin{matrix}0\\ 0\end{matrix}\Big)&=\mbox{\scriptsize $\begin{pmatrix}
 c^2 s^2 & 0 & 0 & 0 & \frac{i c s}{z^2} & 0 & 0 & 0 & 0 & 0 & 0 & 0 & 0 & 0 & 0 & 0 \\
 s^2 z^2 & s^4 & 0 & 0 & i c s & 0 & 0 & 0 & 0 & 0 & 0 & 0 & 0 & 0 & 0 & 0 \\
 -c^2 z^2 & 0 & c^4 & 0 & -i c s & 0 & 0 & 0 & 0 & 0 & 0 & 0 & 0 & 0 & 0 & 0 \\
 -z^4 & -s^2 z^2 & c^2 z^2 & c^2 s^2 & -i c s z^2 & 0 & 0 & 0 & 0 & 0 & 0 & 0 & 0 & 0 & 0 & 0 \\
 0 & 0 & 0 & 0 & -c^2 s^2 & 0 & 0 & 0 & 0 & 0 & 0 & 0 & 0 & 0 & 0 & 0 \\
 i c s z^2 & i c s & -i c s & -\frac{i c s}{z^2} & 0 & -c^2 s^2 & 0 & 0 & 0 & 0 & 0 & 0 & 0 & 0 & 0 & 0 \\
 0 & 0 & 0 & 0 & 0 & 0 & c s^3 & 0 & i s^2 & 0 & \frac{i s^2}{z^2} & 0 & 0 & 0 & 0 & 0 \\
 0 & 0 & 0 & 0 & 0 & 0 & 0 & c s^3 & 0 & 0 & 0 & 0 & 0 & 0 & 0 & 0 \\
 0 & 0 & 0 & 0 & 0 & 0 & 0 & 0 & -c^3 s & 0 & 0 & 0 & 0 & 0 & 0 & 0 \\
 0 & 0 & 0 & 0 & 0 & 0 & 0 & 0 & 0 & -c^2 s^2 & 0 & 0 & 0 & 0 & 0 & 0 \\
 0 & 0 & 0 & 0 & 0 & 0 & 0 & 0 & -c s z^2 & 0 & -c s^3 & 0 & 0 & 0 & 0 & 0 \\
 0 & 0 & 0 & 0 & 0 & 0 & 0 & i c^2 & 0 & 0 & 0 & -c^3 s & 0 & 0 & 0 & -\frac{i c^2}{z^2} \\
 0 & 0 & 0 & 0 & 0 & 0 & 0 & 0 & 0 & 0 & 0 & 0 & -c^2 s^2 & 0 & 0 & 0 \\
 0 & 0 & 0 & 0 & 0 & 0 & 0 & i c^2 z^2 & 0 & 0 & 0 & -c s z^2 & 0 & -c s^3 & 0 & -i c^2 \\
 0 & 0 & 0 & 0 & 0 & 0 & -c s z^2 & 0 & -i s^2 z^2 & 0 & -i s^2 & 0 & 0 & 0 & c^3 s & 0 \\
 0 & 0 & 0 & 0 & 0 & 0 & 0 & -c s z^2 & 0 & 0 & 0 & 0 & 0 & 0 & 0 & c^3 s 
\end{pmatrix}$}\\[10pt]
R\Big(\begin{matrix}1\\ 1\end{matrix}\Big)&=\mbox{\scriptsize $\begin{pmatrix}
c^2 s^2 & \frac{c^2}{z^2} & -\frac{s^2}{z^2} & -\frac{1}{z^4} & 0 & \frac{i c s}{z^2} & 0 & 0 & 0 & 0 & 0 & 0 & 0 & 0 & 0 & 0 \\
 0 & c^4 & 0 & -\frac{c^2}{z^2} & 0 & i c s & 0 & 0 & 0 & 0 & 0 & 0 & 0 & 0 & 0 & 0 \\
 0 & 0 & s^4 & \frac{s^2}{z^2} & 0 & -i c s & 0 & 0 & 0 & 0 & 0 & 0 & 0 & 0 & 0 & 0 \\
 0 & 0 & 0 & c^2 s^2 & 0 & -i c s z^2 & 0 & 0 & 0 & 0 & 0 & 0 & 0 & 0 & 0 & 0 \\
 i c s z^2 & i c s & -i c s & -\frac{i c s}{z^2} & -c^2 s^2 & 0 & 0 & 0 & 0 & 0 & 0 & 0 & 0 & 0 & 0 & 0 \\
 0 & 0 & 0 & 0 & 0 & -c^2 s^2 & 0 & 0 & 0 & 0 & 0 & 0 & 0 & 0 & 0 & 0 \\
 0 & 0 & 0 & 0 & 0 & 0 & c^3 s & 0 & 0 & 0 & 0 & 0 & 0 & 0 & -\frac{c s}{z^2} & 0 \\
 0 & 0 & 0 & 0 & 0 & 0 & 0 & c^3 s & 0 & 0 & 0 & i s^2 & 0 & \frac{i s^2}{z^2} & 0 & -\frac{c s}{z^2} \\
 0 & 0 & 0 & 0 & 0 & 0 & i c^2 & 0 & -c s^3 & 0 & -\frac{c s}{z^2} & 0 & 0 & 0 & -\frac{i c^2}{z^2} & 0 \\
 0 & 0 & 0 & 0 & 0 & 0 & 0 & 0 & 0 & -c^2 s^2 & 0 & 0 & 0 & 0 & 0 & 0 \\
 0 & 0 & 0 & 0 & 0 & 0 & i c^2 z^2 & 0 & 0 & 0 & -c^3 s & 0 & 0 & 0 & -i c^2 & 0 \\
 0 & 0 & 0 & 0 & 0 & 0 & 0 & 0 & 0 & 0 & 0 & -c s^3 & 0 & -\frac{c s}{z^2} & 0 & 0 \\
 0 & 0 & 0 & 0 & 0 & 0 & 0 & 0 & 0 & 0 & 0 & 0 & -c^2 s^2 & 0 & 0 & 0 \\
 0 & 0 & 0 & 0 & 0 & 0 & 0 & 0 & 0 & 0 & 0 & 0 & 0 & -c^3 s & 0 & 0 \\
 0 & 0 & 0 & 0 & 0 & 0 & 0 & 0 & 0 & 0 & 0 & 0 & 0 & 0 & c s^3 & 0 \\
 0 & 0 & 0 & 0 & 0 & 0 & 0 & 0 & 0 & 0 & 0 & -i s^2 z^2 & 0 & -i s^2 & 0 & c s^3
\end{pmatrix}$}\\[10pt]
R\Big(\begin{matrix}0\\ 1\end{matrix}\Big)&=\mbox{\scriptsize $\begin{pmatrix}
0 & 0 & 0 & 0 & 0 & 0 & \frac{c s^2}{z} & 0 & \frac{i s^3}{z} & 0 & \frac{i s}{z^3} & 0 & 0 & 0 & 0 & 0 \\
 0 & 0 & 0 & 0 & 0 & 0 & c s^2 z & 0 & 0 & 0 & \frac{i c^2 s}{z} & 0 & 0 & 0 & 0 & 0 \\
 0 & 0 & 0 & 0 & 0 & 0 & -c z & 0 & -i s^3 z & 0 & -\frac{i s}{z} & 0 & 0 & 0 & \frac{c^3}{z} & 0 \\
 0 & 0 & 0 & 0 & 0 & 0 & -c z^3 & 0 & 0 & 0 & -i c^2 s z & 0 & 0 & 0 & c^3 z & 0 \\
 0 & 0 & 0 & 0 & 0 & 0 & 0 & 0 & -c s^2 z & 0 & -\frac{c s^2}{z} & 0 & 0 & 0 & 0 & 0 \\
 0 & 0 & 0 & 0 & 0 & 0 & i c^2 s z & 0 & 0 & 0 & 0 & 0 & 0 & 0 & -\frac{i c^2 s}{z} & 0 \\
 0 & 0 & 0 & 0 & 0 & 0 & 0 & 0 & 0 & \frac{i c s^2}{z} & 0 & 0 & 0 & 0 & 0 & 0 \\
 s^3 z & \frac{s^3}{z} & 0 & 0 & \frac{i c s^2}{z} & 0 & 0 & 0 & 0 & 0 & 0 & 0 & 0 & 0 & 0 & 0 \\
 0 & 0 & 0 & 0 & 0 & 0 & 0 & 0 & 0 & -\frac{c^2 s}{z} & 0 & 0 & 0 & 0 & 0 & 0 \\
 0 & 0 & 0 & 0 & 0 & 0 & 0 & 0 & 0 & 0 & 0 & 0 & 0 & 0 & 0 & 0 \\
 0 & 0 & 0 & 0 & 0 & 0 & 0 & 0 & 0 & -c^2 s z & 0 & 0 & 0 & 0 & 0 & 0 \\
 i c s^2 z & \frac{i c}{z} & -\frac{i c s^2}{z} & -\frac{i c}{z^3} & 0 & -\frac{c^2 s}{z} & 0 & 0 & 0 & 0 & 0 & 0 & 0 & 0 & 0 & 0 \\
 0 & 0 & 0 & 0 & 0 & 0 & 0 & i c^2 s z & 0 & 0 & 0 & -c s^2 z & 0 & -\frac{c s^2}{z} & 0 & -\frac{i c^2 s}{z} \\
 0 & i c^3 z & 0 & -\frac{i c^3}{z} & 0 & -c^2 s z & 0 & 0 & 0 & 0 & 0 & 0 & 0 & 0 & 0 & 0 \\
 0 & 0 & 0 & 0 & 0 & 0 & 0 & 0 & 0 & -i c s^2 z & 0 & 0 & 0 & 0 & 0 & 0 \\
 -s z^3 & -s z & c^2 s z & \frac{c^2 s}{z} & -i c s^2 z & 0 & 0 & 0 & 0 & 0 & 0 & 0 & 0 & 0 & 0 & 0
\end{pmatrix}$}\\[10pt]
R\Big(\begin{matrix}1\\ 0\end{matrix}\Big)&=\mbox{\scriptsize $\begin{pmatrix}
0 & 0 & 0 & 0 & 0 & 0 & 0 & \frac{c^3}{z} & 0 & 0 & 0 & \frac{i c^2 s}{z} & 0 & 0 & 0 & -\frac{c}{z^3} \\
 0 & 0 & 0 & 0 & 0 & 0 & 0 & c^3 z & 0 & 0 & 0 & i s z & 0 & \frac{i s^3}{z} & 0 & -\frac{c}{z} \\
 0 & 0 & 0 & 0 & 0 & 0 & 0 & 0 & 0 & 0 & 0 & -i c^2 s z & 0 & 0 & 0 & \frac{c s^2}{z} \\
 0 & 0 & 0 & 0 & 0 & 0 & 0 & 0 & 0 & 0 & 0 & -i s z^3 & 0 & -i s^3 z & 0 & c s^2 z \\
 0 & 0 & 0 & 0 & 0 & 0 & 0 & i c^2 s z & 0 & 0 & 0 & 0 & 0 & 0 & 0 & -\frac{i c^2 s}{z} \\
 0 & 0 & 0 & 0 & 0 & 0 & 0 & 0 & 0 & 0 & 0 & -c s^2 z & 0 & -\frac{c s^2}{z} & 0 & 0 \\
 c^2 s z & \frac{c^2 s}{z} & -\frac{s}{z} & -\frac{s}{z^3} & 0 & \frac{i c s^2}{z} & 0 & 0 & 0 & 0 & 0 & 0 & 0 & 0 & 0 & 0 \\
 0 & 0 & 0 & 0 & 0 & 0 & 0 & 0 & 0 & 0 & 0 & 0 & \frac{i c s^2}{z} & 0 & 0 & 0 \\
 i c^3 z & 0 & -\frac{i c^3}{z} & 0 & -\frac{c^2 s}{z} & 0 & 0 & 0 & 0 & 0 & 0 & 0 & 0 & 0 & 0 & 0 \\
 0 & 0 & 0 & 0 & 0 & 0 & i c^2 s z & 0 & -c s^2 z & 0 & -\frac{c s^2}{z} & 0 & 0 & 0 & -\frac{i c^2 s}{z} & 0 \\
 i c z^3 & i c s^2 z & -i c z & -\frac{i c s^2}{z} & -c^2 s z & 0 & 0 & 0 & 0 & 0 & 0 & 0 & 0 & 0 & 0 & 0 \\
 0 & 0 & 0 & 0 & 0 & 0 & 0 & 0 & 0 & 0 & 0 & 0 & -\frac{c^2 s}{z} & 0 & 0 & 0 \\
 0 & 0 & 0 & 0 & 0 & 0 & 0 & 0 & 0 & 0 & 0 & 0 & 0 & 0 & 0 & 0 \\
 0 & 0 & 0 & 0 & 0 & 0 & 0 & 0 & 0 & 0 & 0 & 0 & -c^2 s z & 0 & 0 & 0 \\
 0 & 0 & s^3 z & \frac{s^3}{z} & 0 & -i c s^2 z & 0 & 0 & 0 & 0 & 0 & 0 & 0 & 0 & 0 & 0 \\
 0 & 0 & 0 & 0 & 0 & 0 & 0 & 0 & 0 & 0 & 0 & 0 & -i c s^2 z & 0 & 0 & 0
\end{pmatrix}$}
\end{align}
The matrices {\small $R\Big(\begin{matrix}0\\ 0\end{matrix}\Big)$} and {\small $R\Big(\begin{matrix}1\\ 1\end{matrix}\Big)$} are block diagonal under a direct sum decomposition of the intermediate basis of states 
\bea
\calV=\calV_6\oplus \calV_{10}
\eea
so that 
\bea
R\Big(\begin{matrix}a\\ a\end{matrix}\Big): \quad&\calV_6\to \calV_6, \ \ &\calV_{10}\to \calV_{10}
\eea

\def\R#1#2{R\Big(\begin{matrix}#1\\ #2\end{matrix}\Big)}
\def\smR#1#2{\mbox{\small $R\Big(\begin{matrix}#1\\ #2\end{matrix}\Big)$}}
We show that non-diagonal matrix elements with $\vec a\ne\vec b$ vanish. In this case, the states on the left and right in (\ref{lDr}) are built up by the action of {\small $R\Big(\begin{matrix}a\\ b\end{matrix}\Big)$} on the left and right boundaries $\langle \mbox{left}|$ and $|\mbox{right}\rangle$ with the occurrence of at least one {\small $R\Big(\begin{matrix}a\\ 1\!-\!a\end{matrix}\Big)$} matrix. We begin building up the states by acting with {\small $R\Big(\begin{matrix}a\\ a\end{matrix}\Big)$} on the left and right states. We find that
\bea
v_\text{left}=\langle \mbox{left}|\, \prod_{j=0}^n R\Big(\begin{matrix}a_j\\ a_j\end{matrix}\Big)\in \calV_\text{left},\qquad 
v_\text{right}=\prod_{j=0}^n R\Big(\begin{matrix}a_j\\ a_j\end{matrix}\Big)\,|\mbox{right}\rangle\in\calV_\text{right},\qquad n\ge 0
\eea
where the vector spaces $\calV_\text{left}, \calV_\text{right}$ are given by the linear spans
\begin{align}
\calV_\text{left}&=\Big\langle\Big\{\langle \mbox{left}|,\;\langle \mbox{left}|\smR 00,\;\langle \mbox{left}|\smR 11,\;\langle \mbox{left}|
\mbox{\small $R\Big(\begin{matrix}0\\ 0\end{matrix}\Big)^{\!2}$}\Big\}\Big\rangle\\ 
\calV_\text{right}&=\Big\langle\Big\{ |\mbox{right}\rangle,\; \smR 00 |\mbox{right}\rangle,\; \smR 11 |\mbox{right}\rangle,\; 
\mbox{\small $R\Big(\begin{matrix}0\\ 0\end{matrix}\Big)^{\!2}$}|\mbox{right}\rangle \Big\}\Big\rangle
\end{align}
These spaces are stable under the action of further {\small $R\Big(\begin{matrix}a\\ a\end{matrix}\Big)$} matrices. 
The linear independence of vectors is easily checked by calculating the rank of suitable matrices in Mathematica~\cite{Wolfram}. Since
\bea
v_\text{left} \R {a}{1\!-\!a} v_\text{right}=0
\eea
let assume next that there are at least two {\small $R\Big(\begin{matrix}a\\ 1\!-\!a\end{matrix}\Big)$} matrices. In this case, we similarly find that
\begin{align}
\mbox{}\hspace{-10pt}v'_\text{left}&=v_\text{left} \R{a}{1\!-\!a} \prod_{j=0}^n \R{a_j}{a_j}\in 
\Big\langle\Big\{ \langle \mbox{left}| \R{a}{1\!-\!a},\; 
\langle \mbox{left}|\R 00 \R{a}{1\!-\!a}\Big\}\Big\rangle=\calV'_\text{left},\quad n\ge 0\\
\mbox{}\hspace{-10pt}v'_\text{right}\!&=\prod_{j=0}^n \R{a_j}{a_j} \R{a}{1\!-\!a} v_\text{right} \in 
\Big\langle\Big\{ \R{a}{1\!-\!a}|\mbox{right}\rangle,\,\R{a}{1\!-\!a}\R 00 |\mbox{right}\rangle \Big\}\Big\rangle=\calV'_\text{right},\ \  n\ge 0
\end{align}
where the vector spaces are orthogonal. So next suppose that there are three or more {\small $R\Big(\begin{matrix}a\\ 1\!-\!a\end{matrix}\Big)$} matrices. In this case, we observe that
\bea
v_\text{left} \R{a}{1\!-\!a} \prod_{j=0}^n \R{a_j}{a_j} \R{a}{1\!-\!a}=
\R{a}{1\!-\!a}\prod_{j=0}^n \R{a_j}{a_j} \R{a}{1\!-\!a} v_\text{right} =\vec 0\qquad\qquad
\eea
so that the occurrence of the matrices  {\small $R\Big(\begin{matrix}0\\ 1\end{matrix}\Big)$} and  {\small $R\Big(\begin{matrix}1\\ 0\end{matrix}\Big)$} must alternate along the segment.
Moreover, we observe that
\bea
v_\text{left}^\text{eig}=v'_\text{left} \R{a}{1\!-\!a} \R{1\!-\!a}{a}, \ v'_\text{left}\in\calV'_\text{left},\qquad 
v_\text{right}^\text{eig}=\R{a}{1\!-\!a} \R{1\!-\!a}{a} v'_\text{right},\  v'_\text{right}\in\calV'_\text{right}\qquad
\eea
are simultaneous (respectively left and right) eigenvectors of {\small $R\Big(\begin{matrix}0\\ 0\end{matrix}\Big)$} and 
{\small $R\Big(\begin{matrix}1\\ 1\end{matrix}\Big)$} satisfying the orthogonality
\bea
v_\text{left}^\text{eig}\cdot v_\text{right}^\text{eig}=0,\qquad 
v_\text{left}^\text{eig} \R{a}{1\!-\!a}=\vec 0,\qquad \R{a}{1\!-\!a} v_\text{right}^\text{eig}=\vec 0
\eea
It follows that the only nonzero matrix elements in (\ref{lDr}) are diagonal with $\vec a=\vec b$ and
\begin{align}
\big[\vec D(u)\vec D(u+\lambda)\big]_{\svec a,\svec a}= {}_6\langle \mbox{left}|\, \prod_{j=1}^N R\begin{pmatrix}a_j\\ a_j\end{pmatrix}_{\!\!6}
|\mbox{right}\rangle_6,\qquad a_j=0,1.
\end{align}
where {\small $R\Big(\begin{matrix}a\\ a\end{matrix}\Big)_{\!6}$} denotes the $6\times 6$ diagonal block of {\small $R\Big(\begin{matrix}a\\ a\end{matrix}\Big)$} and
\bea
{}_6\langle \mbox{left}|=(-1,1,1,-1,0,0),\qquad |\mbox{right}\rangle_6=(1,1,1,1,0,0)^T
\eea

Let us now suppose that $\vec a=\vec b=(0,0,0,\ldots,0,0)$ and observe that {\small $R\Big(\begin{matrix}0\\ 0\end{matrix}\Big)$} can be diagonalized by a similarity transformation
\bea
S^{-1}R\!\begin{pmatrix}0\\ 0\end{pmatrix}\!S\!=\!\mbox{\scriptsize$\begin{pmatrix}
s^4&0&0&0&0&0\\ 0&c^4&0&0&0&0\\ 0&0&s^2c^2&0&0&0\\ 
0&0&0&s^2c^2&0&0\\ 0&0&0&0&-s^2c^2&0\\ 0&0&0&0&0&-s^2c^2\end{pmatrix}$}
,\quad 
\setlength{\arraycolsep}{1.pt}
S\!=\!\mbox{\scriptsize $\begin{pmatrix}
 0 & 0 & -\frac{\left(z^4-1\right)^2 \left(z^4+1\right)}{4 z^8} & \frac{z^8-1}{2 z^8} & 0 & \frac{z^4+1}{z^4} \\[3pt]
 -\frac{\left(z^4+1\right)^2}{2 z^4} & 0 & -\frac{\left(z^4-1\right)^2}{2 z^4} & \frac{z^4-1}{z^4} & 0 & \frac{\left(z^4+1\right)^2}{2 z^4} \\[3pt]
 0 & -\frac{\left(z^4+1\right)^2}{2 z^4} & -\frac{\left(z^4-1\right)^2}{2 z^4} & \frac{z^4-1}{z^4} & 0 & \frac{\left(z^4+1\right)^2}{2 z^4} \\[3pt]
 -z^4-1 & -z^4-1 & 0 & \frac{z^8-1}{2 z^4} & 0 & z^4+1 \\[3pt]
 0 & 0 & 0 & 0 & 0 & \frac{z^8-1}{2 z^4} \\[3pt]
 \frac{z^8-1}{2 z^4} & \frac{z^8-1}{2 z^4} & \frac{z^8-1}{2 z^4} & 0 & \frac{z^8-1}{2 z^4} & 0 
\end{pmatrix}$}\qquad
\eea
with
\begin{align}
{}_6\langle \mbox{left}|S&=\mbox{\small $\Big(\frac{z^8-1}{2 z^4},\frac{z^8-1}{2 z^4},-\frac{\left(z^4-1\right)^2 \left(3 z^4-1\right)}{4 z^8},-\frac{\left(z^4-1\right)^3}{2 z^8},0,0\Big)$}\\[8pt] 
S^{-1}|\mbox{right}\rangle_6&=\mbox{\small $\Big(\frac{2 z^4 \left(z^4-1\right)}{\left(z^4+1\right)^3},\frac{2 z^4 \left(z^4-1\right)}{\left(z^4+1\right)^3},-\frac{4 z^4 \left(z^4-1\right)}{\left(z^4+1\right)^3},\frac{2 z^4 \left(5 z^8-2
   z^4+1\right)}{\left(z^4-1\right) \left(z^4+1\right)^3},0,0\Big)$}
\end{align}
Putting everything together, it follows that
\bea
 {}_6\langle \mbox{left}|\, R\Big(\begin{matrix}0\\ 0\end{matrix}\Big)_{\!6}^{\!N}\!
|\mbox{right}\rangle_6=-\tan^2 2u\big[c^{4N}-2(sc)^{2N}+s^{4N}\big]=-\tan^2 2u\, [c^{2N}-s^{2N}]^2
\eea

The last step is to extend this result to all the other diagonal segments. To do this let us define
\bea
\Delta R=\frac{2z^2}{1-z^4}\Big[R\Big(\begin{matrix}0\\ 0\end{matrix}\Big)-R\Big(\begin{matrix}1\\ 1\end{matrix}\Big)\Big]
\eea
We then find, using induction, that
\bea
\prod_{j=1}^{N-1} R\Big(\begin{matrix}a_j\\ a_j\end{matrix}\Big)\, \Delta R\,|\mbox{right}\rangle_6=(-s^2 c^2)^{N-1} (z^{-2},\cos 2u,\cos 2u,z^2,i\sin 2u,i\sin 2u)^T,\quad a_j=0,1\qquad
\eea
It follows that 
\bea
{}_6\langle \mbox{left}|\prod_{j=1}^{N-1} R\Big(\begin{matrix}a_j\\ a_j\end{matrix}\Big)\, \Delta R\,|\mbox{right}\rangle_6=0
\eea
So the weight of the diagonal matrix elements with $\vec b=\vec a$ are independent of $\vec a$ 
\bea
\big[\vec D(u)\vec D(u+\lambda)\big]_{\svec a,\svec a}= {}_6\langle \mbox{left}|\, \prod_{j=1}^N R\Big(\begin{matrix}a_j\\ a_j\end{matrix}\Big)_{\!\!6}\,
|\mbox{right}\rangle_6=-\tan^2 2u\, [c^{2N}-s^{2N}]^2
\eea

\setlength{\arraycolsep}{2.8pt}

\def\dcol{\facegridyellow{1}{2}}
\def\facegridyellow#1#2{
\psframe[fillstyle=solid,fillcolor=yellow,linewidth=0pt](#1,#2)
\psgrid[gridlabels=0pt,subgriddiv=1](0,0)(#1,#2)}
\def\younga#1{
\begin{pspicture}[shift=-3](6,3)
\pspolygon[fillstyle=solid,fillcolor=lightgray](3,0)(4,1)(\numexpr 4-#1\relax,\numexpr 1+#1\relax)(\numexpr 3-#1\relax ,\numexpr #1\relax)(3,0)
\psline[linewidth=1pt,linestyle=dotted](0,3)(1,2)
\psline[linewidth=1pt,linestyle=dotted](5,2)(6,3)
\psline[linewidth=1pt](1,2)(3,0)
\psline[linewidth=1pt](3,0)(5,2)
\psline[linewidth=1pt](2,1)(4,3)
\psline[linewidth=1pt](3,2)(4,1)
\psline[linewidth=1pt](4,3)(5,2)
\end{pspicture}}
\def\youngb#1{
\begin{pspicture}[shift=-3](6,3)
\pspolygon[fillstyle=solid,fillcolor=lightgray](3,0)(4,1)(\numexpr 4-#1\relax,\numexpr 1+#1\relax)(\numexpr 3-#1\relax ,\numexpr #1\relax)(3,0)
\psline[linewidth=1pt,linestyle=dotted](0,3)(1,2)
\psline[linewidth=1pt,linestyle=dotted](5,2)(6,3)
\psline[linewidth=1pt](1,2)(3,0)
\psline[linewidth=1pt](3,0)(5,2)
\psline[linewidth=1pt](2,1)(4,3)
\psline[linewidth=1pt](2,3)(4,1)
\psline[linewidth=1pt](4,3)(5,2)
\psline[linewidth=1pt](1,2)(2,3)
\end{pspicture}}
\def\youngc#1{
\begin{pspicture}[shift=-3](6,3)
\pspolygon[fillstyle=solid,fillcolor=lightgray](3,0)(4,1)(\numexpr 4-#1\relax,\numexpr 1+#1\relax)(\numexpr 3-#1\relax ,\numexpr #1\relax)(3,0)
\psline[linewidth=1pt,linestyle=dotted](0,3)(1,2)
\psline[linewidth=1pt,linestyle=dotted](5,2)(6,3)
\psline[linewidth=1pt](1,2)(3,0)
\psline[linewidth=1pt](3,0)(5,2)
\psline[linewidth=1pt](2,1)(4,3)
\psline[linewidth=1pt](2,3)(4,1)
\psline[linewidth=1pt](3,4)(5,2)
\psline[linewidth=1pt](1,2)(3,4)
\end{pspicture}}

\section{Skew $q$-Binomials}
\label{SkewqBinom}
The skew $q$-binomials, related to generalized $q$-Narayana numbers (\ref{qNara}), are \cite{PR2007,PR2007b}
\bea
\begin{array}{rl}
\sbin{M}{m,n}_q\!\!\!\!&=\;{\gauss{M}{m}}{\gauss{M}{n}}-q^{n-m+1}{\gauss{M}{m-1}}{\gauss{M}{n+1}}\\[12pt]
&=\;q^{-M+n}
  \Big({\gauss{M}{m}}{\gauss{M+1}{n+1}}-{\gauss{M+1}{m}}{\gauss{M}{n+1}}\Big),\qquad 0\le m\le n\le M
  \end{array}
\eea
At $q=1$, the skew binomials {\scriptsize $\sbin{M}{m,m}_{q=1}$} are determinants of ordinary binomials
\bea
\begin{array}{c}
\mbox{\color{blue}Binomials}\\
\begin{array}{cccccc}
1\;&&&&&\\
1\;&1&&&\\
1\;&2&1&&&\\
1\;&3&3&1&&\\
1\;&\raisebox{-3pt}{\color{yellow}\rule{12pt}{12pt}}\hspace{-9pt}4&\raisebox{-3pt}{\color{yellow}\rule{12pt}{12pt}}\hspace{-9pt}6&4&1&\\
1\;&\raisebox{-3pt}{\color{yellow}\rule{12pt}{12pt}}\hspace{-9pt}5&\raisebox{-3pt}{\color{yellow}\rule{16pt}{12pt}}\hspace{-14pt}10&10&5&1
\end{array}
\end{array}
\hspace{.5in}
\begin{array}{c}
\mbox{\color{blue}Skew Binomials $(n=m)$}\\
\begin{array}{cccccc}
1&&&&&\\
1&1&&&\\
1&3&1&&&\\
1&6&6&1&&\\
1&\raisebox{-3pt}{\color{yellow}\rule{16pt}{12pt}}\hspace{-14pt}10&20&10&1&\\
1&15&50&50&15&1
\end{array}
\end{array}
\hspace{.4in}
\begin{array}{c}
\mbox{\color{blue}Catalan}\\
\begin{array}{c}
1\\
2\\
5\\
14\\
42\\
132
\end{array}
\end{array}
\eea
\vspace{-.2in}
The skew $q$-binomials are enumerated by double column diagrams with dominance

\psset{unit=.5cm}
\bea
\begin{pspicture}(0,-1)(9,7.7)
\psset{unit=.7cm}
\multirput(0,0)(2,0){5}{\dcol}
\multirput(2,2.5)(2,0){3}{\dcol}
\psline[linecolor=red,linewidth=1.5pt](0,0)(1,1)
\psline[linecolor=red,linewidth=1.5pt](2,0)(3,2)
\psline[linecolor=red,linewidth=1.5pt](4,0)(5,2)
\psline[linecolor=red,linewidth=1.5pt](6,2)(7,2)
\psline[linecolor=red,linewidth=1.5pt](8,2)(9,2)
\rput(0,-.5){\psline[linecolor=red,linewidth=1.5pt](2,4)(3,4)
\psline[linecolor=red,linewidth=1.5pt](4,4)(5,5)
\psline[linecolor=red,linewidth=1.5pt](6,4)(7,5)}
\pscircle*(0,0){.12}
\pscircle*(1,0){.12}
\pscircle*(1,1){.12}
\pscircle*(2,0){.12}
\pscircle*(3,0){.12}
\pscircle*(3,2){.12}
\rput(0,-.5){\pscircle*(2,4){.12}
\pscircle*(3,3){.12}
\pscircle*(3,4){.12}}
\pscircle*(4,0){.12}
\pscircle*(5,1){.12}
\pscircle*(5,2){.12}
\rput(0,-.5){\pscircle*(4,4){.12}
\pscircle*(5,3){.12}
\pscircle*(5,5){.12}}
\pscircle*(6,2){.12}
\pscircle*(7,0){.12}
\pscircle*(7,2){.12}
\rput(0,-.5){\pscircle*(6,4){.12}
\pscircle*(7,4){.12}
\pscircle*(7,5){.12}}
\pscircle*(8,2){.12}
\pscircle*(9,1){.12}
\pscircle*(9,2){.12}
\multirput(1.5,-.5)(2,0){4}{$+$}
\rput(.5,-.5){$1$}
\rput(2.5,-.5){$2q$}
\rput(4.5,-.5){$2q^2$}
\rput(6.5,-.5){$2q^3$}
\rput(8.5,-.5){$q^4$}
\rput(9.5,-.5){$=$}
\rput(10.8,-.5){$\sbin{3}{1,2}_q$}
\end{pspicture}
\eea

A partition $\lambda$ is equivalent to a Young diagram $Y$. A skew Young diagram $Y_2/Y_1$ is equivalent to the pair $(Y_1,Y_2)$ with $Y_1\subseteq Y_2$. Let us define
\bea
{\begin{array}{c}
E(Y)=\mbox{Energy}=\{\mbox{\# of boxes in the Young digram $Y$}\}\\[2pt]
Y_{m,n}=\{\mbox{$m\times n$ rectangular Young diagram}\}
\end{array}}
\eea
A skew $q$-binomial can be written as an energy weighted sum over skew Young diagrams
\bea
{\Big\{{M\atop {m,n}}\Big\}_q\rule{0pt}{0pt}=\;q^{(m-n)n}\hspace{-20pt}\sum_{{Y_1\subseteq Y_2}\atop{\emptyset\subseteq Y_1\rule{0pt}{0pt}\subseteq Y_{M-m,m}\atop Y_{n-m,n}\subseteq Y_2\subseteq Y_{M-m,n}}} \hspace{-20pt}q^{E(Y_1)+E(Y_2)},\qquad 0\le m\le n\le M}
\eea
The bijection is implemented by interpreting the left and right column (particle) configurations in the double column diagrams as Maya diagrams and using the standard bijection between Maya diagrams and Young diagrams. For example, shading $Y_1$, gives\\
\psset{unit=.3cm}
\setlength{\unitlength=.3cm}
\bea
\Big\{{3\atop 1,2}\Big\}_q=\ q^{-2}\left[\;\younga{0}
\ \younga{1} \hspace{-6.5\unitlength}\raisebox{4\unitlength}{\youngb{0}}
\ \youngb{1} \hspace{-6.5\unitlength}\raisebox{4\unitlength}{\youngc{0}}
\ \youngb{2} \hspace{-6.5\unitlength}\raisebox{4\unitlength}{\youngc{1}}
\ \youngc{2}\;\right],
\quad {\emptyset\subseteq Y_1\subseteq Y_{2,1}\atop Y_{1,2}\rule{0pt}{20pt}\subseteq Y_2\subseteq Y_{2,2}}
\eea
$
\hspace{4cm}1\hspace{.5cm}+\hspace{.5cm}2q\hspace{.5cm}+\hspace{.5cm}2q^2\hspace{.45cm}+\hspace{.45cm}2q^3\hspace{.5cm}+\hspace{.5cm}q^4$\hfill

\goodbreak 


\end{document}